\documentclass[12pt]{article}

\usepackage{epsf,epsfig}
\usepackage{graphics}
\usepackage{bbold}
\usepackage{dsfont}
\usepackage{cite}
\usepackage{slashed}
\usepackage{amsmath,amssymb,booktabs}
\def\slash#1{#1 \hskip-0.45em /}

\def\be{\begin{equation}}
\def\ee{\end{equation}}
\def\bea{\begin{eqnarray}}
\def\eea{\end{eqnarray}}

\def\rd{d}
\def\nn{\nonumber}

\def\slash#1{#1 \hskip-0.45em /}

\begin{document}
\thispagestyle{empty}

\begin{flushright}
{\small
TUM-HEP-846/12\\
TTK-12-29\\
SFB/CPP-12-45\\
1209.5897 [hep-ph]\\
JHEP08(2013)010\\[0.1cm]
August 5, 2013}
\end{flushright}

\vspace{0.8cm}
\begin{center}
\Large\bf\boldmath
The muon anomalous magnetic moment in the Randall-Sundrum model
\unboldmath
\end{center}

\vspace{1cm}
\begin{center}
{\sc M.~Beneke$^{a,b}$},
{\sc P.~Dey$^{a,b}$} and 
{\sc J.~Rohrwild$^{a,b}$}\\[5mm]
{\sl ${}^a$Physik Department T31,\\
James-Franck-Stra\ss e 1, Technische Universit\"at M\"unchen,\\
D--85748 Garching, Germany\\
\vspace{0.3cm}
${}^b$Institut f\"ur Theoretische Teilchenphysik und 
Kosmologie,\\
RWTH Aachen University, D--52056 Aachen, Germany}\\[0.5cm]
\end{center}

\vspace{1cm}
\begin{abstract}
\vspace{0.2cm}\noindent
We calculate the anomalous magnetic 
moment of the muon in the minimal Ran\-dall-Sundrum model with  
standard model fields in five-dimensional (5D) warped space and a 
brane-localized Higgs. We use a fully 5D framework to 
compute the one-loop matching coefficients of the effective theory 
at the electroweak scale. The extra contribution to the anomalous 
magnetic moment from the model-independent gauge-boson exchange 
contributions is 
\[
\Delta a_\mu \approx 8.8\cdot 10^{-11}\,\times (1\,\mbox{TeV}/T)^2,
\] 
where $1/T$ denotes the location of the TeV brane in conformal 
coordinates, and is  related to the mass of the lowest gauge boson 
KK excitation by $M_{\rm KK}\approx 2.5\,T$.  The result constitutes the first complete 
determination of the gauge-boson contribution to $g-2$
and is robust against the variation of the bulk fermion masses 
and 5D Yukawa coupling. We also determine the strongly 
model-parameter dependent effect of Higgs-exchange diagrams.
\end{abstract}

\newpage
\setcounter{page}{1}

\newpage
\allowdisplaybreaks

%%%%%%%%%%%%%%%%%%%%%%%%%%%%%%%%%%%%%%%%%%%%%%%%%%%%%%%%%%%%%%%%%%%%%%%%%

\section{Introduction}
\label{sec:intro}

The idea \cite{Randall:1999ee} that our four-dimensional world is one
of the boundaries of a slice of strongly curved five-dimensional (5D) 
Anti-de-Sitter space provides an attractive approach to the
gauge-gravity and flavour hierarchy problems of the Standard Model
(SM). In conformal coordinates the metric of the 5D bulk is 
\begin{equation}
ds^2 = \left(\frac{1}{k z}\right)^2\left(\eta_{\mu\nu} dx^\mu dx^\nu-
dz^2\right), 
\label{metric}
\end{equation}
where $k\sim M_{\rm Pl} \sim 10^{19}\,$GeV is of order of the Planck 
scale $M_{\rm Pl}$, 
while the metric on the four-dimensional boundaries located at 
$z=1/k$ and $z=1/T$ is flat. This set-up provides a solution to the 
gauge-gravity hierarchy problem, since, if the proper 
distance between the two branes, $1/k\times\ln(k/T)$, is only a few 
times the Planck length, then $T$ can still be of order of the TeV scale. 
The original proposal \cite{Randall:1999ee} considered only gravity  
propagating in the bulk, but it was soon realized that the SM fields 
may be 5D, too~\cite{Davoudiasl:1999tf, Pomarol:1999ad,
  Grossman:1999ra,hepph:9912498,Gherghetta:2000qt}. An exception is
the Higgs field, which in the 
simplest set-up 
is required to be localized on the TeV brane $z=1/T$ to solve the 
hierarchy problem. Allowing the SM fermions to propagate into the 
fifth dimension opens up the possibility to explain the quark and 
lepton mass, and the quark mixing-angle hierarchies  
\cite{Grossman:1999ra,Gherghetta:2000qt,Huber:2000ie} 
without postulating large hierarchies or small coupling 
constants in the 5D theory.

The model predicts Kaluza-Klein (KK) excitations of the SM particles 
with typical KK mass $M_{\rm KK}\sim 2.5 \,T$ for the first excitation. 
The collider physics and flavour phenomenology of these KK excitations 
has been extensively studied. Direct searches at the Large Hadron
Collider lead to lower limits on $M_{\rm KK}$ similar to those 
for Drell-Yan like 
single production of heavy resonances, which are now 
in the 1 to 2 TeV region. Indirect constraints from electroweak 
precision observables and flavour require higher KK masses, 
but may be avoided by adding custodial protection 
or flavour structure. The overwhelming number of studies is 
based on investigating the implications of tree-level 
production or tree-level exchange of the KK excitations.

Comparatively little is known about the Randall-Sundrum (RS) model at 
the loop level due to the intricacies of quantum field theory in curved 
space-times. It appears that contrary to 5D theories with a flat
compact extra dimension, the running of gauge couplings is only 
logarithmic \cite{Pomarol:2000hp,Randall:2001gb,Agashe:2002bx,Goldberger:2003mi}. 
Divergences associated with the fifth dimension show up in 
divergent KK sums. The interpretation of these divergences is not 
as straightforward as in Minkowski space. A precise regularization 
scheme, required to be able to calculate finite renormalizations, 
has not yet been developed. At the same time, there are a number of 
processes of much interest for searching for deviations of 
the SM, which arise only at the loop level; among them are the 
anomalous magnetic moment of the muon, and electromagnetic dipole 
transitions in the lepton ($\mu\to e\gamma$) and the quark 
sectors ($b\to s\gamma$), which have been considered in the 
RS model in \cite{Davoudiasl:2000my} and 
\cite{hepph:0408134,Agashe:2006iy,Csaki:2010aj,Blanke:2012tv,Delaunay:2012cz}, 
respectively. Higgs production through gluon-gluon fusion has been discussed in 
\cite{Casagrande:2010si,Azatov:2010pf,Carena:2012fk}.

The systematics of renormalization and the calculation of 
loop processes is probably easier to approach from the 
5D perspective \cite{Randall:2001gb,Csaki:2010aj} instead of 
the KK decomposition of the dependence on the fifth coordinate. In 
this paper we set up the propagator and vertex Feynman rules 
for the 5D quantum field theory in a slice of Anti-de-Sitter (AdS) 
space (in this 
part there is significant overlap with \cite{Csaki:2010aj}) 
and then revisit the muon anomalous magnetic moment as our first 
application.  There is only one previous calculation of $(g-2)_\mu$ in 
the RS model \cite{Davoudiasl:2000my}, which is based on summing 
over towers of KK excitations. Since  $(g-2)_\mu$ is not 
ultraviolet (UV) sensitive, it is well suited to demonstrate 
the workings of the 5D formalism without the complications of 
divergences from the Planck scale. The main result of the present 
paper is a complete calculation of $(g-2)_\mu$, since, as will 
be explained, our result differs considerably 
from  \cite{Davoudiasl:2000my}.

The outline of the paper is as follows. In Sec.~\ref{sec:thsetup} 
we set up notations by defining the 5D theory and explain our 
strategy for calculating the anomalous magnetic moment. We match 
the 5D theory to an effective 4D theory, in which the scale 
$T$ is integrated out, keeping the relevant SU(2)$\times$U(1) 
invariant dimension-6 operators
that must be added to the SM. After expressing the Lagrangian 
in terms of the fields in the electroweak vacuum, we 
calculate the non-standard contribution to  $(g-2)_\mu$ 
in the 4D effective theory. The matching coefficients of the 
dimension-6 operators are computed in Sec.~\ref{sec:matching}.  
We work with 5D fields and avoid multiple sums over KK modes 
by integrating out directly the fifth dimension. 
We identify two relevant contributions, related to 
the electromagnetic dipole transition and a four-fermion operator.  
The actual integrations must be performed numerically. 
We focus on the contributions to the muon anomalous magnetic moment
from gauge-boson exchange. At the end of Sec.~\ref{sec:matching} 
we also consider genuine Higgs-exchange diagrams, whose 
calculation requires taking an appropriate limit of 
a model with a slightly brane-delocalized Higgs field. 
We argue that in models with anarchic Yukawa matrices 
Higgs-exchange is always small relative to gauge-boson exchange 
due to constraints from lepton-flavour-violating observables.
The numerical result for the gauge-boson contributions 
to  $(g-2)_\mu$ 
in the single-flavour approximation is shown 
in Sec.~\ref{sec:result}. We also discuss the dependence of the 
result on the bulk mass and 5D Yukawa coupling and the generalization 
to the case with three lepton flavours, and we compare our result   
to previous ones. We conclude in Sec.~\ref{sec:conclude}. 
The Feynman rules of the 5D theory are summarized in 
Appendices~\ref{ap:5Drules} and \ref{App:Expansions}, and in 
Appendix~\ref{ap:diags} we provide the explicit expressions 
of all diagrams for reference.

%%%%%%%%%%%%%%%%%%%%%%%%%%%%%%%%%%%%%%%%%%%%%%%%%%%%%%%%%%%%%%%%%%%%%%%%%

\section{\boldmath 
From the 5D theory to $(g-2)_\mu$}
\label{sec:thsetup}

In this section we define the 5D theory, and then discuss the 
4D effective Lagrangian at scales below $T$ relevant to electroweak 
and flavour physics when $M_{\rm EW}\ll T$. We then calculate the 
muon anomalous magnetic moment 
\begin{equation}
a_\mu = \frac{g_\mu-2}{2} = a_\mu^{\rm SM} + \Delta a_\mu,
\end{equation}
in the effective theory, that is, 
we express the non-standard contributions $\Delta a_\mu$ in terms of the 
matching coefficients of higher-dimension operators in the 
effective Lagrangian.

\subsection{The 5D theory}

We adopt the simplest set-up for the SM in the RS geometry
(\ref{metric}), in which all fields except the Higgs doublet 
can propagate in the bulk, and no further structure is added. 
While not the most attractive model for phenomenology since the 
scale $T$ must be quite large due to the lack of custodial 
protection, it provides the appropriate starting point for the 
study of loop processes. 
Quarks and the entire strong interaction sector are not 
relevant to leptonic processes at one-loop. In the following 
we specify only the SU(2)$\times$U(1) interactions and leave out the
quark sector.

The 5D theory is defined by the action 
\begin{eqnarray}
S_{\rm (5D)} &=& \int d^4x \int\nolimits_{1/k}^{1/T} dz\,\sqrt{G}\;
\Bigg\{ \!- \frac{1}{4} F^{MN}F_{MN} -\frac{1}{4} W^{a,MN}W_{MN}^a 
\nonumber\\
&& \hspace*{0cm}
+\,
\sum_{\psi=E,L} \left(e^M_m \left[ \frac{i}{2} \, \bar\psi_i \,\Gamma^m 
(D_M-\overleftarrow{D}_M) \psi_i\right] - 
M_{\psi_i} \bar{\psi}_i \psi_i \right)
\Bigg\} + \,S_{\rm GF+ ghost}
\nonumber  \\ 
&&\hspace*{0cm}
+\,\int d^4x \,
 \bigg\{(D^{\mu} \Phi)^\dagger D_\mu\Phi - V(\Phi) 
 - \left(\frac{T}{k}\right)^{\!3} 
 \left[y_{ij}^{\rm (5D)} (\bar L_i\Phi) E_j + \mbox{h.c.}\right]\bigg\},
\label{5Daction}
\end{eqnarray}
where $G = 1/(k z)^{10} $ is the determinant of the 5D metric  and 
$e^M_m = k z \,\mbox{diag}\,(1,1,1,1,1)$ the inverse 
vielbein.\footnote{The spin connection term in the fermion action 
does not contribute and has been dropped. The gauge-fixing action 
is discussed in Appendix~\ref{ap:5Drules}.} 
The potential is given by
\begin{equation}
V(\Phi) = - \mu^2_{\rm (5D)}  \left(\frac{T}{k}\right)^{\!2} 
\Phi^\dagger\Phi+\frac{\lambda}{4}  \,(\Phi^\dagger\Phi)^2\,.
\end{equation} 
The covariant derivative is defined as usual by 
\begin{equation}
D_M=\partial_M-i g_5^\prime \frac{Y}{2} B_M-ig_5 T^a W^a_M.
\end{equation}
$Y$ is the hypercharge of the field and $T^a$ the SU(2) 
generator in the given field representation 
($\tau^a/2$ with $\tau^a$ the Pauli matrices for the
doublets $L_i$, $\Phi$, and 0 for the singlets $E_i$). The  
hypercharge and SU(2) field strengths read
\begin{equation}
F_{NM}=\partial_N B_M- \partial_M B_N 
\qquad W_{NM}^a=\partial_N W_M^a- \partial_M W_N^a +
g_5 \epsilon^{abc} W_N^b W_M^c \,.
\end{equation}
The Higgs doublet is localized at $z=1/T$ and described by 
a four-dimensional field. The powers of $(T/k)$ in the Higgs action
arise from making the kinetic term canonical when the $z$-integration 
in the original 5D action is eliminated using the $\delta(z-1/T)$ 
factor. The 5D lepton fields in the lepton-Higgs Yukawa 
interaction are therefore evaluated at $z=1/T$.  Latin indices 
from the middle of the alphabet refer to lepton flavour. 

The dimensionful parameters 
$g_5, g^\prime_5, M_{\psi_i}, y_{ij}^{\rm (5D)}$ 
of the Lagrangian are all of order $M_{\rm Pl}$. It is conventional 
to introduce the dimensionless quantities $c_{\psi_i}=M_{\psi_i}/k$ 
for the bulk fermion masses, 
whose precise values determine the hierarchical masses of the 
known leptons. The 5D action is given in the flavour basis that 
renders the bulk mass matrix diagonal, which can always be arranged by
a field rotation. It is not possible in general to simultaneously 
diagonalize the 5D Yukawa coupling matrix $y_{ij}^{\rm (5D)}$. 
Unlike the SM, the 5D theory violates lepton flavour conservation 
and CP-symmetry  in the lepton sector.

The Feynman rules for the Higgs propagator and vertices involving 
the Higgs field can be read off directly from (\ref{5Daction}) in the 
standard way. The rules for the 5D fields in the bulk of AdS 
are discussed and summarized in Appendix~\ref{ap:5Drules}.

\subsection{Integrating out the KK scale}
\label{sec:match}

We assume that the scale $T$ is much larger than the masses of 
the SM electroweak gauge bosons and the Higgs boson. In this case 
we may integrate out the fifth dimension, that is, the towers of 
KK states to obtain an effective four-dimensional theory. Since 
the states with masses below $T$ are precisely represented by the 
SM fields, the four-dimensional theory is the SM plus 
SU(3)$\times$SU(2)$\times$U(1) invariant higher-dimension 
operators. 

An issue that arises here is that the 5D theory is non-renormalizable 
and must itself be defined as an effective theory below a scale 
$\Lambda$ that should be at least a few times the Planck scale. It 
is generally believed that in the mixed representation the 
four-dimensional loop momenta should be cut-off at a value 
$\Lambda(z)$ that depends on the position $z$ in the fifth dimension. 
If $\Lambda(1/k)$ is a few times the Planck scale, then the cut-off 
$\Lambda(1/T)$ relevant to processes dominated by physics near the 
TeV brane should be near the TeV scale. This would render the 
naive application of the 5D formalism problematic, since it encodes 
the sum over all KK states rather than including only the few 
below the cut-off $\Lambda(1/T)$. However, for a finite quantity 
such as the anomalous magnetic moment of the muon, the KK sum 
must converge, and the effect of including the entire tower 
relative to the truncation is of order $T^2/\Lambda(1/T)^2$, which 
is the generic size of corrections expected from the UV completion 
of the RS model.

The dominant non-standard effects in the expansion in $M_{\rm EW}/T$ are 
associated with dimension-6 operators. We write 
\begin{equation}
{\cal L}_{\rm RS}^{\rm (5D)} \quad\longrightarrow \quad 
{\cal L}_{\rm eff} = {\cal L}_{\rm SM} + \frac{1}{T^2} 
\sum_i c_i {\cal O}_i.
\end{equation}
Since the matching coefficients $c_i$ are dominated by 
distances $\ll 1/T$, the Higgs bilinear term in $V(\Phi)$ can be 
treated as a perturbation, and the $c_i$ can be computed in the 
theory with unbroken electroweak gauge symmetry.

The effective Lagrangian is a special case of the general 
SU(3)$\times$SU(2)$\times$U(1) invariant theory with SM field 
content including dim-6 operators~\cite{CERN-TH-4254/85}. 
Since the anomalous magnetic moment is generated only at 
the one-loop level, we distinguish two types of 
operators.

\vskip0.3cm
{\em Class-0} --- operators in ${\cal L}_{\rm eff}$ that contribute
to $a_\mu$ at tree-level, but whose short-distance coefficients are 
generated in the RS model only at the one-loop order. The only relevant 
terms are 
\begin{equation} 
\sum_i c_i {\cal O}_i{}_{|\,\rm class-0} =
a_{B,ij} \,\bar L_i\Phi \sigma_{\mu\nu} E_j B^{\mu\nu} + 
a_{W,ij} \,\bar L_i\tau^a \Phi \sigma_{\mu\nu} E_j \,W^{a,\mu\nu} + 
\mbox{h.c.}
\label{dim60}
\end{equation}
Note that we use the same notation for the 5D and 4D fermion and 
gauge fields. It is understood that the fields in the 4D effective
Lagrangian correspond to the zero modes in the KK decomposition 
of the 5D fields. The general effective Lagrangian also 
contains the chirality changing operators $(\bar L D_\mu E)D^\mu\Phi$ and 
$(\bar L \overleftarrow{D}_\mu E)D^\mu\Phi$, but they  do not generate 
anomalous magnetic couplings of the photon at tree level.
After replacing $\Phi$ by its vacuum expectation value, the 
operators are $(\bar L_2 \partial_\mu E) [D^\mu]_{22}$ and 
$(\bar L_2 \overleftarrow{\partial}_\mu E) [D^\mu]_{22}$. 
Since $[D^\mu]_{22}$ contains only the $Z$-boson, this contributes 
to the anomalous magnetic couplings of the $Z$-boson but not the 
photon. The dimensionless coefficients $a_{B,W}$ are of order 
$g_5^3 y_{ij}^{\rm (5D)} k^{5/2}$, where here $g_5$ denotes a generic 
5D gauge coupling. Since the anomalous magnetic moment is 
generated only at one loop, we may use the tree-level relation 
\begin{equation}
\frac{1}{g^2(\mu)} = \frac{\ln(k/T)}{g_5^2 k}
\end{equation}
between the 5D coupling and the 4D coupling at a scale $\mu$ of 
order $T$. Corrections to this relation are of order 
$1/(4\pi^2)$, which counts a higher-order effect in the weak-coupling 
expansion.
\vskip0.3cm
{\em Class-1} --- operators in ${\cal L}_{\rm eff}$ that contribute
to $a_\mu$ only at the one-loop level, but which are generated in the 
RS model at tree-level. The relevant terms are
\begin{eqnarray}
\sum_i c_i {\cal O}_i{}_{|\,\rm class-1} &=& 
b_{ij} \,(\bar L_i\gamma^\mu L_i)(\bar E_j\gamma_\mu E_j) 
+ c_{1,i} \,(\bar E_i\gamma_\mu E_i) (\Phi^\dagger iD^\mu \Phi)
\nonumber\\[0.0cm]
&& + \,c_{2,i} \,(\bar L_i\gamma_\mu L_i) (\Phi^\dagger iD^\mu \Phi)
+ c_{3,i} \,(\bar L_i\gamma^\mu \tau^a L_i) (
\Phi^\dagger \overleftrightarrow{i \tau^a D_\mu} \Phi)\;,
\label{dim61}
\end{eqnarray}
where $\overleftrightarrow{i \tau^a D_\mu}=1/2 \, 
(i \tau^a D_\mu - i \overleftarrow D_\mu \tau^a)$.
They are generated at tree-level by the exchange of the 4D vector 
components of the KK gauge bosons.  Although the 5D gauge couplings 
to fermions are flavour-diagonal and universal, flavour-dependence 
of the coefficient functions arises through the different zero-mode 
profiles of the 4D fermions.  The operators 
$(\bar L_i D_\mu E_j)D^\mu\Phi$ and 
$(\bar L_i \overleftarrow{D}_\mu E_j)D^\mu\Phi$ discussed 
already above might contribute to the photon anomalous magnetic 
moment at the one-loop level, but inspection shows that it cannot 
be generated at tree-level from integrating out the KK scale.
Another class-1 operator that could possibly contribute to $g-2$ is 
$\bar L_i \Phi E_j \Phi^\dagger \Phi$ (and its hermitian conjugate).
It can only be generated at tree level by diagrams with three Yukawa 
couplings. 
In scenarios with an (approximately) brane localized Higgs field 
these diagrams are non-vanishing only when one includes the 
``wrong-chirality'' Higgs couplings of the fermions \cite{Azatov:2009na}.  
For simple Higgs profiles these terms can be 
calculated fully analytically in the limit that the width of the 
Higgs profile around the TeV brane is taken to zero. We 
will discuss these terms separately in Sec.~\ref{subsec:Higgs} 
and therefore do not include $\bar L_i \Phi E_j \Phi^\dagger \Phi$ 
in (\ref{dim61}).
KK gauge boson exchange also generates further 
four-fermion operators with only left-handed and only right-handed
fields, but these do not contribute to the anomalous magnetic moment. 
The mixed left-right lepton-quark operator $(\bar L E) (\bar Q U)$, 
on the other hand, does contribute at one loop, 
but it is not generated in the RS model. Other class-1 operators 
that one might write down contribute to $a_\mu$ only indirectly, 
through a modification of the relation of the SM parameters to 
those of the 5D theory. When expressing the SM contribution to $a_\mu$
in terms of SM parameters, this effect is already 
accounted for.  Thus (\ref{dim61}) represents the complete list of relevant 
operators. The coefficient functions $b, c_{1,2,3}$ are of order 
$g_5^2 k$. In particular, we note the presence of
chirality-preserving fermion-Higgs interactions proportional 
to 5D gauge rather than Yukawa couplings. 

\vskip0.3cm
We now express the effective Lagrangian in terms of the fields 
in the electroweak vacuum, in which the SU(2)$\times$U(1) gauge 
symmetry is spontaneously broken to electromagnetism. 
Thus,  we replace
\begin{equation}
\Phi\to \left(
\begin{array}{c} \phi^+ \\\frac{1}{\sqrt{2}}  (v+H+iG) \end{array}\right),
\end{equation}
and
\begin{equation}
D_\mu\to \partial_\mu - ie Q A_\mu -i\frac{g}{c_W}(T^3-s^2_W Q)Z_\mu
-i\frac{g}{\sqrt{2}}(T^1+iT^2)W^+_\mu 
-i\frac{g}{\sqrt{2}}(T^1-iT^2)W^-_\mu,
\label{rep1}
\end{equation}
where $c_W\,(s_W)$ is the cosine (sine) of the Weinberg angle, with 
$c_W^2 = g^2/(g^2+g^{\prime\,2})$. For the fermions it is 
convenient to transform to the mass basis, in which the effective 
Yukawa interaction $-y_{ij} \bar L_i\Phi E_j$ 
of the 4D fields is diagonal. Tree-level 
matching results in the 4D Yukawa coupling
\begin{eqnarray}
y_{ij} &=& \left(\frac{T}{k}\right)^{\!3} f^{(0)}_{L_i}(1/T)  
g^{(0)}_{E_j}(1/T) \,y_{ij}^{\rm (5D)}  
\nonumber\\
&=& \sqrt{\frac{1-2 c_{L_i}}{1-(T/k)^{1-2 c_{L_i}}}}
\,\sqrt{\frac{1+2 c_{E_j}}{1-(T/k)^{1+2 c_{E_j}}}}
\,y_{ij}^{\rm (5D)}k\,
\label{4dyukawa}
\end{eqnarray}
where $f^{(0)}_{L_i}(z)$ and $g^{(0)}_{E_j}(z)$ are the fermion zero-mode profiles, 
see appendix~\ref{ap:5Drules}. The transition to the mass basis 
is given by the unitary rotations 
\begin{equation}
\label{eq:changeBasis}
L_i \to U_{ij} L_j, \qquad 
E_i \to V_{ij} E_j
\end{equation}
of the 4D fields in ${\cal L}_{\rm eff}$.  To go to the broken theory,
\begin{equation}
E_i \to V_{ij} P_R \psi_j,\qquad
L_i \to U_{ij}  P_L  \!\left(
\begin{array}{c} \nu_j \\ \psi_j \end{array}\right),
\end{equation}
where $\psi_i$ is the Dirac spinor field for the massive 
leptons ($i=1,2,3$ corresponding to electron, muon, tau)
and $\nu_i$ is the corresponding neutrino spinor field. 
$P_{L/R} = (1\mp\gamma_5)/2$ are the chiral projectors.
Inserting these substitutions into (\ref{dim60}), (\ref{dim61}),
generates numerous different operators composed of 
SM fields. We neglect operators which cannot contribute 
to the magnetic and electric dipole form factors or only start 
contributing at two loops.    
The effective Lagrangian for the computation of the 
anomalous magnetic (and electric) moments, 
as well as charged lepton-flavour violating processes at 
one loop is then given by
\begin{eqnarray}
\sum_i c_i {\cal O}_i &\to&  
\frac{\alpha_{ij}+\alpha_{ji}^*}{2} \frac{v}{\sqrt{2}} \,
\bar\psi_i\sigma_{\mu\nu} \psi_j F^{\mu\nu} + 
\frac{\alpha_{ij}-\alpha_{ji}^*}{2 i}\,\frac{v}{\sqrt{2}} \,
\bar\psi_i\sigma_{\mu\nu} i\gamma_5 \psi_j F^{\mu\nu}
\nonumber\\[-0.15cm]
&& + \,\beta_{ijkl} \, (\bar\psi_i\gamma^\mu P_L \psi_j) 
(\bar \psi_k\gamma_\mu P_R\psi_l) 
\nonumber\\[0.05cm]
&& + \,\gamma_{1,ij} \,\frac{v}{2} \,
(\bar\psi_i P_L \gamma_\mu \psi_j) (i\partial^\mu H)
+ [\gamma_{2,ij}+\gamma_{3,ij}] \,\frac{v}{2} \,
(\bar\psi_i P_R \gamma_\mu \psi_j) (i\partial^\mu H)
\nonumber \\[0.05cm]
&& +\,\gamma_{3,ij}\,\frac{v}{\sqrt{2}}(\bar\psi_i P_R \gamma_\mu \nu_j) (-i\partial^\mu \phi^-)
   +\gamma_{3,ij}\,\frac{v}{\sqrt{2}}(\bar\psi_i P_R \gamma_\mu \nu_j) ( e A^\mu \phi^-)
\nonumber \\&& +\, \mbox{ h.c. of previous line}
\label{dim6broken}
\end{eqnarray}
with $F_{\mu\nu}$ the photon field strength tensor. 
The couplings are
\begin{eqnarray}
\alpha_{ij} &=& [U^\dagger a V]_{ij},
\nonumber\\[0.1cm]
\beta_{ijkl} &=& \sum_{m,n} \,
[U^\dagger]_{im} U_{mj} [V^\dagger]_{kn} V_{nl} \,b_{mn} ,
\nonumber\\
\gamma_{1,ij} &=& \sum_m \,[V^\dagger]_{im} V_{mj} \,c_{1,m} ,
\nonumber\\
\gamma_{x,ij} &=& \sum_m \,[U^\dagger]_{im} U_{mj} \,c_{x,m} \quad
(x=2,3)
\label{couplingdefs}
\end{eqnarray}
with $a_{ij}=c_W a_{B,ij} - s_W a_{W,ij}$.   Many further terms can 
in principle appear on the right-hand side of (\ref{dim6broken}) upon 
replacing the fields in ${\cal O}_i$ by those of the broken theory. 
Inspection shows that these additional terms are not relevant, since they 
contribute only to the form factor $F_1$, see (\ref{formfactor}) below, 
or to $F_2$ and the magnetic moment only beyond 
one loop.

\subsection{The muon anomalous magnetic moment}
\label{sec:lecalc}

\begin{figure} 
\begin{center} 
\vskip0.3cm
\includegraphics[width=14cm]{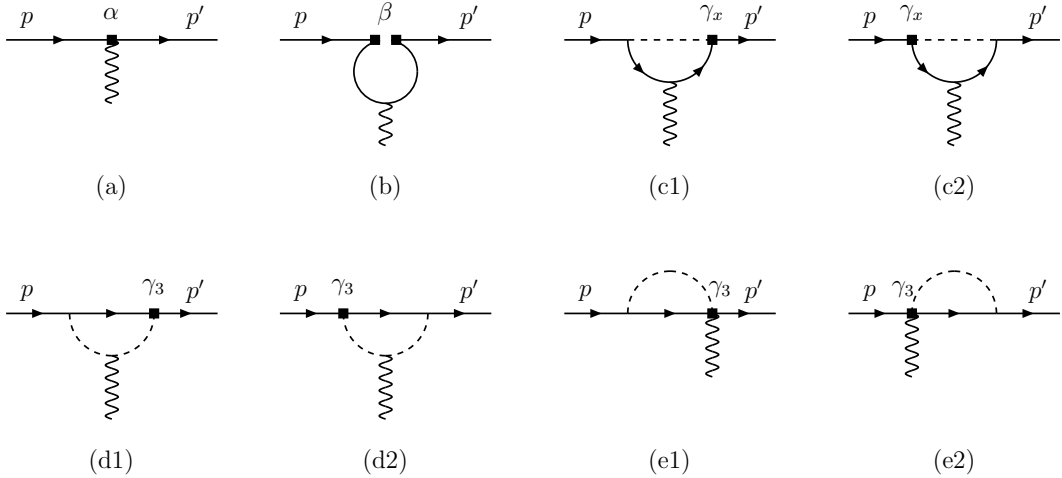}  
\end{center} 
\caption{Diagrams for $a_\mu$ from (\ref{dim6broken}).}  
\label{fig:4Ddiags}  
\end{figure}

The muon-photon vertex function is given by 
\begin{equation}
\Gamma^\mu(p,p^\prime) = i e Q_\mu\,\bar u(p^\prime,s^\prime) 
\left[\gamma^\mu F_1(q^2) +\frac{i\sigma^{\mu\nu}q_\nu}{2 m_\mu}
F_2(q^2) + \ldots\right] u(p,s)
\label{formfactor}
\end{equation}
where the ellipses denote parity-violating terms, and 
$q=p^\prime-p$. The factor $i eQ_\mu$
with $Q_\mu=-1$ corresponds to our definition of the gauge 
covariant derivative and the standard normalization of the charge form
factor $F_1(0)=1$.  The anomalous magnetic moment is defined by 
\begin{equation}
a_\mu = \frac{(g-2)_\mu}{2} = F_2(0).
\end{equation}
It is straightforward to compute the extra contribution $\Delta a_\mu$
from the effective Lagrangian (\ref{dim6broken}). Since $\alpha$ 
is a one-loop quantity, while $\beta$ and $\gamma_x$ are tree-level, 
we need to compute the diagrams shown in figure~\ref{fig:4Ddiags}.

%%%%%%%%%%%%%%%%%%%%%%%%%%%%%%%%%%%%%%%%%%%%%%%%%%%%%%%%%%%%%%%
%%%%%%%%%%%%%%%%%%%%%%%%%%%%%%%%%%%%%%%%%%%%%%%%%%%%%%%%%%%%%%%

The tree-level diagram (a) with an insertion of the class-0
electromagnetic dipole operator results in 
\begin{equation}
\Gamma^\mu(p,p^\prime) = \frac{1}{T^2}\,\mbox{Re}\,(\alpha_{22})\,
\,\frac{v}{\sqrt{2}} \,2 i \,\bar u(p^\prime,s^\prime) 
i\sigma^{\mu\nu}q_\nu u(p,s).
\label{alphares}
\end{equation}
The imaginary part of the $\alpha_{ij}$ couplings generates an 
electric dipole moment. 

There are two possible contractions of the four-fermion operator 
(diagram (b)), which differ by the exchange of $P_L \leftrightarrow 
P_R$ and $\beta_{2kk2} \leftrightarrow \beta_{k22k}$. The loop 
integral is ultraviolet (UV) divergent by power counting and must 
be regularized. The divergence is related to factorization of 
the contributions from the KK scale $T$ and the low-energy scales 
$m_\ell$, which is the only scale present in diagram (b) from the 
external momenta and massive lepton propagators. 
We adopt dimensional regularization 
with $d=4-2\varepsilon$. The final result is finite, and arises 
from the UV $1/\varepsilon$ pole that multiplies a numerator 
of ${\cal O}(\varepsilon)$. It depends on the 
scheme and we use the ``naive dimensional regularization'' (NDR) 
scheme with anti-commuting $\gamma_5$.
The scheme dependence  of this ultraviolet-sensitive term is compensated 
by a corresponding dependence in the 
calculation of the matching coefficients as we discuss later. The situation is 
similar to the calculation of the $b\to s\gamma$ transition 
with the weak effective Lagrangian method.

The contribution to the vertex function from the four-fermion 
operator in the effective Lagrangian is 
\begin{eqnarray}
\Gamma^\mu(p,p^\prime) &=& \frac{i \beta_{2kk2}}{T^2} \,
i e Q_{\ell_k} \,\mu^{2\varepsilon} \!\int\!\frac{d^dl}{(2\pi)^d}\,
\frac{\bar u(p^\prime,s^\prime)\gamma^\rho P_L i(\slash{l}
  +m_{\ell_k})
\gamma^\mu i(\slash{l}-\slash{q}+m_{\ell_k})\gamma_\rho P_Ru(p,s)}
{(l^2-m_{\ell_k}^2)((l-q)^2-m_{\ell_k}^2)}
\nonumber\\
&& +\, (\beta_{2kk2} \leftrightarrow \beta_{k22k}, \,P_L \leftrightarrow 
P_R)
\nonumber\\[0.2cm]
&=& -i e Q_\mu \,\frac{m_{\ell_k}}{16\pi^2 T^2}\,
\bar u(p^\prime,s^\prime) 
i\sigma^{\mu\nu}q_\nu \left[\beta_{2kk2}+\beta_{k22k} 
+(\beta_{2kk2}-\beta_{k22k})\gamma_5 \right]u(p,s)
\label{betares}
\end{eqnarray}
where a sum over internal lepton flavours $k=1,2,3$ is understood, 
and the limit $d\to 4$ has been taken.
The $\gamma_5$ term is related to the electric dipole moment; the
remaining terms are the desired contributions to the anomalous
magnetic moment of the muon.

The Higgs derivative interactions from  (\ref{dim6broken}) 
contribute via the diagrams (c1), (c2) and (d1), (d2) of
figure~\ref{fig:4Ddiags}. Diagram (c1) with an insertion 
of the $\gamma_2$ coupling is given by
\begin{eqnarray}
\Gamma^\mu(p,p^\prime) &=& \frac{i \gamma_{2,22} v}{2 T^2} \,
i e Q_{\mu} \,\frac{i y_\mu}{\sqrt{2}} \,
\nonumber\\
&& \times\,
\mu^{2\varepsilon} \!\int\!\frac{d^dl}{(2\pi)^d}\,
\frac{(+i)\,\bar u(p^\prime,s^\prime)P_R \,\slash{l}  
i(\slash{p}^\prime - \slash{l} +m_{\mu})
\gamma^\mu i(\slash{p}-\slash{l}+m_{\mu}) u(p,s)}
{(l^2-m_H^2) ((p-l)^2-m_{\mu}^2)((p^\prime-l)^2-m_{\mu}^2)}.\quad
\end{eqnarray}
Here $y_\mu = \sqrt{2} m_\mu/v$ is the small standard model muon
Yukawa coupling. The prefactor is proportional to the muon mass 
$m_\mu =y_\mu v/\sqrt{2}$, since one of the vertices is the SM 
Higgs-fermion vertex. We may therefore neglect the muon mass in 
the integrand, and use the on-shell conditions to obtain 
\begin{eqnarray}
\Gamma^\mu(p,p^\prime) &=& -\frac{i \gamma_{2,22} m_\mu}{2 T^2} \,
i e Q_{\mu} \,
\mu^{2\varepsilon} \!\int\!\frac{d^dl}{(2\pi)^d}\,
\frac{\bar u(p^\prime,s^\prime) P_R 
\gamma^\mu (-\slash{l}) u(p,s)}
{(l^2-m_H^2) (l^2-2 p\cdot l)}.
\end{eqnarray}
This vanishes, since the loop integral can only result in 
$\slash{p} u(p,s)=0$. Diagram (c2) and the insertions of the 
$\gamma_{1,3}$ interactions vanish in an analogous way. The reason 
for this can be deduced without calculations from the 
operator $\bar \psi  P_{L/R} \gamma_\mu\psi \partial^\mu H$, 
which, up to a total derivative, can be converted into 
$\partial^\mu(\bar \psi P_{L/R}\gamma_\mu \psi) H$. In the NDR scheme,
by the field equations, this operator is proportional to lepton 
masses, or small coupling constants, and hence of higher-order 
in our approximations. The corresponding arguments 
cannot be applied in the unbroken theory, since there the operator 
contains two Higgs fields. 

Diagrams (d1) and (d2) involve an internal charged Higgs and 
a neutrino line. For diagram (d1) we find  
\begin{eqnarray}\label{deltares}
\Gamma^\mu(p,p^\prime) &=& \frac{i \gamma_{3,22} v}{\sqrt{2} T^2} 
\,(i e) \,(-i y_{\mu}) 
\nonumber\\
&& \times\,
\mu^{2\varepsilon} \!\int\!\frac{d^dl}{(2\pi)^d}\,
\frac{i^2\,\bar u(p^\prime,s^\prime)\,
P_R (\slash{l}-\slash{p}^\prime)i(\slash{l}+m_\nu) P_R
u(p,s)\;(2l-p-p^\prime)^\mu}
{((p-l)^2-m_\phi^2)((p^\prime-l)^2-m_\phi^2) (l^2-m_{\nu}^2)}\;.
\quad
\end{eqnarray}
The prefactor is proportional to the muon mass, hence we can neglect 
the muon and neutrino mass in the remainder of the expression.  
Introducing the Feynman parameter 
$x$ ($\bar x=1-x$) to combine denominators leads to 
\begin{eqnarray}\label{ContributionHiggs}
\Gamma^\mu(p,p^\prime) &=&  -\frac{ e \gamma_{3,22} m_\mu}{T^2} \,
\nonumber\\
&& \times\,
\mu^{2\varepsilon} \!\int\!\frac{d^dl}{(2\pi)^d}\,\int_0^1dx 
\frac{\bar u(p^\prime,s^\prime)P_R u(p,s)\;(p+p^\prime-2l)^\mu}
{((l-x\,p-\bar x\,p^\prime)^2-m_\phi^2 +x p^2+
\bar x {p^\prime}^2 -(xp +\bar x p^\prime)^2)^2}
\nonumber \\[0.1cm]
&=&  -\frac{ e  \gamma_{3,22} m_\mu }{T^2}  \,
\mu^{2\varepsilon}\!\int\!\frac{d^dl}{(2\pi)^d}\int_0^1\!dx
\frac{\bar u(p^\prime)P_R u(p)[(\bar x-x)p^\mu+
(x-\bar x)p^{\prime \mu}]}
{(l^2-m_\phi^2 + x\bar x q^2)^{2}}
\nonumber \\[0.1cm]
&=&0\,,
\end{eqnarray}
where  in the last step we used that the integrand is antisymmetric 
under the exchange $x\leftrightarrow \bar x$.
For later purposes, we note that the zero arises due to a cancellation 
between a finite term and an ultraviolet-sensitive term  that arises as
the product $1/\varepsilon \times \varepsilon$ of a UV divergence and 
a numerator of ${\cal O}(\varepsilon)$. To see this, we assume $l\gg p, 
p^\prime$, expand the integrand of \eqref{deltares} 
to first order in the external 
momenta, and extract the $(p+p^\prime)^\mu$ structure that 
is relevant to the magnetic moment.  This results in 
\begin{eqnarray}
\Gamma^\mu(p,p^\prime)&\stackrel{\rm UV}{\longrightarrow}& 
\frac{i\gamma_{3,22} v}{\sqrt{2}T^2} \,
(i e) \,(-i y_{\mu}) \,i\,
\nonumber \\
&& \times
\mu^{2\varepsilon}\!\int\!\frac{d^dl}{(2\pi)^d} 
 \frac{\bar u(p^\prime,s^\prime) P_R u(p,s)\,(p+p^\prime-2l)^\mu}
{(l^2-m_\phi^2)^2}\left(1+\frac{2p\cdot l+2p^\prime \cdot l}
{l^2-m_\phi^2}\right)
\nonumber \\
&\longrightarrow&-\frac{i e \gamma_{3,22}m_{\mu}}{T^2} \,
  \frac{\mu^{2\varepsilon}}{(4\pi)^{d/2}}\,
\left[\left(1-\frac{4}{d}\right)\times \frac{\Gamma(\varepsilon)}
{m_\phi^{2\varepsilon}}\,(p+p^\prime)^\mu 
+ \frac{4}{d} \frac{\Gamma(1+\varepsilon)}{2m_\phi^{2\varepsilon}}
\,(p+p^\prime)^\mu   \right]
\nonumber \\
&& \times \bar u(p^\prime,s^\prime)P_R u(p,s)
\nonumber\\[0.2cm]
&=&
\frac{i e \gamma_{3,22}m_\mu }{(4\pi)^{2} T^2} \,
\bar u(p^\prime,s^\prime)P_R  (p+p^\prime)^\mu  u(p,s) 
\times \left( \frac{\varepsilon}{2}\times \frac{1}{\varepsilon} -  
\frac12 \right) + {\cal O}(\varepsilon).
\label{ChargedHiggsOperatorsUV}
\end{eqnarray}
The total contribution is, of course, zero; but the contribution due 
to the UV pole is finite despite the superficial divergence of the integral. 
This will be important when verifying the scheme independence 
in Sec.~\ref{subsec:scheme}. An analogous result holds for (d2). 

Diagrams (e1) and (e2) follow from the insertion of the operator
$\bar \psi_i P_R \gamma^\mu \nu_j \, (e A_\mu\phi^-)$ 
in (\ref{dim6broken}) and its hermitian conjugate, respectively.
None of the two contributes to $(g-2)_\mu$ as the 
loop only depends on a single external four-momentum. We can use
on-shell condition to convert each appearance of $\slashed{p}^{(\prime)}$
to a lepton mass. The only possible Lorentz structure is then
$\bar u(p^\prime,s^\prime)\gamma^\mu P_{R/L}u(p,s) A_\mu$, 
which contributes only to the $F_1$ form factor.

The muon anomalous magnetic moment follows by adding the 
non-vanishing matrix elements of the dipole and four-fermion operator, 
(\ref{alphares}) and (\ref{betares}),  respectively, resulting in 
\begin{eqnarray}
\Delta a_\mu &=& \frac{1}{T^2}\,\mbox{Re}\,(\alpha_{22})\,
\,\frac{1}{-e}\,\frac{4 m_\mu v}{\sqrt{2}} 
+ \sum_{k=1,2,3} (-1) \,\frac{2m_\mu m_{\ell_k}}{16\pi^2 T^2}\,
\left[\beta_{2kk2}+\beta_{k22k} \right]
\nonumber\\
&=& -\,\frac{4 m_\mu^2}{T^2}\left(
\frac{\mbox{Re}\,(\alpha_{22})}{y_\mu e}
+\sum_{k=1,2,3} \frac{1}{16\pi^2}\,
\frac{m_{\ell_k}}{m_\mu}\,
\mbox{Re}\,(\beta_{2kk2})
\right).
\label{amures}
\end{eqnarray}
Here $k=1,2,3$ refers to electron, muon and tau
leptons, respectively. In passing to the last line of (\ref{amures}), 
we used that $b_{mn}$, being generated by gauge interactions,
is real. It therefore follows from the definitions
(\ref{couplingdefs}) that $\beta_{k22k}=\beta_{2kk2}^*$, and hence 
$\beta_{2kk2}+\beta_{k22k}$ is real as it should be.

%%%%%%%%%%%%%%%%%%%%%%%%%%%%%%%%%%%%%%%%%%%%%%%%%%%%%%%%%%%%%%%%%%%%%%%%
\section{Calculation of matching 
coefficients}
\label{sec:matching}

%%%%%%%%%%%%%%%%%%%%%%%%%%%%%%%%%%%%%%%%%%%%%%%%%%%%%%%%%%%%%%%
\begin{figure}[p] 
\begin{center} 
\vskip-0.0cm
\includegraphics[width=14.5cm]{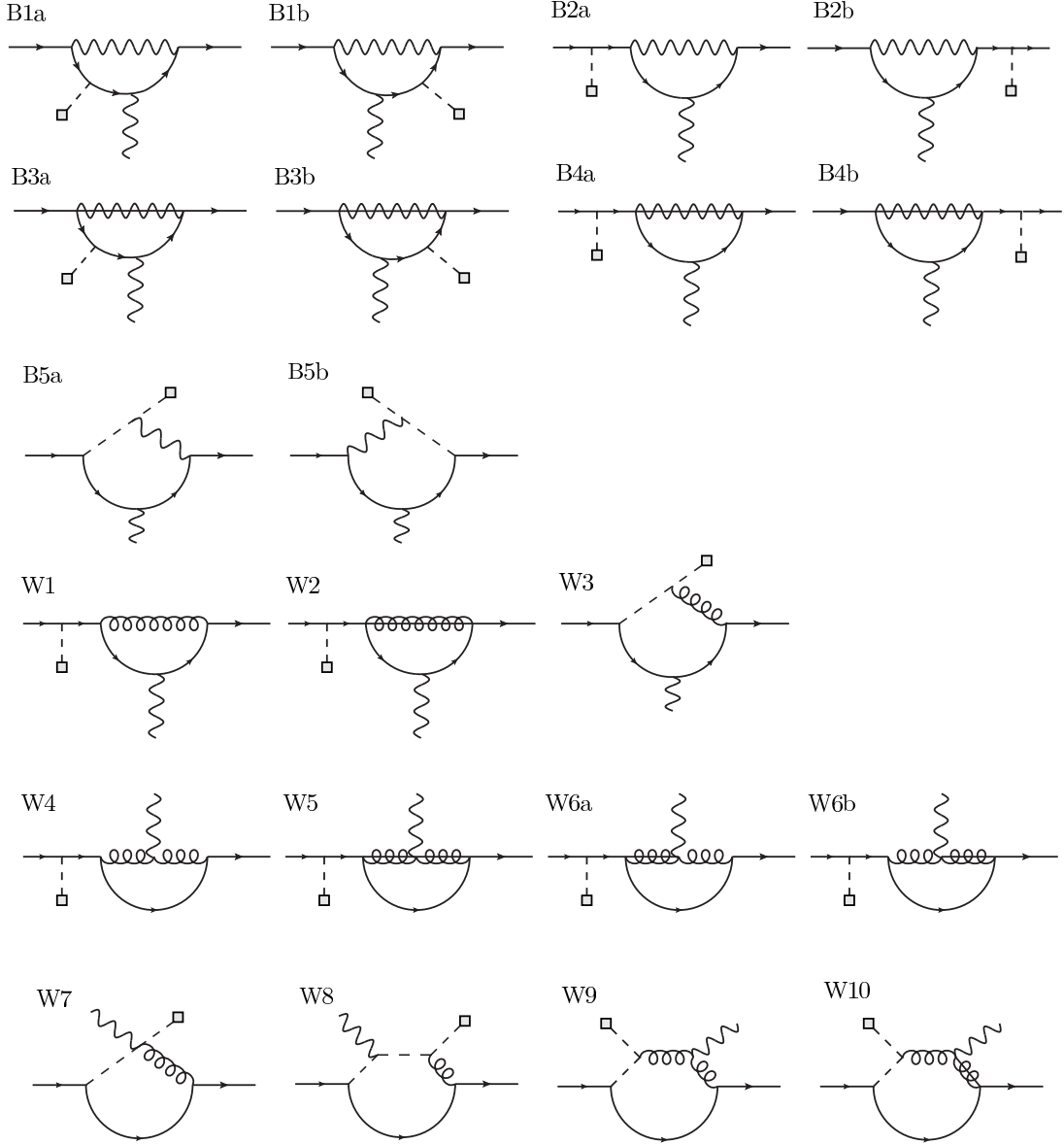}  
\end{center} 
\caption{Diagrams contributing to $\Delta a_\mu$, more precisely the matching
  coefficients of the class-0 operators. Solid lines refer to leptons,
with the right external line belonging to the doublet $L_i$, 
the left one to $E_j$. Wavy lines denote hypercharge gauge bosons and 
the external photon, curly lines SU(2) W-bosons. A solid-wavy 
(solid-curly) line refers to the scalar fifth component  
of the gauge field. Dashed lines denote Higgs bosons, including the 
external Higgs field (grey box). Vertices involving Higgs fields 
are localized at $1/T$, all other vertices are integrated over 
position in the fifth dimension.}  
\label{fig:diagsall}  
\end{figure}  
%%%%%%%%%%%%%%%%%%%%%%%%%%%%%%%%%%%%%%%%%%%%%%%%%%%%%%%%%%%%%%%

We now turn to the calculation of the matching coefficients in the 
RS model (\ref{5Daction}). The diagrams relevant to the anomalous 
magnetic moment in the full 5D theory with unbroken gauge symmetry 
are shown in figure~\ref{fig:diagsall}. 
The figure does not include genuine Higgs-exchange diagrams, 
which involve three Yukawa interactions (rather than one Yukawa and 
two gauge interactions as in figure~\ref{fig:diagsall}), and 
which we discuss separately in Sec.~\ref{subsec:Higgs}.

In a direct calculation of $a_\mu$ in the 5D theory, 
the diagrams should be represented in terms of the fields and 
interactions in the expansion around the electroweak vacuum of the 
spontaneously broken theory, which depend on the 
scales $k$, $T$, $M_{\rm EW}$, $m_{\ell_k}$. In drawing 
the diagrams in figure~\ref{fig:diagsall} we already accounted for 
the fact that we only need to extract 
the contributions from the scales $k$ and $T$, which gives 
directly the short-distance 
coefficients of the SU(3)$\times$SU(2)$\times$U(1) 
invariant effective Lagrangian. The diagrams in the 
unbroken theory do not represent the infrared physics 
from the scales near and below the electroweak scale $M_{\rm EW}$ correctly, 
but the incorrect infrared contribution cancels in the matching 
procedure. The low-momentum contributions from diagrams 
with 5D non-zero-mode propagators have already been included through the 
calculations of the previous subsection. Note that we do 
not consider diagrams with graviton exchange; see 
\cite{Davoudiasl:2000my, Niehoff13} for a discussion of the graviton contribution.

The diagrams are understood in the mixed representation with an
integration over four-dimensional loop momentum $l$ and the vertex 
positions in the fifth dimension. 
A generic diagram such as diagram B1a in figure~\ref{fig:diagsall} 
can have three different contributions:
\begin{itemize}
\item All 5D propagators propagate the zero mode. The loop integral 
does not contain the short-distance scales $T,k$ explicitly, and 
is purely long-distance. This corresponds to a contribution to the 
SM anomalous magnetic moment $a_{\mu}^{\rm SM}$. There is no need 
to perform this computation here, so we subtract this contribution.
\item At least one of the 5D propagators propagates a KK mode, but the
  loop momentum $l$ satisfies $l\ll T$. In this case, the subgraph 
  consisting of propagators with KK modes can be contracted to a
  point. The reduced loop diagram must be computed with the 4D
  effective Lagrangian. This corresponds to tree-level matching of 
  class-1 operators and the one-loop operator matrix element 
 calculations performed in the 
  previous subsection. An explicit diagrammatic analysis shows 
  that the relevant subgraphs in figure~\ref{fig:diagsall} correspond 
  precisely to the operators (\ref{dim61}), and neglected operators of  
dimension higher than six. In diagram B1a, for example, the only relevant
contribution occurs when the gauge boson is a KK mode and all 
internal fermions are zero modes, which corresponds to the insertion 
of a local four-fermion operator as in diagram (b) of 
figure~\ref{fig:4Ddiags}.  The fermion-Higgs effective vertex and diagram 
(d1) are related exclusively to diagram W8. 
If the gauge boson and at least one of the fermion propagators 
refers to a KK mode and the loop momentum is small, the contribution 
is suppressed and corresponds to a higher-dimension operator. If the 
gauge boson is the zero mode, then the flatness of the gauge 
zero mode and the orthogonality condition for the fermion 
modes forces all fermion lines to propagate zero modes as well. This is 
the SM contribution discussed above.
\item The loop momentum is of order $l\sim T$ or larger. 
This contribution goes into the one-loop matching coefficients of the
class-0 operators.
\end{itemize}
In the following we describe the computation of the matching coefficients 
mentioned in the second 
and third item. 

\subsection{Four-fermion operator}

%%%%%%%%%%%%%%%%%%%%%%%%%%%%%%%%%%%%%%%%%%%%%%%%%%%%%%%%%%%%%%%
\begin{figure} 
\begin{center} 
\vskip0.3cm
\includegraphics[width=4cm]{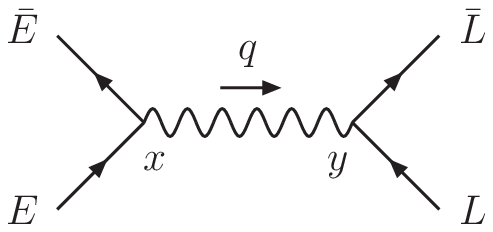}  
\end{center} 
\caption{Hypercharge boson exchange that generates the 
four-fermion operator.}  
\label{fig:fourfermion}  
\end{figure}  
%%%%%%%%%%%%%%%%%%%%%%%%%%%%%%%%%%%%%%%%%%%%%%%%%%%%%%%%%%%%%%%

The tree-level matching calculation is very similar to the one 
that generates flavour-changing four-quark operators from 
KK gluon exchange\cite{hepph:0408134,
  arXiv:08041954,arXiv:08091073, arXiv:08074937,arXiv:08113678}. 
Here we perform the calculation in the 5D 
picture, which makes it particularly simple (provided the 5D
propagators are known).

The relevant diagram is shown in figure~\ref{fig:fourfermion}.
Only hypercharge vector gauge boson exchange can generate the 
$(\bar L_i\gamma^\mu L_i)(\bar E_j\gamma_\mu E_j)$ operator at tree level, 
since SU(2) gauge fields do not interact with $E_i$. The scalar 
fifth component of the 5D gauge boson cannot be exchanged, 
since the two external zero-mode fermions at the vertex 
have the same handedness. From the tree diagram shown in 
figure~\ref{fig:fourfermion}, we obtain
\begin{eqnarray}
b_{ij} &=& -i \,(i g_5^{\prime})^2 \,\frac{Y_L}{2}\frac{Y_E}{2} \,
T^2 \int_{1/k}^{1/T} \!\!dx dy \,
\frac{{f_{L_i}^{(0)}}^2\!(x)}{(k x)^4}\,
\frac{{g_{E_j}^{(0)}}^2\!(y)}{(k y)^4}\,
\Delta_\perp^{\rm ZMS}(q=0,x,y).
\label{bijstart}
\end{eqnarray}
Here $Y_L=-1$, $Y_E=-2$ denote the hypercharges of the leptons, 
and $\Delta_\perp$ is the $\eta_{\mu\nu}$ component of the 5D gauge boson
propagator given in (\ref{gaugeprop1}) of the appendix 
(see also \cite{Randall:2001gb}). We subtracted the zero mode 
from the propagator, since this corresponds to a SM contribution to 
$a_\mu$. 

Because of the large mass of the KK excitations, the external 
lepton momenta of order $m_{\mu}$ can be set to zero.  
The zero-momentum limit of the gauge-boson propagator 
(\ref{gaugeprop1}) is 
\begin{eqnarray}
\label{gaugepropZeroMom}
\Delta_\perp(q,x,y) &\stackrel{q\to 0}{=}& 
\Theta(x-y) \,\frac{i k}{\ln\frac{k}{T}} 
\,\bigg(-\frac{1}{q^2} + \frac{1}{4}\,
\bigg\{\frac{1/T^2-1/k^2}{\ln\frac{k}{T}} - x^2-y^2
+ 2 x^2 \ln(x T)\nonumber\\
&& \hspace*{0cm} + \, 
2 y^2 \ln(y T)+2 y^2 \ln\frac{k}{T}
\bigg\} + {\cal O}(q^2) \bigg) 
+ (x\leftrightarrow y),
\end{eqnarray}
which agrees with a similar expression obtained in \cite{arXiv:08074937}
from the explicit summation over all KK excitations. 
The $1/q^2$ term is evidently the zero mode. Subtracting it,
neglecting $1/k^2$ relative to $1/T^2$ in the first term in curly 
brackets, since $\epsilon\equiv 
T/k\sim 10^{-16}$ is very small, 
and employing the normalization (\ref{OrthonormFermion}) 
of the zero modes, we obtain
\begin{eqnarray}
b_{ij} &=& - \frac{1}{4} {g^{\prime}}^2 \,\frac{Y_L}{2}\frac{Y_E}{2} \, T^2
\,\Bigg(\frac{1}{T^2 \ln\frac{k}{T}} 
\nonumber\\
&& \hspace*{-0cm} +\,
\int_{1/k}^{1/T} \!\!dx \,
\frac{{f_{L_i}^{(0)}}^2\!(x)}{(k x)^4}\,
x^2 \left(2  \ln(x T) -1\right) +
\int_{1/k}^{1/T} \!\!dy \,
\frac{{g_{E_j}^{(0)}}^2\!(y)}{(k y)^4}\,
y^2 \left(2  \ln(y T) -1\right) 
\nonumber\\
&& \hspace*{-0cm} +\,
2 \ln\frac{k}{T} \int_{1/k}^{1/T} \!\!dx dy \,
\frac{{f_{L_i}^{(0)}}^2\!(x)}{(k x)^4}\,
\frac{{g_{E_j}^{(0)}}^2\!(y)}{(k y)^4}\,
\mbox{min}\,(x^2,y^2) 
\Bigg).
\label{b1}
\end{eqnarray}
The remaining integrals over the fifth-dimension coordinates are
dominated by $x,y\sim 1/T$ up to terms suppressed by some power of 
$T/k$ provided the bulk mass parameters satisfy $c_{L_i}<3/2$ and 
$c_{E_j}>-3/2$, which will be assumed. 
We can therefore set the lower integration limit to 0 and 
use the explicit form of the fermion zero modes from 
(\ref{FermionZeroModesSinglet}),
(\ref{FermionZeroModesDoublet}) in the appendix to 
find 
\begin{equation}
b_{ij} = b_0 + b_1(c_{L_i}) + b_1(-c_{E_j}) 
+ b_2(c_{L_i},c_{E_j}) 
\end{equation}
with
\begin{eqnarray}
&& b_0 = - \frac{ {g^{\prime}}^2}{8}\,\frac{1}{\ln(1/\epsilon)} ,
\nonumber\\
&& b_1(c) = - \frac{ {g^{\prime}}^2}{8}\,\frac{(5-2 c)(1-2 c)}{(3-2
  c)^2} \,\frac{\epsilon^{2 c-1}}{1-\epsilon^{2 c-1}},
\\
 && b_2(c_L,c_E) = 
 - \frac{ {g^{\prime}}^2}{4}\,
\frac{(1-2 c_L) (1+2 c_E)(3-c_L+c_E)}
{(3-2 c_L) (3+2 c_E)(2-c_L+c_E)}
 \,\ln\frac{1}{\epsilon}\,
 \frac{\epsilon^{2 c_L-1}}{1-\epsilon^{2 c_L-1}}
 \,\frac{\epsilon^{-2 c_R-1}}{1-\epsilon^{-2 c_R-1}}.
\nonumber
\end{eqnarray}
Substituting this result for $b_{ij}$ into (\ref{couplingdefs}) 
and exploiting the unitarity of the flavour rotation matrices 
to the mass basis, we obtain
\begin{eqnarray}
\beta_{ijkl} &=& b_0 \delta_{ij}\delta_{kl} 
+ \delta_{kl} \sum_m b_1(c_{L_m}) \,[U^\dagger]_{im} U_{mj} 
+ \delta_{ij} \sum_n b_1(-c_{E_n}) \,[V^\dagger]_{kn} V_{nl} 
\nonumber\\[-0.1cm]
&&
+ \,\sum_{m,n} b_2(c_{L_m},c_{E_n}) \,
[U^\dagger]_{im} U_{mj} [V^\dagger]_{kn} V_{nl}.
\end{eqnarray}
The first term does not change lepton flavour number $F$,  
the next two may change $\Delta F=\pm 1$, and only the term in the 
second line can produce $\Delta F=\pm 2$ transitions. All terms 
are formally required to compute $\sum_k m_{\ell_k}/m_\mu\times  
\beta_{2kk2}$ relevant to the 
anomalous magnetic moment.

A particularly simple result is obtained in 
the ``single-flavour approximation'', where we ignore the 
presence of other leptons than the muon. This corresponds to 
replacing $\sum_k m_{\ell_k}/m_\mu\times  
\beta_{2kk2} \to \beta_{2222} \equiv\beta$, and 
$U_{ij}=V_{ij}=\delta_{ij}$, such that 
\begin{equation}
\beta = b_0 + b_1(c_{L}) + b_1(-c_{E})+
b_2(c_{L},c_{E}) \equiv 
- \frac{ {g^{\prime}}^2}{8} \frac{1}{\ln(1/\epsilon)}\,
f(\epsilon,c_L,c_E).
\label{bsingflavour}
\end{equation}
Setting further 
$c_L=-c_E=1/2+r$ ($r>0$), we find 
\begin{equation}
f = 1 - \frac{\epsilon^{2 r}}{1-\epsilon^{2 r}}\,\ln\frac{1}{\epsilon} 
\,\frac{2 r (2-r)}{(1-r)^2} + 
\left[ \frac{\epsilon^{2 r}}{1-\epsilon^{2
      r}}\,\ln\frac{1}{\epsilon}\right]^2 
\,\frac{4 r^2}{(1-r) (1-2 r)}.
\label{ffunc}
\end{equation} 
According to (\ref{4dyukawa}), $r$ is related to the muon mass by 
\begin{equation}
\frac{2 r \epsilon^{2 r}}{1-\epsilon^{2 r}} = 
\frac{y_\mu}{y_\mu^{\rm (5D)} k}
= \frac{\sqrt{2} m_\mu}{y_\mu^{\rm (5D)} k v}. 
\end{equation}
With $\epsilon=10^{-16}$ and $y_\mu^{\rm (5D)} k=1$, we find 
$r=0.0749$ and $f=0.95$. In general, when 
$y_\mu^{\rm (5D)} k$ is of order 1 and $\epsilon$ within a few
orders of magnitude near $10^{-16}$, the deviation of 
$f$ from 1 is only a few percent, and the flavour-conserving 
four-fermion matching
coefficient is dominated by the $r$-independent term $b_0$. 
In this case the total contribution (\ref{bsingflavour}) 
is suppressed by the large logarithm $\ln(1/\epsilon)$ and the small 
hypercharge gauge coupling.

\subsection{Fermion-Higgs operator}
\label{higgfermionoperators}
%%%%%%%%%%%%%%%%%%%%%%%%%%%%%%%%%%%%%%%%%%%%%%%%%%%%%%%%%%%%%%%
\begin{figure} 
\begin{center} 
\vskip0.3cm
\includegraphics[width=4cm]{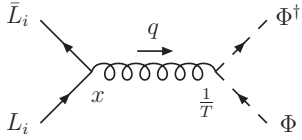}  
\end{center} 
\caption{Weak isospin boson exchange that generates the 
         fermion-Higgs operator.}  
\label{fig:fermionHiggs}  
\end{figure}  
%%%%%%%%%%%%%%%%%%%%%%%%%%%%%%%%%%%%%%%%%%%%%%%%%%%%%%%%%%%%%%%
For completeness we also give the matching coefficient $c_{3,i}$ 
for the fermion-Higgs operator $(\bar L_i\gamma^\mu \tau^a L_i) (
\Phi^\dagger \overleftrightarrow{i \tau^a D_\mu} \Phi)$, 
which gives rise to diagram (d1) in figure~\ref{fig:4Ddiags}.
Only SU(2) gauge-boson exchange shown in figure \ref{fig:fermionHiggs} 
can generate an operator with a single charged scalar at tree level. 
The fifth component of the 5D gauge boson cannot contribute due to the 
boundary conditions. We find
\begin{align}
\label{deltacharged}
c_{3,i}= - i \, \frac{(ig_5)^2}{2} T^2 
\int_{1/k}^{1/T} dx\,\frac{{f_{L_i}^{(0)}}^2\!(x)}{(kx)^4}
\Delta_\perp^{ZMS}(q=0,x,1/T)\,. 
\end{align}
Inserting \eqref{gaugepropZeroMom} into  \eqref{deltacharged} we obtain
\begin{align}
c_{3,i}=& - \frac{g^2}{8} T^2 \,
\Bigg(\frac{1/T^2-1/k^2}{\ln\frac{k}{T}} -\frac{1}{T^2} 
+\int_{1/k}^{1/T} dx\, \frac{{f_{L_i}^{(0)}}^2\!(x)}{(kx)^4} 
\,x^2 \big(2 \ln(kx)-1\big)
\Bigg).
\end{align}
 As above, we neglect $1/k^2$ relative to $1/T^2$, and keep 
only the leading term in $\epsilon$.  Using the 
normalization of the fermion modes simplifies the expression to 
\begin{align}
  c_{3,i}= \frac{g^2}{8} \,\Bigg(1-\frac{1}{\ln1/\epsilon}  
    - \left[ \frac{(1-2c_{L_i})(5-2c_{L_i})}{(3-2c_{L_i})^2} 
    -\frac{2(1-2c_{L_i})\ln1/\epsilon}{(3-2c_{L_i})} \right]
    \frac{\epsilon^{2c_{L_i}-1}}{1-\epsilon^{2c_{L_i}-1}}
\Bigg).
\label{c3coeff}
\end{align}
Note that $c_{3,i}$ is about  $10^2$ times larger than $b_{ij}$, since 
it is proportional to the SU(2) gauge coupling and since its largest term 
does not feature the suppression factor $\frac{1}{\ln1/\epsilon}\sim 1/35$ 
that is present in $b_{ij}$. For $c_{L_i}$ larger than but not too close 
to $1/2$ the dependence of $c_{3,i}$ on the model parameters is 
rather mild, since the 1 in the round brackets dominates.

\subsection{Electromagnetic dipole operator}
\label{EDMoperator}
The matching coefficient of the electromagnetic dipole operator 
is more difficult to compute, since it comes from the 5D one-loop 
diagrams in figure~\ref{fig:diagsall}. We are interested 
only in the electromagnetic dipole transition, so we take the external 
gauge boson to be the superposition $c_W B^\mu + s_W W^{3,\mu}$ 
corresponding to the photon. We also 
anticipate that the Higgs doublet $\Phi$ in the operators 
$\bar L_i\Phi \sigma_{\mu\nu} E_j B^{\mu\nu}$ and 
$\bar L_i\tau^a \Phi \sigma_{\mu\nu} E_j \,W^{a,\mu\nu}$ 
will be replaced by its vacuum expectation value and therefore 
extract only the weak isospin component that corresponds to 
charged lepton $\ell_i\to\ell_j\gamma$ transitions after electroweak 
symmetry breaking. These simplifications have already been employed 
to omit some additional diagrams with the photon coupling to the 
Higgs-doublet field. To compute the matching coefficient 
we apply the following strategy:
\begin{itemize}
\item Subtract the zero mode from every internal 5D gauge boson
  propagator. It can be shown that selecting the gauge-boson 
  zero modes forces the fermion propagators to only propagate 
  zero modes. This follows from the flatness of the gauge boson 
  and orthogonality of the fermion mode functions. Thus, the 
  subtracted terms correspond precisely to the 
  Standard Model contribution to $(g-2)_\mu$. 
\item Expand the diagram in the external fermion momenta $p$, 
  $p^\prime$ and neglect the (tachyonic) mass in the SU(2) 
  Higgs propagator.\footnote{Recall that the computation of the matching
  coefficients is done in the unbroken gauge theory.} This ensures 
  that we pick up only the contribution from loop momenta $l\sim T$ 
  as appropriate for a matching coefficient. Some of the one-loop 
  diagrams have tree-level subgraphs corresponding 
  to propagating KK excitations while the loop momentum is of order of the 
  electroweak or muon mass scale. These contributions are 
  excluded by the above expansion; they have already been 
  accounted for by the other dimension-6 operators with coefficient 
  functions   $b$, $c_i$ and the one-loop diagrams in 
  Sec.~\ref{sec:lecalc} as discussed above.
\end{itemize}

\subsubsection{Sample diagram}

We now illustrate this procedure by discussing the calculation 
of diagram B1a from figure~\ref{fig:diagsall} in further detail. 
With the 5D Feynman rules from appendix~\ref{ap:5Drules} we find for 
this diagram the expression
\begin{eqnarray}
  {\rm {\bf B1a}} &=& (ig'_5)^2(i e_5 Q_\mu)
\left(\frac{-iT^3}{k^3}\right) 
y^{\rm (5D)}_{ij}\,\frac{Y_L}{2}\frac{Y_E}{2} \int^{1/T}_{1/k}
  \!\!\!\frac{dx}{(kx)^4} \int^{1/T}_{1/k}
  \!\!\!\frac{dy}{(ky)^4} \int^{1/T}_{1/k}
  \!\!\!\frac{dz}{(kz)^4}\int \!\frac{d^4l}{(2\pi)^4} \nonumber \\
  &&  f^{(0)}_{L_i}(z) f^{(0)}_\gamma(y) g^{(0)}_{E_j}(x)\,\epsilon^{*\mu} 
  \Delta^{\rho\nu}_{\mbox{\tiny ZMS}}(l,x,z)
  \nonumber \\[0.1cm]
  &&  \bar{L_{i}}(p^\prime)P_{R} \gamma_\rho
  \Delta^{L}_{i}({p^\prime-l},z,y) \gamma_\mu
  \Delta^{L}_{i}({p-l},y,1/T) \Delta^{E}_{j}({p-l},1/T,x)
  \gamma_\nu P_{R} E_{j}(p),\qquad
\label{diag1text}
\end{eqnarray}
which contains the 5D propagators of the fermions and the 
zero-mode subtracted hypercharge gauge boson, and the 
wave functions $f^{(0)}_{L_i}(z)\bar{L_{i}}(p^\prime)$, 
$g^{(0)}_{E_j}(x)E_{j}(p)$, $f^{(0)}_\gamma(y)\epsilon^{*\mu}$ 
of the external states, which are the zero modes of the 
5D fields. The positions of the three vertices in the 5th dimension, 
$x$, $y$, $z$, and the four-dimensional loop momentum $l$ 
are integrated over. The electromagnetic charge prefactor arises, 
because the coupling of the external photon in isospin space is 
$[\frac{Y_L}{2} \mathds{1}+\frac{\tau^3}{2}]_{22}=Q_\mu=-1$.

We decompose each fermion propagator into its four chiral 
components using (\ref{FermionPropagator}). Most of the 
64 possible terms vanish due to the projectors $P_R$ in 
(\ref{diag1text}) and the brane boundary 
conditions $g_{L_i}(1/T)=f_{E_j}(1/T)=0$, 
see (\ref{FermionBCforE}), (\ref{FermionBCforL}). 
The two remaining terms can be deduced from 
(\ref{KKdecFermionPropagator}), which results in 
\begin{eqnarray}
  {\rm {\bf B1a}} &=& \frac{g'^2_5 e_5 Q_\mu Y_{L} Y_{E}
    y^{\rm(5D)}_{{ij}} T^3}{4 k^3} \int^{1/T}_{1/k}
  \!\!\!\frac{dx}{(kx)^4} \int^{1/T}_{1/k}
  \!\!\!\frac{dy}{(ky)^4} \int^{1/T}_{1/k}
  \!\!\!\frac{dz}{(kz)^4}\int
  \!\!\! \frac{d^4l}{(2\pi)^4}  \nonumber \\
  &&  f^{(0)}_{L_i}(z) 
  f^{(0)}_\gamma(y) g^{(0)}_{E_j}(x) \epsilon^{*\mu}  \Delta^{\rho\nu}_{\mbox{\tiny ZMS}}(l,x,z)
  \nonumber \\
  && \bar{L_{i}}(p^\prime) \Big[ F^+_{L_i}({\hat{p}^{\,\prime}},z,y) F^+_{L_i}({\hat{p}},y,1/T) F^-_{E_j}({\hat{p}},1/T,x) \left\{ \gamma_\rho (\not\!p^{\,\prime}-\!\!\!\not l) \gamma_\mu \gamma_\nu\right\}(p-l)^2 + \nonumber \\  
  && d^+F^-_{L_i}({\hat{p}^{\,\prime}},z,y) d^-F^+_{L_i}({\hat{p}},y,1/T) F^-_{E_j}({\hat{p}},1/T,x) 
  \left\{ \gamma_\rho \gamma_\mu (\not\!p-\!\!\!\not l) \gamma_\nu
  \right\} \Big] P_{R} E_{{j}}(p),
\label{diag1text2}
\end{eqnarray}
where $\hat{p} = p-l$, $\hat{p}^{\,\prime} = p^\prime-l$.
In the KK picture the first term in square brackets corresponds to 
picking up three momentum factors from the propagator numerators, while
the second contains two KK mass factors. In appendix 
\ref{ap:diags} we provide the expressions corresponding to 
(\ref{diag1text}), (\ref{diag1text2}) for all 21 one-particle 
irreducible diagrams shown 
in figure~\ref{fig:diagsall}.

We note that at this point the integral over the coordinate $y$ of 
the external photon vertex could be performed analytically. Since the 
photon zero mode is constant and combines with 
$e_5  f^{(0)}_\gamma(y) = e$ to the dimensionless 4D electric 
charge, the remaining $y$-integral is 
\begin{equation}
 \int^{1/T}_{1/k}
  \!\!\!\frac{dy}{(ky)^4}
\, F^+_{L_i}({\hat{p}^{\,\prime}},z,y) F^+_{L_i}({\hat{p}},y,1/T)
= \sum_n
\frac{(-i)^2 f_{L_i}^{(n)}(z)  f_{L_i}^{(n)}(1/T)}
{({\hat{p}^{\,\prime}}{}^2-m_n^2)({\hat{p}}^2-m_n^2) }
\end{equation} 
and a similar expression for the mass terms. The key point is that 
the $y$-integral is the orthogonality relation of the fermion 
mode functions. Thus, the KK number is not changed at the external 
photon vertex and the remaining sum involves only two 
mode functions. 
The KK sum can be expressed in terms of Bessel functions. However,  
the result is algebraically complicated, so we do not make use of 
this simplification in the numerical evaluation of the diagrams.

Returning to (\ref{diag1text2}) we perform the expansion in the 
small external momenta $p$, $p^\prime$ and pick up the terms linear 
in the momenta, since these produce the electromagnetic dipole 
structure $i \sigma^{\mu\nu} q_\nu$ after application of the Gordon 
identity. Since the various $F$-functions depend only on the square 
of the four-momentum argument, this expansion can be done by 
using, e.g.,  
\begin{equation}
F^+_{L_i}(p^{\,\prime}-l,z,y) = 
F^+_{L_i}(l,z,y) -2 p^\prime \cdot l\,
\frac{\partial}{\partial l^2}F^+_{L_i}(l,z,y) 
+\ldots .
\end{equation}
The derivatives of all required propagator functions are listed in 
appendix \ref{App:Expansions}. In Feynman gauge (where the gauge boson
propagator is proportional to $\eta^{\rho\nu}$), the first term 
in square brackets in  (\ref{diag1text2})  turns into 
\begin{eqnarray}
&& F^+_{L_i}(l,z,y) F^+_{L_i}(l,y,1/T) F^-_{E_j}(l,1/T,x) 
\left\{ \gamma^\nu \!\!\not\!p^{\,\prime} \gamma_\mu
  \gamma_\nu \,l^2 + 
 \gamma^\nu \!\!\!\not l \gamma_\mu \gamma_\nu \,2 p\cdot l
\right\}
\nonumber\\
&&+\,  \gamma^\nu (-\!\!\!\not l) \gamma_\mu \gamma_\nu \,l^2
\,\bigg\{-2 p^\prime\cdot l 
\left(\frac{\partial}{\partial l^2}F^+_{L_i}(l,z,y)\right)
 F^+_{L_i}(l,y,1/T) F^-_{E_j}(l,1/T,x) 
\nonumber\\
&& \hspace*{1cm} -\,2 p\cdot l\,F^+_{L_i}(l,z,y)\,
 \frac{\partial}{\partial l^2} \left[ F^+_{L_i}(l,y,1/T)
   F^-_{E_j}(l,1/T,x) 
\right]
\bigg\}\,,
\end{eqnarray}
where terms odd in the loop momentum $l$ have already been dropped, since 
they integrate to zero. 
The angular integrations can now be done trivially. In the above 
integral we simply replace 
$l^\alpha l^\beta\to \eta^{\alpha\beta}\,l^2/4$ and perform 
the Dirac algebra. Employing the on-shell conditions, 
\begin{equation}
\gamma^\nu \!\!\not\!p^{\,\prime} \gamma_\mu \gamma_\nu 
\to 4 p^{\,\prime}_\mu,
\qquad
 \gamma^\nu \!\!\!\not l \gamma_\mu \gamma_\nu \,2 p\cdot l
\to 2 p_\mu\,l^2
\end{equation}
etc. What remains is a scalar integrand that depends on 
$l^2$ only, and the bulk coordinates $x,y,z$. The integral can be evaluated 
numerically after performing the Wick rotation $l^2\to -l^2$. 
The propagators given explicitly in the appendix refer to these 
Wick-rotated (``Euclidean'') propagators. Diagrams involving 
internal Higgs lines are easier to evaluate, since the number 
of independent bulk coordinates to be integrated reduces to two or
even one, due to the brane localization of the Higgs field.

For the numerical evaluation the loop integral needs to be ultraviolet
and infrared finite. We find that the integration converges for 
large $l$, as expected since the leading one-loop expression for 
the anomalous magnetic moment should not require UV renormalization. 
The diagrams are also IR finite with the exception of the 
internal insertion diagrams B1a and B1b, for which only the sum is 
finite, and the non-abelian diagram W8. The IR divergence arises when 
all internal fermion modes are zero modes. 
There is no IR divergence in the internal insertion 
diagrams B3a, B3b with scalar hypercharge boson exchange, since 
the chirality-flip at the fermion-$B_5$ vertex forbids the 
propagation of zero modes in these diagrams. The existence of 
an IR divergence in B1a, B1b and W8 implies that the purely four-dimensional 
treatment described above potentially misses terms of the
 form $1/\epsilon\times \epsilon$. We discuss this further below.

\subsubsection{External insertions}
\label{subsec:external}

Another comment is necessary on the diagrams with an external 
Higgs insertion, such as B2a. The fermion propagator that connects 
the external Higgs vertex to the internal gauge vertex is (in this
example)  
\begin{equation}
\Delta_i^L(p,x,1/T)P_R = -F_{L_i}^+(p,x,1/T) \slash{p} P_R + 
d^-F_{L_i}^+(p,x,1/T) P_R\,.
\end{equation}
If this expression contained a relevant $1/p^2$ contribution from the zero
mode, the diagram would be long-distance sensitive, and the expansion 
in the small external momenta would be invalid. The second term 
is a mass term, see (\ref{KKdecFermionPropagator}), 
to which the zero mode cannot contribute, since 
$m_0=0$. Hence this term can be expanded in $p$, and since it depends
only on $p^2$, we can simply set $p=0$ to linear order. The zero 
mode is present in the first term, which contains 
$i \slash{p}/p^2 \times f_{L_i}^{(0)}(x) f_{L_i}^{(0)}(1/T)$. 
If the one-particle pole at $p^2=0$ remains in the final answer,
then this part of the external Higgs insertion into a zero mode needs 
to be amputated; it 
corresponds to the first term in the sum of tree diagrams that 
sums to the SM lepton mass matrix. After amputation 
the remaining short-distance 
contribution is a one-loop correction to the chirality preserving 
$\bar L_i L_i\gamma$ vertex. The general $\bar L_i L_i\gamma$ vertex 
function with off-shell zero-mode fermions can be decomposed 
into ``on-shell'' and ``off-shell'' terms as follows:
\begin{equation}
\label{OffShellVertex}
\Lambda^\mu = \Lambda^\mu_{\rm on}
                   + \slash{p}' \Lambda^\mu_{\rm off,\,p'} 
                   + \Lambda^\mu_{\rm off,\,p} \,\slash{p}\,.
\end{equation}
The first piece constitutes the correction to the on-shell 
$\bar L_i L_i\gamma$ vertex and is not relevant to the anomalous magnetic 
moment. It constitutes the long-distance piece that must be amputated. The
off-shell term with a $\slash{p}'$ vanishes, since diagram B2a 
has an on-shell external $p^\prime$ line on the right.
However, the $\Lambda^\mu_{\rm off,\,p} \,\slash{p}$ term 
cancels the propagator factor $\slash{p}/p^2$ from the internal 
fermion line, and represents a short-distance contribution that 
must be added to the matching coefficient of the dipole operator 
and hence contributes to $(g-2)_\mu$.\footnote{See \cite{Beneke:2004rc} for a 
related discussion in a different context.} 

It follows that in all external insertion 
diagrams\footnote{ 
B2a, B4a, W1, W2, W4, W5, W6a, W6b in 
figure~\ref{fig:diagsall}. Diagrams B2b, B4b are treated similarly, 
with the appropriate modifications for an insertion into a line 
with momentum $p^\prime$.} 
we have two different contributions to consider.
The first one can be obtained by replacing
\begin{eqnarray}
&& \Delta_i^L(p,x,1/T) \to 
d^-F_{L_i}^+(p=0,x,1/T) = \sum_{n\not=0}\frac{-i}{m_n} 
g_{L_i}^{(n)}(x) f_{L_i}^{(n)}(1/T)
\nonumber\\
&&\hspace*{1cm}
= -i\,\frac{k^4 x^{5/2}}{T^{3/2}}\,
\frac{(kx)^{1/2-c_{L_i}}-(kx)^{c_{L_i}-1/2}}
{\left(\frac{T}{k}\right)^{\!c_{L_i}-1/2}-
\left(\frac{k}{T}\right)^{\!c_{L_i}-1/2}}, 
\label{extfirstterm}
\end{eqnarray}
where the last expression follows from the explicit expression for 
the propagator functions given in the appendix. We note that 
the KK mode contribution to $d^-F_{L_i}^+(p,x,1/T)$ is not 
suppressed due to the large KK masses of order $T$. Thus, the 
external insertion diagrams are not suppressed relative to the 
internal insertions, contrary to what has been assumed in 
the previous literature, where the external insertions have been 
neglected.\footnote{See, however, the arXiv version \cite{Csaki:2010ajv3} 
of \cite{Csaki:2010aj}.} 

The second contribution stems from the off-shell vertex terms and 
explicitly contains the zero-mode propagator. It can be obtained via
the replacement  
\begin{eqnarray}
&& \Lambda^\mu \Delta_i^L(p,x,1/T) \to 
\Lambda^\mu_{\rm off,\,p} \,\slash{p} \times  
i \slash{p}/p^2 
\times f_{L_i}^{(0)}(x) f_{L_i}^{(0)}(1/T) 
= i\,\Lambda^\mu_{\rm off,\,p} \, 
f_{L_i}^{(0)}(x) f_{L_i}^{(0)}(1/T)\,,
\qquad
\end{eqnarray}
where the internal zero-mode pole is cancelled. We will refer to these 
second contributions as ``off-shell'' terms. They require the 
separate computation of the $\Lambda^\mu_{\rm off,\,p}$ structure of the 
$\bar L_i L_i\gamma$ vertex, which can be obtained from the expansion 
in the small external momenta. The expansion must now be performed to 
second order, since we need the $\slash{p} p^\mu$,  
$\slash{p} p^{\prime \mu}$ structures in the vertex 
function $\Lambda^\mu$. 

The second contribution appears to be suppressed by a  
power of the lepton mass matrix due to the two 
additional fermion zero mode profiles. The expansion of $\Lambda^\mu$ 
to second order brings an additional factor $1/T$ from the scale of 
the loop momentum. The ratio of the second to the first 
contribution can therefore be estimated as 
\begin{equation}
\label{offshellestimate}
\frac{\frac{i}{T} f_{L_i}^{(0)}(x) f_{L_i}^{(0)}(1/T)}
{d^-F_{L_i}^+(p=0,x,1/T)} \approx 
 - \left[\sqrt{\frac{1-2 c_{L_i}}{1-(T/k)^{1-2 c_{L_i}}}}\,\right]^2\,,
\end{equation}
where we put $x=1/T$ for the second estimate. 
For symmetric bulk mass parameters $c_{E_i} = - c_{L_i}$ such that 
$f_{L_i}^{(0)}(x) = g_{E_i}^{(0)}(x)$, and anarchic 
Yukawa couplings this is indeed of order $m_{\ell_i}/v$, 
see (\ref{4dyukawa}). For the other special choice $c_L\to 0.5$ we find
\begin{equation}
\frac{\frac{i}{T} f^{(0)}_{L_i}(x)f^{(0)}_{L_i}(1/T)}
{d^- F^+_{L_i}(p=0,x,1/T)}\to - \frac{1}{T x \log(kx)}\;.
\end{equation}
For $x$ close to $1/T$ this approaches $-1/\log(1/\epsilon) \approx -1/35$,  
and we obtain a lepton-mass independent suppression. Thus, the 
``off-shell'' terms are expected to be subleading for standard choices 
of the bulk mass parameters, which is indeed found in the exact numerical 
evaluation discussed in Sec.~\ref{subsec:singleflavour}.
 
\subsubsection{Scheme independence}
\label{subsec:scheme}

We mentioned above that the naive four-dimensional calculation 
of the sum of the individually IR divergent diagrams 
B1a+B1b might be wrong, since it could miss terms of the 
form $1/\varepsilon \times \varepsilon$. A similar issue appears in 
the finite result (\ref{betares}) for diagram (b) from 
figure~\ref{fig:4Ddiags}, for which we would obtain zero, 
if the Dirac algebra were performed naively in four dimensions. 
The finite results of both calculations depend on the treatment 
of $\gamma_5$ in $d$ dimensions. In the following we show 
algebraically that the scheme dependence cancels if the 
same scheme is applied consistently to both parts, and that 
the finite result agrees with what we obtained above.

Both parts are limits of a diagram in the full 5D theory with fields 
expanded around the electroweak vacuum and massive zero-mode 
fermions. This diagram is finite (as far as the $\sigma^{\mu\nu} 
q_\nu$  structure is concerned). The divergences arise only when 
this diagram is split into a short-distance contribution 
from the KK scale, diagrams 
B1a+B1b, and a long-distance contribution from the electroweak 
and lepton mass scale, represented by the four-fermion operator 
insertion diagram. Before factorizing the full-theory 
diagram we can freely anti-commute $\gamma_5$. 
The convention that corresponds to (\ref{betares}) amounts to 
eliminating all chiral projectors except for two placed 
at the operator vertex $\gamma^\rho P_L \otimes \gamma_\rho P_R$.

For consistency, the same convention must be applied to the starting 
expression for the diagrams B1a+B1b, which therefore reads
\begin{eqnarray}
\Gamma^\mu(p,p^\prime) &=& 
(ieQ_\mu)(ig'_5)^2\left(\frac{-iT^3}{k^3}\right) 
\frac{v}{\sqrt{2}} \,[U^\dagger]_{2m}
y^{\rm (5D)}_{mn}V_{n2} ~ \frac{Y_L}{2}\frac{Y_E}{2} 
\, f^{(0)}_{L_m}(1/T)g^{(0)}_{E_n}(1/T)
\nonumber\\
&& \int^{1/T}_{1/k}\!\!dx dz \,
\frac{{f_{L_m}^{(0)}}^2\!(x)}{(k x)^4}\,
\frac{{g_{E_n}^{(0)}}^2\!(z)}{(k z)^4}\,
\,\mu^{2 \varepsilon}
\int \!\!\frac{d^dl}{(2\pi)^d}\,\Delta^{\rho\nu}_{\mbox{\tiny ZMS}}(l,x,z)
\nonumber \\
  &&  i^3\,\bar{L}_m(p^\prime)
  \gamma_\rho P_L \frac{\slash p^{\,\prime}-\slash l}{(p^\prime-l)^2}
  \left[\gamma^\mu \frac{\slash p-\slash l}{(p-l)^2}
+\frac{\slash p^{\,\prime}-\slash l}{(p^\prime-l)^2}\gamma^\mu \right]
  \frac{\slash p-\slash l}{(p-l)^2} \gamma_\nu P_R E_{n}(p)\,.\qquad
\label{ddimB1}
\end{eqnarray}
To obtain this result, we write down 
the explicit expressions from appendix~\ref{ap:diags} and multiply 
it by $[U^\dagger]_{2i} V_{j2}$ so that the external states 
correspond to the SM muons in the mass eigenbasis. All fermion 
propagators can be replaced by zero-mode propagators, since only 
this term is IR divergent, and the chiral projectors are placed as 
discussed above. Performing the integration over the 
bulk coordinate $y$ of the photon vertex, we arrive at (\ref{ddimB1}).

The IR divergence comes from 
$l\ll T$, hence we can set $l=0$ in the zero-mode subtracted 
gauge-boson propagator. Moreover, only the $\eta^{\rho\nu}$ 
structure contributes to the IR divergence. After these
simplifications, we can identify the expression (\ref{bijstart}) 
for $b_{ij}$. We perform the expansion in external momenta 
and simplify the Dirac algebra as much as possible. When the Dirac 
structure is multiplied by a $1/\varepsilon$ pole we do {\em not}  
anti-commute $\gamma_5$ with $\gamma_\alpha$, but since the 
external momenta are four-dimensional we may use that 
$\slash{p}^{(\prime)}\gamma_5 = - \gamma_5\,\slash{p}^{(\prime)}$. 
The result of all this is that the IR sensitive contribution 
in the matching coefficient of the electromagnetic dipole 
operator is given by
\begin{eqnarray}
\Gamma^\mu_{\mbox{\scriptsize IR}}(p,p^\prime) &=&
i e Q_\mu \frac{m_{l_k} \beta_{2kk2}}{16\pi^2 T^2} 
\,
\bar u(p^\prime,s^\prime)\Big[2\,(p+p^\prime)^\mu P_R 
- \frac{1}{\varepsilon}\, \gamma^\rho P_L 
i\sigma^{\mu\nu}q_\nu \gamma_\rho P_R
\Big] u(p,s).\quad
\label{IRcontribution}
\end{eqnarray}
The second term in brackets cannot be further reduced without
assumptions on $\gamma_5$. Since 
$ \gamma^\rho P_L i\sigma^{\mu\nu}q_\nu \gamma_\rho P_R$
vanishes in four dimensions, 
the entire expression is finite, but scheme-dependent.

In the one-loop matrix element calculation of the four-fermion 
operator that led to (\ref{betares}) we assumed the 
naive dimensional regularization (NDR) scheme to obtain the 
result in the last line. We repeat the 
calculation in an arbitrary scheme, leaving out the 
$(\beta_{2kk2} \leftrightarrow \beta_{k22k}, \,P_L \leftrightarrow 
P_R)$ term, which is related  to the matching coefficient 
of the electromagnetic dipole operator with $L$ and $E$ exchanged.
It is easy to see that the 
ultraviolet $1/\varepsilon$ pole in the one-loop matrix element 
(\ref{betares}) that is relevant to the electromagnetic 
dipole transition is obtained correctly by approximating 
$(\slash{l} +m_{\ell_k})
\gamma^\mu (\slash{l}-\slash{q}+m_{\ell_k}) \to -m_{\ell_k} 
\gamma^\mu\slash{q}$. Dropping terms proportional to $q^\mu$, 
we obtain 
\begin{eqnarray}
\Gamma^\mu_{\mbox{\scriptsize UV}}(p,p^\prime) &=&
i e Q_\mu \frac{m_{l_k} \beta_{2kk2}}{16\pi^2 T^2} 
\,\frac{1}{\varepsilon}\,
\bar u(p^\prime,s^\prime) \Big[
\gamma^\rho P_L i\sigma^{\mu\nu}q_\nu \gamma_\rho P_R
\Big]u(p,s).
\label{UVcontribution}
\end{eqnarray}

Now we can make the following two observations:
\begin{itemize}
\item In the NDR scheme (anti-commuting $\gamma_5$), between 
$\bar u(p^\prime,s^\prime) [\ldots]u(p,s)$,
\begin{equation}
\gamma^\rho P_L i\sigma^{\mu\nu}q_\nu \gamma_\rho P_R = 
2\varepsilon  \,(p+p^\prime)^\mu P_R = 
-2\varepsilon\,i\sigma^{\mu\nu}q_\nu P_R \,,
\end{equation}
where the Gordon identity is used in the last step, and the term not 
related to the electromagnetic dipole structure is dropped.
Then (\ref{IRcontribution}) is zero and (\ref{UVcontribution}) 
coincides with (\ref{betares}). Thus we reproduce the previous results.
\item In an arbitrary scheme the scheme-dependent terms drop out when 
summing $\Gamma^\mu_{\mbox{\scriptsize IR}}(p,p^\prime)$ and 
$\Gamma^\mu_{\mbox{\scriptsize UV}}(p,p^\prime)$,
and the result agrees with the one obtained before in the NDR scheme.
\end{itemize}
This proves that the result is scheme-independent, and that the naive 
four-dimensional treatment of diagrams B1a, B1b fortuitously 
gives the correct result.

A similar issue arises for the IR-sensitive diagram W8, but in this case 
it turns out that the naive four-dimensional calculation needs to be 
corrected by a finite term of the form $1/\varepsilon \times \varepsilon$ that is 
present when  working consistently in $4-2 \varepsilon$ dimensions.  
This finite term is scheme-dependent in general, and the scheme-dependence 
is compensated by the effective theory diagram (d1) in 
figure~\ref{fig:4Ddiags}.  The zero result (\ref{deltares}) for diagram (d1)   
corresponds to a full-theory diagram in which the charged-Higgs 
vertex in (\ref{dim6broken}) is written as 
$\bar \psi_iP_R\gamma_\mu\nu_j$ rather than 
$\bar \psi_i \gamma_\mu P_L\nu_j$. We must apply this convention
consistently to diagram W8 as well. The IR sensitive part of W8 can be 
obtained from \eqref{diag16} by replacing the internal fermion by the zero 
mode. Then, rotating the external states to the mass eigenbasis 
by multiplying  \eqref{diag16} by $[U^\dagger]_{2 i} V_{j2}$, 
the relevant part of W8 simplifies to 
\begin{align}
 \Gamma^\mu(p,p^\prime) =&
 (ig_5)^2 (ie_5)
\left(\frac{\tau^a}{2} \left[\frac{Y_{\Phi}}{2}+\frac{\tau^3}{2}\right]
\frac{\tau^a}{2}\right)_{\!\!22} 
\left(\frac{-iT^3}{k^3}\right)
\frac{[U^\dagger]_{2i}  vy^{\rm (5D)}_{ij} V_{j2}}{\sqrt{2}} 
 f^{(0)}_{L_i}(1/T)g^{(0)}_{E_j}(1/T)
\nonumber\\
 &
\times \mu^{2 \varepsilon}\int \!\!\frac{d^dl}{(2\pi)^d}\, 
\int^{1/T}_{1/k}\!\!dx \,
 \frac{{f_{L_i}^{(0)}}^2(x)}{(k x)^4} 
 \Delta^{\rho\nu}_{\mbox{\tiny ZMS}}(p^\prime-l,x,1/T) 
f^{(0)}_{\gamma}(1/T)
\nonumber\\
 &
\times i^3 \,\bar L(p^\prime) P_R 
\frac{ (p^\prime-l)_\nu\gamma_\rho\slashed{l}( p+ p^\prime-2 l)^\mu}
{(p-l)^2 (p^\prime-l)^2 l^2} P_R E(p)\,.
\end{align}
Exploiting  $q_\nu  \,\Delta^{\rho\nu}_{\mbox{\tiny ZMS}}(q,x,1/T) 
= q_\rho \,\Delta_{\parallel}^{\mbox{\tiny ZMS}}(q,x,1/T,\xi)$ , 
which follows from (\ref{gaugepropansatz}),
\begin{align}
 \frac{T^3}{k^3}\frac{y^{\rm (5D)}_{ij} v}{\sqrt{2}} 
f^{(0)}_{L_i}(1/T)g^{(0)}_{E_j}(1/T)
   = \sum_k m_k\;U_{ik}\mathds{1}_{kh} V^{\dagger}_{hj}\,,
\end{align}
and $e_5 f^{(0)}_{\gamma}(x)=e$,  this simplifies further to 
\begin{eqnarray}
\Gamma^\mu(p,p^\prime) &=&
g_5^2 (ie)\, \frac{1}{2}\, m_\mu \sum_i \,[U^\dagger]_{2i} U_{i2} 
\,\mu^{2 \varepsilon}\int \!\!\frac{d^dl}{(2\pi)^d}\, 
\int^{1/T}_{1/k}\!\!dx \,
 \frac{{f_{L_i}^{(0)}}^2(x)}{(k x)^4} 
 \Delta_{\parallel}^{\mbox{\tiny ZMS}}(p^\prime-l,x,1/T,\xi)
\nonumber\\
 &&
\times \bar L(p^\prime) P_R 
\frac{(\slashed{p}^\prime-\slashed{l}) \slashed{l}( p+ p^\prime-2 l)^\mu}
{(p-l)^2 (p^\prime-l)^2 l^2} P_R E(p)\,.
\end{eqnarray}
Since the difference to the naive four-dimensional calculation arises only 
from $l\ll T$, we apply a cut-off $l\ll \Lambda\ll T$ to the loop momentum. 
This allows us to set the gauge-boson propagator momentum to zero 
without generating a spurious UV divergence, and 
to identify the coefficient function $c_{3,i}$, since  
$\Delta_{\parallel}^{\mbox{\tiny ZMS}}(0,x,1/T,\xi)= 
\Delta^{\mbox{\tiny ZMS}}_{\perp}(0,x,1/T) $. Expansion 
to linear order in $p$, $p^\prime$ results in 
\begin{eqnarray}
\Gamma^\mu(p,p^\prime) &=&
(-e)\, \frac{m_\mu }{T^2}\sum_i \,c_{3,i} [U^\dagger]_{2i} U_{i2} 
\,\mu^{2 \varepsilon}\int\limits^{\Lambda} \!\!\frac{d^dl}{(2\pi)^d}\, 
\frac{1}{l^4}\left[(p+p^\prime)^\mu-
\frac{4  l^\mu (p+p^\prime)\cdot l}{l^2}\right]
\nonumber\\
 &&
\times \bar L(p^\prime) P_R E(p)\,.
\end{eqnarray}
The difference between the correct and naive treatment comes from the 
second term in square brackets, where we must use 
$l^\mu l^\nu \to 1/d\times l^2 \eta_{\mu\nu}$ instead 
of $1/4$ as done in the four-dimensional treatment of the 
integrand.\footnote{In this case the square bracket vanishes, which 
explains the absence of an explicit IR divergence.} Using the 
definition (\ref{couplingdefs}) of $\gamma_{3,ij}$, we therefore find that 
the difference between the $d$-dimensional and four-dimensional 
result that must be added to the naive four-dimensional 
treatment of W8 is
\begin{eqnarray}
\Delta\Gamma^\mu(p,p^\prime) &=&
(-e)\, \frac{\gamma_{3,22} m_\mu }{T^2}\, 
 \left(-\frac{4}{d}+1\right) \,
\mu^{2 \varepsilon}\int\limits^{\Lambda} \!\!\frac{d^dl}{(2\pi)^d}\, 
\frac{1}{l^4}\times \bar L(p^\prime) P_R (p+p^\prime)^\mu E(p)
\nonumber\\
 &= &
ieQ_\mu\, \frac{\gamma_{3,22} m_\mu }{32 \pi^2 T^2}\, 
\times \bar L(p^\prime) P_R (p+p^\prime)^\mu E(p) 
+ {\cal O}(\varepsilon)\,.
\label{IRTermInW8}
\end{eqnarray}
The result is independent of the arbitrary cut-off as it should be 
and is part of $a_{W,ij}$, the coefficient of the class-0 operator 
$\bar L_i\tau^a \Phi \sigma_{\mu\nu} E_j \,W^{a,\mu\nu}$.\footnote{
One could choose a scheme where W8 agrees with the 
purely four-dimensional treatment. In this scheme we would find 
a non-zero contribution from the one-loop diagrams (d1) (and (d2)).
Our choice corresponds to the NDR scheme.}
We can check (\ref{IRTermInW8}) by noting that it should 
be minus the UV sensitive $\varepsilon\times 1/\varepsilon$ term in 
the matrix element calculation that led to (\ref{ChargedHiggsOperatorsUV}).  
Since $Q_\mu=-1$ this is indeed the case.

In principle the issue of scheme-dependence could also occur for 
diagrams B5a, B5b and W3, related to the effective theory diagrams 
(c1), (c2). However, it can be checked that in this case 
our convention for the position of the chiral projection operators 
in (\ref{dim6broken}) implies that diagrams (c1) and (c2) are always 
zero and accordingly there are no $\varepsilon\times 1/\varepsilon$ terms 
in B5a, B5b and W3.

\subsubsection{Gauge invariance}
\label{gaugeinvariance}

To check the gauge independence of the matching coefficient, 
we performed the calculation in 5D $R_\xi$ gauge, and verified that our 
numerical result is independent of $\xi$ for a range 
of values of $\xi$. 
We also checked analytically that the $\xi$-dependent terms cancel. The 
proof is too lengthy to be presented here explicitly, but we find it 
instructive to discuss the structure of the argument and the key algebraic 
identities.

It is clear that the subsets of diagrams with hypercharge and SU(2) 
gauge boson exchange must be separately gauge-independent. It is also 
evident that the five diagrams Bxa and the other five hypercharge exchange 
diagrams Bxb must be separately invariant, since the dependence on 
the bulk mass parameters $c_{L_i}$ and $c_{E_j}$ is different whether 
the Higgs coupling to the fermion is to the left or to the right of 
the one of the external photon. Hence we consider first B1a to B5a, 
see figure~\ref{fig:diagsall}.

The gauge-parameter dependence arises from the scalar gauge boson 
propagator $\Delta_5$ (see (\ref{gaugeprop5def}), (\ref{gaugeprop5})) 
and the longitudinal part 
\begin{equation}
\frac{l^\rho l^\nu}{l^2}\,\Delta_{\parallel}(l,z,z^\prime,\xi) 
= \frac{l^\rho l^\nu}{l^2}\,\Delta_{\perp}(l/\sqrt{\xi},z,z^\prime)
\end{equation}
of the vector gauge boson propagator. In diagrams 
B1a, B2a and B5a we therefore 
replace $\Delta^{\rho\nu}(l,z,z^\prime)$ by the previous expression. 
Using (\ref{eq:gaugemoderelation}), 
the mode representation of the propagator,  
and the completeness relation 
$\sum_n f_{B_5}^{(n)}(z) f_{B_5}^{(n)}(z^\prime) = 
kz\,\delta(z-z^\prime)$ we derive the identity
\begin{equation}
\Delta_5(l,z,z^\prime) = \frac{ikz}{l^2}\,\delta(z-z^\prime) 
-\frac{1}{l^2}\,\partial_z\partial_{z^\prime}
\Delta_\perp(l/\sqrt{\xi},z,z^\prime),
\label{del5remove}
\end{equation}
which is used to eliminate the propagator of the 
scalar component of the gauge boson in diagrams B3a and B4a. 
At this point all gauge-dependence is in the expression 
$\Delta_\perp(l/\sqrt{\xi},z,z^\prime)$, which appears in all 
five diagrams. The delta-function term in (\ref{del5remove}) 
results in a $\xi$-independent expression that can be 
ignored.\footnote{This is not true in the non-abelian diagrams 
with more than one internal gauge boson line.} 

To proceed we need 
an identity that relates the different diagram topologies 
-- internal insertions (B1a, B3a), external insertions (B2a, B4a), 
and Higgs (B5a) diagrams. This is obtained as follows. 
After using (\ref{del5remove}), we integrate by parts the bulk 
coordinate derivatives on $\Delta_\perp$, which yields the corresponding 
derivatives on the two fermion propagators (B3a), or 
one propagator and one external zero mode (B4a) adjacent to 
the gauge boson vertex. Now we can use successively identities such as 
\begin{eqnarray}
&&\partial_z\left(\frac{1}{(k z)^4}\,\Delta_i^L(p_1,x,z) 
\gamma_5 \Delta_i^L(p_2,z,y)\right) 
= 
\frac{1}{(k z)^4}\,\Delta_i^L(p_1,x,z) \left(\slash{p}_1-\slash{p}_2\right)
\Delta_i^L(p_2,z,y) 
\nonumber\\[0.2cm] 
&& \hspace*{1cm} +\, 
i \Delta_i^L(p_1,x,z) \delta(z-y) - i \Delta_i^L(p_2,z,y) \delta(x-z)
\label{fermionrel}
\end{eqnarray}
which follow from the defining equations for the fermion propagator 
(\ref{InverseDiracOp}), (\ref{defDiracOp}).
If one of the propagators on the left-hand side of (\ref{fermionrel}) 
is an external zero mode, the corresponding delta-function and momentum 
term is absent. When this is used in B3a and B4a, the first 
term on the right-hand side of (\ref{fermionrel}) cancels precisely 
the gauge-dependent terms of B1a and B2a, respectively, leaving over 
only the delta-function terms and the Higgs diagram B5a. 

Some of the delta-function terms vanish, since the remaining integral 
is independent of $p^\prime$ ($p$), such that the loop integral must 
result in $\slash{p} E(p)_j=0$ ($\bar L_i(p^\prime)\slash{p}^\prime=0$). 
The surviving two terms correspond to diagrams with one deleted propagator. 
Inspection shows that these two terms and the Higgs diagram B5a are 
identical up to a different hypercharge prefactor. The prefactors 
are such that we obtain 
\begin{equation}
Y_L\,\mbox{[B3a]} - Y_E \,\mbox{[B4a]}  - Y_\phi \,\mbox{[B5a]} = 
-1+2-1 = 0,
\end{equation}
which proves the gauge-parameter independence of the abelian diagrams 
B1a to B5a. The algebra works out in the same way for the symmetric 
diagrams B1b to B5b.

The proof of gauge invariance of the SU(2) gauge boson contributions 
though similar is more involved and we only make a few remarks. First, 
the three abelian diagram topologies W1 to W3 are by themselves 
gauge-independent, and the proof proceeds as above. The remaining 
eight genuinely non-abelian diagrams W4 to W10 can 
also be split into two groups using the structure of the 
vertices and propagators. It is useful to begin with diagram 
W5 containing two internal scalar W bosons, 
with $\gamma \,W_5W_5$ vertex $(p-l)_\mu+(p^\prime-l)_\mu$, and to 
proceed by treating the two terms separately. To show the gauge 
cancellation, identities such as 
\begin{equation}
\int_{1/k}^{1/T}\frac{dy}{k y}\,
\left[\partial_y\Delta_\parallel(k,z,y,\xi)\right]
\partial_y\Delta_\perp(k,y,x) = 
- \int_{1/k}^{1/T}\frac{dy}{k y}\,
\Delta_\parallel(k,z,y,\xi) \left[y\partial_y\frac{1}{y}\partial_y
\Delta_\perp(k,y,x)\right]
\end{equation}
must be used, where the derivative term on the right-hand-side 
can subsequently 
be eliminated by the equation of motion (\ref{1stEqGaugeProp}) 
for the gauge-boson propagator, and the boundary terms in the 
partial integration vanish due to the orbifold boundary conditions on 
the branes. To simplify the proof of gauge-invariance 
of the genuinely non-abelian diagrams we assumed that the external 
photon has physical transverse polarization, 
$\epsilon^*\cdot q =0$.

\subsubsection{Anapole moment and one-particle reducible diagrams}

The calculation of the diagrams of figure~\ref{fig:diagsall} 
does not produce the expected form factor 
\begin{equation}
(p+p^\prime)^\mu \,\bar u(p^\prime,s^\prime) P_R \,u(p,s)\,,
\end{equation}
which can be converted into the anomalous magnetic moment structure 
via the Gordon identity.
Instead, we find different coefficients of $p^\mu$ and 
$p^{\prime\mu}$, that is, there is an extra $q^\mu$ term.
More precisely, we find a nonzero coefficient $C_q$ for 
the structure $\bar L \,q^\mu P_R E$ arising only from the
genuinely non-abelian diagrams W4 to W10 in figure~\ref{fig:diagsall}. 
It can be shown that the same diagrams, but with an incoming doublet and
outgoing SU(2) singlet field lead to the appearance of the structure
$\bar E \,q^\mu P_L L$ with coefficient $-C_q$. Thus, the 
vertex function (\ref{formfactor}) contains the parity-violating 
term
\begin{equation}
\Gamma^\mu(p,p^\prime) \supset i e Q_\mu\,C_q \,
\bar u(p^\prime,s^\prime) q^\mu\gamma_5 u(p,s).
\label{qformfactor}
\end{equation}
We find analytically that the overall dependence of $C_q$ on 
the external lepton mass is
 \begin{equation}
   C_q=2 m_\ell C^{\rm red}_q
 \end{equation}
where $C^{\rm red}_q$ is independent of the mass of external states. 
Therefore (\ref{qformfactor}) can be rewritten as 
\begin{equation}
\Gamma^\mu(p,p^\prime) \supset i e Q_\mu\,C_q^{\rm red} \,
\bar u(p^\prime,s^\prime)  \slashed{q} q^\mu\gamma_5 u(p,s).
\label{qformfactor2}
\end{equation}
The presence of such a term does not necessarily violate the
Ward identity for the vertex function. 
The most general vertex function  of on-shell fermions compatible with 
U(1)$_{\rm em}$ gauge invariance is given by (see, 
e.g.,~\cite{Musolf:1990sa,Bernreuther:1996dh})
\begin{eqnarray}
\Gamma^\mu(p,p^\prime) &=& i e Q_\mu\,\bar u(p^\prime,s^\prime) 
\bigg[\gamma^\mu F_1(q^2) +\frac{i\sigma^{\mu\nu}q_\nu}{2 m_\mu}
F_2(q^2) +\frac{i\sigma^{\mu\nu}q_\nu}{2 m_\mu}\,\gamma_5\,
F_3(q^2) 
\nonumber\\
&& \, + \big(q^2 \gamma^\mu -\slashed{q}q^\mu\big)\gamma_5
\,F_4(q^2) \bigg] u(p,s)
\label{formfactorextended}
\end{eqnarray}
The form factor $F_4$ is related to the so-called anapole moment of the 
lepton. Contrary to the magnetic and electric dipole moment the 
anapole moment is gauge-dependent \cite{Musolf:1990sa} and 
by itself not a physical observable. It is, however, non-zero in the 
Standard Model \cite{Dombey:1980eu}. Since we dropped the 
$q^2 \gamma_\mu\gamma_5$ terms in the course of our calculation, 
the left-over $q^\mu$ terms (\ref{qformfactor2}) should be interpreted 
as the RS contribution to the anapole moment of the muon. They are 
irrelevant for the dipole moment calculation. 
The $q^\mu \gamma_5$ term can be removed either analytically or 
by averaging the coefficients of $p^\mu$ and ${p^\prime}^\mu$ to 
extract the coefficient of $(p+p^\prime)^\mu$.

%%%%%%%%%%%%%%%%%%%%%%%%%%%%%%%%%%%%%%%%%%%%%%%%%%%%%%%%%%%%%%%
\begin{figure}[t] 
\begin{center} 
\vskip0.3cm
\includegraphics[width=8cm]{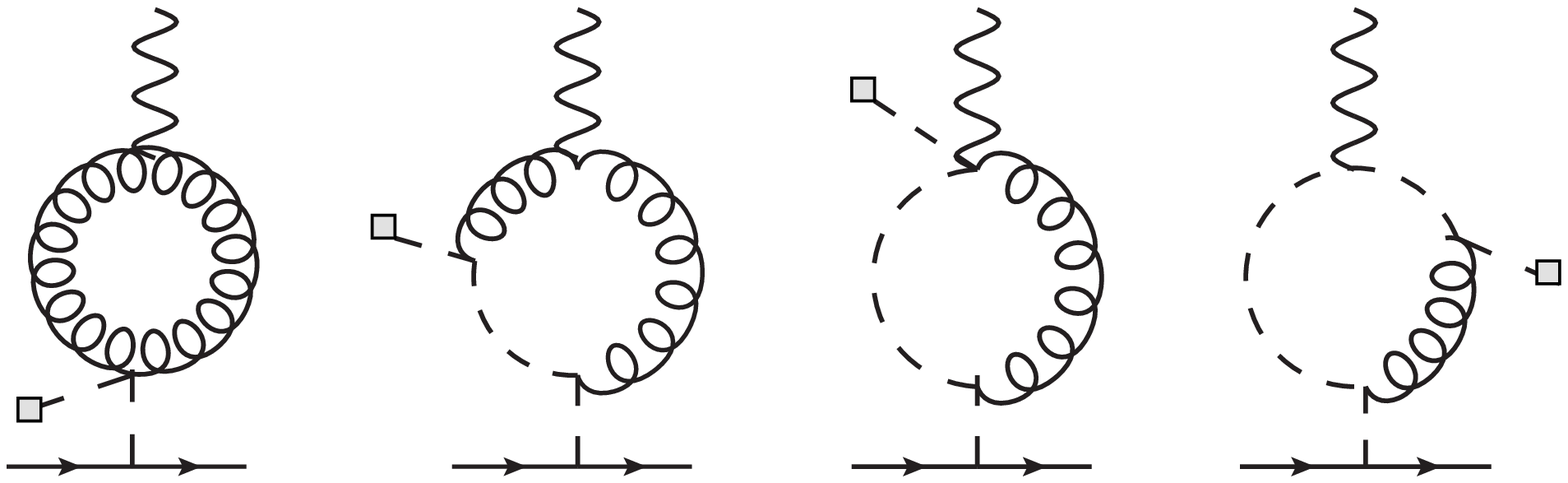}  
\includegraphics[width=11cm]{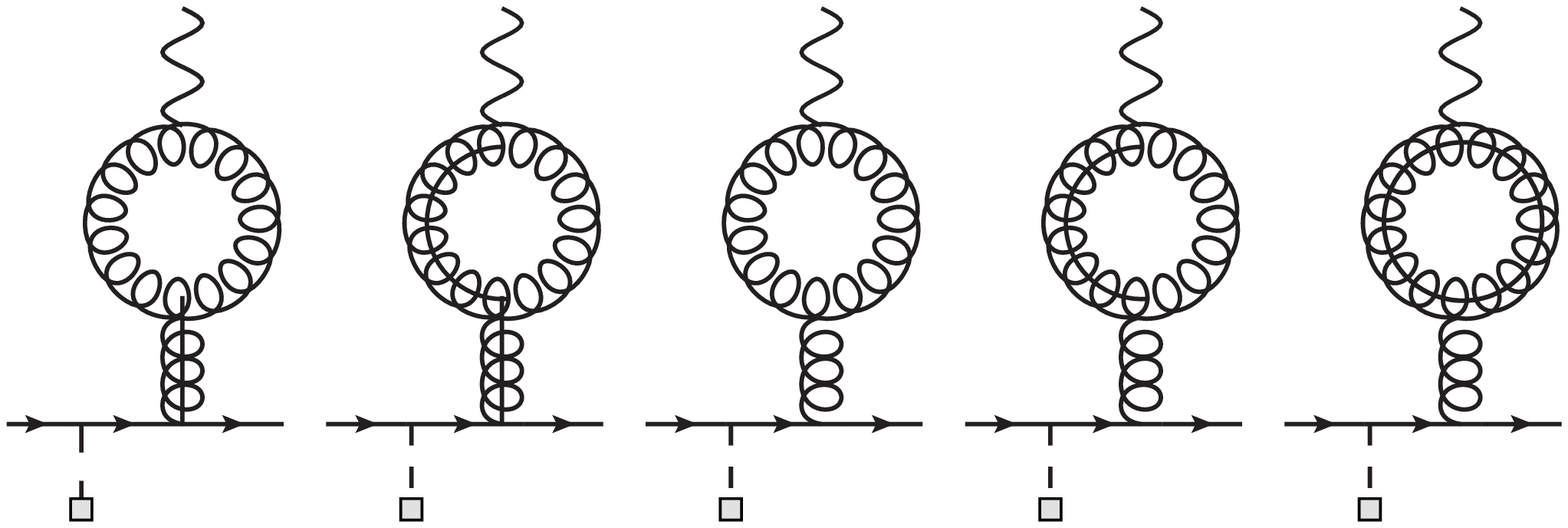}  
\end{center} 
\caption{One-particle reducible diagrams with 
short-distance contributions.}  
\label{fig:1PRdiags}  
\end{figure}  
%%%%%%%%%%%%%%%%%%%%%%%%%%%%%%%%%%%%%%%%%%%%%%%%%%%%%%%%%%%%%%%

In fact, there is an entire class of one-particle reducible (1PR) diagrams 
shown in figure \ref{fig:1PRdiags}, which generates only 
terms $\propto q^\mu \gamma_5$. 
The diagrams in the first row give a short-distance contribution if 
the internal Higgs propagator connecting to the fermion line 
is cancelled. In this case
the contributions are finite and have exactly the same structure as
the $q^\mu$ term arising from the diagrams of figure~\ref{fig:diagsall} 
discussed above. The diagrams in the second row of figure 
\ref{fig:1PRdiags} are different in that they are not only gauge 
dependent but also suffer from UV divergences. This is not a 
problem per se as the anapole moment is known to have both features
already in the SM \cite{Musolf:1990sa}.  In any case, we do not have to 
consider the contributions from the self-energy like 1PR diagrams 
further, since we are not interested in the anapole moment.  

\subsubsection{Numerical evaluation}
\label{subsec:numerical}

The evaluation of the Wilson coefficient of the electromagnetic dipole 
operator requires the numerical evaluation of up to four-dimensional
integrals. The integration variables are typically three 5D 
coordinates and the modulus of the loop momentum. 
For $l \gg T$ the Bessel functions in the propagators 
behave like exponentials and it is possible to determine the 
asymptotic behaviour of the integrand analytically. The numerical 
integration of the integrand in the high-momentum region 
is time-consuming and introduces significant uncertainties. 
The origin of this problem is large cancellations 
between the different Bessel functions in the propagators.
Thus, we perform the 5D position integrals over the whole (length)
of the fifth dimension, but limit the integral over the modulus 
of the momentum to the region below some momentum cut-off. 
In the high-momentum region we replace the integrand by its 
leading asymptotic behaviour, and integrate it from the cut-off 
to $l =\infty$. 
In practice a cut-off of $\mathcal{O}(100T)$ is already sufficiently high.  
In order to estimate the uncertainty associated with this 
approximation we vary the cut-off in the interval from $60 T$ to $150 T$;
the typical uncertainties are of the order of a few per mille.

Some of the propagator functions feature discontinuous behaviour 
if two bulk coordinates coincide (see~appendix~\ref{App:Propagators}). 
Both speed and 
precision of the evaluation can be improved by explicitly splitting the 
integration domain into sectors where the integrand is continuous.
This approach appears to be slightly more efficient than using a 
coordinate transformation to make the discontinuities manifest by 
mapping them 
onto the coordinate axes. Separating the different regions leads to 
a reduction of uncertainties by about a factor of 2 to 10, 
depending on the amount of 
discontinuous propagators (and thus regions) in the diagram. 

The calculation is performed in general covariant gauge. 
Since the gauge invariant subsets of
Feynman diagrams are known (see~Sec.~\ref{gaugeinvariance})
 a numerical verification of gauge
invariance provides a powerful check for our code.   
Furthermore, the error estimates provided by 
standard integration routines can be checked by using 
the small spurious residual dependence of $\Delta a_\mu$ 
on the gauge parameter $\xi$. Very large gauge parameters
are numerically difficult to handle, but the limit $\xi \to \infty$, 
i.e. unitary gauge, can be performed analytically on the level 
of propagators and gives consistent results.\footnote{It is not 
generally true that the  $\xi \to \infty$ limit can be taken before 
integration over loop momentum. However, for the anomalous 
magnetic moment terms we find that they are already gauge-parameter 
independent after integrating over bulk positions, at 
fixed value of 4D loop momentum $l$.}
For the error estimates we typically choose 
$\xi=1$ and $\xi=4$ as reference gauge parameters. 
This choice already changes the contributions 
of individual diagrams significantly and provides a reliable check
on the gauge-independence of the sum within numerical uncertainties. 
For instance, the result for the gauge invariant set of diagrams 
with an internal hypercharge boson (B1-B5b) 
varies by one to three percent 
when changing the the gauge parameter by a factor four, while the 
individual diagrams experience changes of order one.
A similar behaviour is found for the genuinely non-abelian diagrams (W4-W10);
here the variation of the sum does not exceed two percent. The abelian $W$
diagrams also form a gauge invariant subset. However, the sum of 
W1, W2 and W3 is over two orders of magnitude smaller than the 
individual diagrams. This large cancellation makes the result numerically
unstable and prohibits a check of the  $\xi$ independence. 
Due to this cancellation the  abelian $W$ diagrams are not 
important for the determination of $\Delta a_\mu$.

It should be noted that the gauge invariance proof 
in Sec.~\ref{gaugeinvariance} does not discriminate between 
``on-shell'' and ``off-shell'' terms as it takes into account the 
full propagator of the external fermion. That is, only the 
sum of both sets of terms needs to be gauge invariant. Indeed we 
find numerically that the ``off-shell'' terms alone are $\xi$-dependent, 
but we checked analytically that the gauge-dependent terms are of the 
form $(p+p')^\mu$ and hence can combine with the other 
terms to a gauge-independent result. Due to the
relative smallness of the ``off-shell'' terms 
(see also Sec.~\ref{subsec:singleflavour})
it is not possible to verify within the numerical accuracy that the on-shell
terms contain a tiny residual gauge-dependence that cancels this 
$\xi$-dependence. The analytic proof, however, shows that this must 
be the case. 

As a further check we work with two separate 
implementations of our  evaluation strategy, which both rely 
on Mathematica to handle the numerical integrations.  
After combining the uncertainty due to the extrapolation and 
the error estimate for the integration itself, we determine 
$\Delta a_\mu$ with a typical uncertainty at the percent level.
The overall evaluation time generally depends on the specific choice
of the 5D input parameters and the desired precision;
a typical runtime for the results presented in Sec.~\ref{sec:result}
is $18\,$hours on an Intel i7-950 3.4~GHz processor.  

\subsection{Genuine Higgs-exchange contributions}
\label{subsec:Higgs}

%%%%%%%%%%%%%%%%%%%%%%%%%%%%%%%%%%%%%%%%%%%%%%%%%%%%%%%%%%%%%%%
\begin{figure} 
\begin{center} 
\vskip0.3cm
\includegraphics[width=9cm]{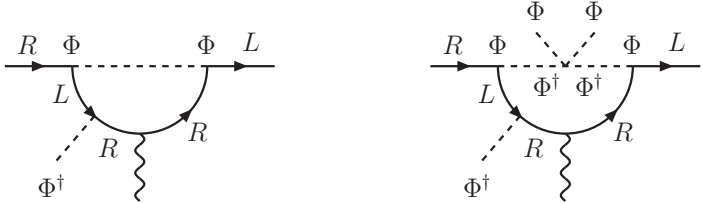}  
\end{center} 
\caption{Higgs boson exchange diagrams. The diagram on the left does 
not exist, since there is no $\Phi\Phi$ propagator. The leading 
contributions come from higher-dimension operators generated by 
diagrams such as the right one.}  
\label{fig:impossibleHiggs}  
\end{figure}  
%%%%%%%%%%%%%%%%%%%%%%%%%%%%%%%%%%%%%%%%%%%%%%%%%%%%%%%%%%%%%%

We note that the diagrams in figure~\ref{fig:diagsall} contain 
internal Higgs lines, but they do not correspond to the usual 
Higgs exchange contributions. For example, the 
SM Higgs contribution to $(g-2)_\mu$ is not part of the zero mode 
contributions of these diagrams. 

The Higgs exchange diagrams with an internal insertion, 
such as shown in figure~\ref{fig:impossibleHiggs} left, 
cannot exist at the level of dimension-six operators 
$\bar L\Phi\sigma_{\mu\nu} E F^{\mu\nu}$, since 
the internal Higgs propagator would correspond to the 
(non-existing) $\Phi\Phi$ contraction of the Higgs doublet, 
while the external line would be $\Phi^\dagger$ rather than 
$\Phi$. The leading Higgs exchange contribution with all 
Higgs interactions in the loop are generated at one loop by diagrams 
such as in figure~\ref{fig:impossibleHiggs} right. 
This corresponds to dimension-8 operators of the form 
\begin{equation}
\frac{1}{T^4}\times 
[y^{\rm (5D)}y^{\rm (5D)\dagger}y^{\rm (5D)}]_{ij}
\,\bar L_i\Phi\sigma_{\mu\nu} E_j F^{\mu\nu} \Phi^\dagger\Phi\,,
\label{dim8Higgs}
\end{equation}
and implies an additional $v^2/T^2$ suppression relative to 
the gauge-boson contribution. The 
loop integral is infrared divergent due to the two (massless) 
Higgs propagators, which results in an 
additional logarithm of $T$, similar to the 
$\ln \,(m_H/m_\mu)$ in the SM Higgs contribution.

In the present paper, we do not 
address dimension-8 operators. The above operator, generated by 
Higgs exchange, despite being suppressed, may, however, be of interest 
for flavour-changing processes, since it depends on the 
flavour structure $yy^\dagger y$ of Yukawa couplings, whereas the 
dimension-6 operators are proportional to a single 
Yukawa matrix~\cite{Csaki:2010aj}. However, there are contributions 
involving  $yy^\dagger y$ already at the dimension-6 level 
as we discuss next.

\subsubsection{Wrong-chirality Higgs couplings}
\label{WrongChiralityHiggsCouplings}

\begin{figure} 
\begin{center} 
\vskip0.3cm
\includegraphics[width=15cm]{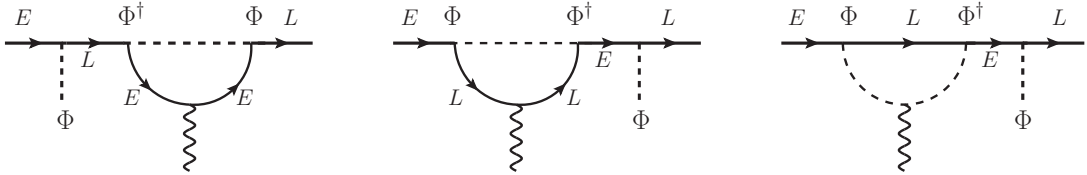}  
\end{center} 
\caption{Non-vanishing Higgs-boson exchange diagrams. The diagrams require
either a wrong-chirality Higgs coupling or the cancellation of the 
external propagator.}  
\label{fig:NonZeroHiggs}  
\end{figure}

The diagrams in figure \ref{fig:NonZeroHiggs} have an external Higgs 
insertion, which allows for an internal Higgs propagator. 
However, they contain either chiral components of the 
brane-to-brane fermion propagator that naively vanish because of 
$f_{E_j}^{(n)}(1/T) = g_{L_i}^{(n)}(1/T)=0$, or require the external fermion 
propagator to be an on-shell zero-mode. We now discuss that in 
both cases there is a non-vanishing contribution to the matching 
coefficient of the dimension-six electromagnetic dipole operator, 
but argue that it is numerically small relative to the gauge-boson 
exchange term. 

%%%%%%%%%%%%%%%%%%%%%%%%%%%%%%%%%%%%%%%%%%%%%%%%%%%%%%%%%%%%%%%

%%%%%%%%%%%%%%%%%%%%%%%%%%%%%%%%%%%%%%%

A sharp delta-function localized Higgs profile leads to ambiguities 
in the interactions with 5D fields. In the KK-decomposed theory, after 
electroweak symmetry breaking, this becomes evident, when one solves 
the mode equations for the fermions \cite{Csaki:2003sh}. 
In the 5D unbroken theory the problem arises from the discontinuities 
of the 5D propagators at coincident points, when these points 
approach the TeV brane coordinate $z=1/T$.
To avoid these ambiguities it is required to define the brane localized 
minimal RS model by the limit of a model with a regularized Higgs profile 
with a small width $\delta/T$, where $\delta \ll 1$ as discussed in 
\cite{Azatov:2010pf, Carena:2012fk, Delaunay:2012cz}.  
Since the fermion modes  $f_{E_j}^{(n)}(z)$ and $g_{L_i}^{(n)}(z)$
vanish only directly on the brane, the finite width of the Higgs 
profile leads to a non-vanishing coupling of the left-handed singlet 
and right-handed doublet modes to the Higgs field, which 
is referred to as wrong-chirality Higgs coupling (WCHC). The treatment 
of these couplings is subtle, since it involves very high 
KK excitations with $n\sim 1/\delta$, such that the limits of mode number to
infinity and Higgs regulator to zero do not necessarily 
commute \cite{Carena:2012fk}. 
Since we use 5D propagators, all KK modes are already summed, that is, we 
implicitly take the limit of the Higgs regulator $\delta$ to zero after 
the sum over all KK modes. We comment below on how the presence of 
a cut-off on the scale of validity of the RS model might affect our 
result. 

It is convenient to consider a simple step-function Higgs profile
of the form 
\begin{align}
\label{higgsprofile}
 \Phi(x,z)= \Phi(x)\,\frac{T}{\delta}\,\Theta(z-(1-\delta)/T)
\end{align}  
with a dimensionless parameter $\delta\ll 1$. This choice for the profile 
and, up to small corrections of order $v^2/T^2$,
also for the Higgs vacuum expectation value allows for an analytical 
calculation of the Higgs-exchange diagrams. Note that we consider 
(\ref{higgsprofile}) as a regularization of the minimal RS model, 
implying $\delta\to 0$ in the end, and hence do not need to address 
the question which dynamics might 
generate this profile as a solution to the field equations. 

We calculate the diagrams shown in figure~\ref{fig:NonZeroHiggs} 
starting from an expression, where all three Higgs vertices are 
delocalized. In the limit $\delta\to 0$, we find that the resulting 
contribution to the matching coefficient $a_{ij}$ of the 
electromagnetic dipole operator 
takes the simple expression
\begin{equation}
\frac{1}{T^2}\,a_{ij}^{\rm WCHC} = \frac{c\, e}{16\pi^2}\frac{1}{T^2} \times 
\frac{T^3}{k^4}\, f_{L_i}^{(0)}(1/T) [Y Y^\dagger Y]_{ij}
g_{E_j}^{(0)}(1/T)\,,
\label{aijWCHC}
\end{equation}
where $c=-\frac{1}{12}$ is a constant that can be computed analytically, and 
$Y_{ij} = y^{(\rm 5D)}_{ij} k$ is the dimensionless 5D Yukawa matrix. The 
expression after the $\times$ symbol is dimensionless and 
depends on the 5D mass parameters $c_{E_j}$, $c_{L_i}$ and scales 
$k$, $T$ only through the 4D Yukawa matrix $y_{ij}$, since 
\begin{equation}
 f_{L_i}^{(0)}(1/T)g_{E_j}^{(0)}(1/T) = 
\frac{k^4}{T^3}\,\frac{y_{ij}}{Y_{ij}}\,,
\end{equation}
see (\ref{4dyukawa}).
The presence of three Yukawa matrices makes these 
diagrams particularly important for lepton-flavour changing observables. 
The details of the calculation and the flavour aspects will be 
presented in \cite{inprep}.

%%%%%%%%%%%%%%%%%%%%%%%%%%%%%%%%
%%%%%%%%%%%%%%%%%%%%%%%%%%%%%%%%

The contribution of \eqref{aijWCHC} to $g_\mu-2$ is
more model-dependent than the gauge-boson contributions 
as there is no general argument that connects 
the size of the elements of $Y$  to those of the product 
$Y Y^\dagger Y$. In the model with all entries of $Y$ of the
same order, and assuming no cancellations that introduce structure 
to the product of three anarchic Yukawa matrices, we can use 
the current bound on the $\mu \to e \gamma$ decay rate \cite{Adam:2011ch}, 
which involves exactly the same diagrams, to constrain the 
product of Yukawa factors and $T^{-2}$. This leads to 
a rough upper limit on the size of the Higgs-exchange 
contribution to $g_\mu-2$, 
\begin{align}
\Delta a_\mu^{\rm WCHC}\leq 2 \cdot 10^{-12}\,,
\label{wchcestimate}
\end{align}
which is independent of the KK scale $T$. Thus,  
Higgs exchange is always irrelevant phenomenologically unless there is a 
significant enhancement of the Yukawa coupling factor relevant 
to the muon anomalous magnetic moment relative 
to the anarchic estimate. 
If we abandon the assumption of anarchy as might be suggested by 
the strong constraint from lepton flavour violation, we cannot 
use the $\mu \to e \gamma$  decay rate to constrain the Higgs contribution to 
$a_\mu$ as the two observables are not governed by the same model parameters.
However, the requirement
that the Higgs contribution does not exceed the present experimental 
measurement of $\Delta a_\mu$ still imposes a constraint on the 
product of Yukawa couplings as will be discussed in the 
next section.

Since the RS models should be defined as an effective 
theory with cut-off $\Lambda\gg T$, it may be argued that the KK sum should 
be truncated once the KK masses exceed $\Lambda$, and that the loop 
momentum in the diagrams of figure  \ref{fig:NonZeroHiggs} 
should be restricted to values smaller than the cut-off. 
When the brane-localization limit $\delta\to 0$ is taken at 
fixed $\Lambda$, the wrong-chirality Higgs-coupling contribution 
to $g_\mu-2$ vanishes. Alternatively, we may interpret brane localization as 
$\delta \sim T/\Lambda \ll 1$, in which case there is no simple 
analytic result for $a_{ij}^{\rm WCHC}$, but the magnitude is 
similar to (\ref{aijWCHC}). Thus, the rough bound (\ref{wchcestimate}) 
should remain valid independent of the precise relation between $\delta$ 
and $\Lambda$ and the order of limits. The 
different treatments of the order of limit amounts to a different 
definition of the meaning of ``brane localization'' and hence the model
itself. The Higgs-exchange contribution (contrary to the gauge boson exchange 
contribution discussed later) is therefore model-dependent. 
A similar
situation arises in the calculation of Higgs 
production \cite{Azatov:2010pf,Carena:2012fk}.

There is another Higgs contribution to electromagnetic dipole transitions 
at order $1/T^2$ that arises from the class-1 operator 
$h_{ij} \bar L_i\Phi E_j \Phi^\dagger\Phi + \mbox{h.c.}$, which was not 
included in (\ref{dim61}), since $h_{ij}$ is non-zero only when the 
wrong-chirality Higgs couplings are taken into account. Tree-level 
matching with the step-function Higgs profile (\ref{higgsprofile}) 
gives the coefficient function (see also \cite{Azatov:2009na})
\begin{equation}
\frac{1}{T^2}\,h_{ij} = \frac{1}{3 T^2} \times 
\frac{T^3}{k^4}\, f_{L_i}^{(0)}(1/T) [Y Y^\dagger Y]_{ij}
g_{E_j}^{(0)}(1/T)\,,
\label{hijWCHC}
\end{equation}
which differs from (\ref{aijWCHC}) only by the absence of the electromagnetic 
coupling and the loop factor. When two of the Higgs fields in 
$\bar L_i\Phi E_j \Phi^\dagger\Phi$ are put to their vacuum expectation 
values, this operator modifies the Yukawa couplings and leads to 
flavour-changing couplings of the zero-mode fermions to the Higgs boson. 
Inserting this vertex into the Higgs-exchange contribution to the 
electromagentic dipole transition similar to diagrams c and d 
of figure~\ref{fig:4Ddiags}, we find that the result is suppressed 
relative to (\ref{aijWCHC}) by a factor of [lepton mass]$^2/m_H^2$, 
where $m_H$ is the physical Higgs mass. The additional lepton mass 
factors arise from the 4D Yukawa coupling at one of the Higgs-fermion vertices 
and the need for a helicity flip in the loop. Thus, the Higgs-exchange 
contribution to the anomalous magnetic moment and to radiative 
lepton flavour violating transtions from loop momentum $k\sim m_H$ 
is strongly suppressed relative to the contribution (\ref{aijWCHC}) 
that is generated at the KK scale. 

%%%%%%%%%%%%%%%%%%%%%%%%%%%%%%%%
%%%%%%%%%%%%%%%%%%%%%%%%%%%%%%%%

\subsubsection{Off-shell Higgs diagrams}
\label{OffShellHiggs}

Each of the three diagrams in figure~\ref{fig:NonZeroHiggs} also
contains ``off-shell'' contributions of the type discussed in 
Sec.~\ref{subsec:external}, where the external zero-mode propagator 
is cancelled, resulting in a local short-distance contribution. 
Since the zero mode always has the right chirality, the ``off-shell'' 
terms do not involve wrong-chirality Higgs couplings, and 
we can work with a brane-localized Higgs field from the start. 
Then the diagrams involve at most one bulk coordinate integral from 
the photon vertex. This integral can be taken analytically, as 
well as the remaining integral over the loop momentum. Here it 
proves useful to use the KK decomposition of the internal fermion 
propagators. The zero mode has to be subtracted from the internal 
propagators, since their contribution cannot be associated with 
the KK scale $T$. This subtraction also renders the loop integral 
infrared finite.

The dependence on lepton flavour of the off-shell terms 
is of the form\footnote{The expression given holds for the 
first diagram in figure~\ref{fig:NonZeroHiggs}.} 
\begin{equation}
\sum_k f_{L_i}^{(0)}(1/T)  Y_{ih} [Y^\dagger]_{hk} Y_{kj} 
[f^{(0)}_{L_k}(1/T)]^2  g_{E_j}^{(0)}(1/T) 
% F^-_{E_h}(0,1/T,1/T,c_E)
. 
\vspace*{-0.4cm}
\end{equation}
Due to the extra factor $[f^{(0)}_{L_k}(1/T)]^2$ there is now 
a strong dependence on the bulk mass parameters in addition to the  
explicit form of the Yukawa matrices, which makes 
it difficult to give a precise estimate for the off-shell Higgs
contribution to $(g-2)_\mu$. For the symmetric choice of bulk mass 
parameters, $c_{E_k} = -c_{L_k}$, we expect the product 
$[f^{(0)}_{L_k}(1/T)]^2$ to count as a factor of lepton mass $m_{\ell_k}$; 
in this case the ``off-shell'' contribution is small relative 
to the WCHC contribution.
In general, it can be of similar size. In principle 
this would allow for a flavour-specific cancellation between the 
two contributions, that is, a cancellation for $\mu\to e\gamma$, but not for 
the flavour-diagonal quantity $(g-2)_\mu$, in which case the 
bound (\ref{wchcestimate}) is invalidated. 
However, if we assume that the bulk mass parameters and anarchic 
Yukawa matrices do not conspire in this way, the sum of all three 
``off-shell'' Higgs diagrams is two orders of magnitude smaller than the 
experimental uncertainty on $a_\mu$, and hence negligible.

\subsection{Wrong-chirality Higgs couplings 
 in gauge boson diagrams}

Naturally the question arises whether there exist contributions with   
WCHCs in the gauge boson exchange diagrams of figure~\ref{fig:diagsall}. 
Inspection shows that only the internal insertion diagrams B1a/b and B3a/b
contain non-vanishing wrong-chirality Higgs-fermion vertices, once the 
Higgs profile is regulated by a finite width. The following analysis 
demonstrates, however, that these contribution vanish as ${\cal O}(\delta)$, 
when the Higgs profile regulator $\delta$ is removed. This should be 
contrasted to the case of the Higgs-exchange diagrams discussed above, 
where the finite contribution (\ref{aijWCHC}) survives in the 
$\delta\to 0$ limit.

We analyze the loop momentum integrand in the two regions 
$l\ll T/\delta$ and $l \gg T/\delta$ according to whether the 
loop momentum is much smaller (larger) than the width of the 
Higgs profile at the TeV brane. We assume $\delta \ll 1$ and 
hence $T/\delta \gg T$. For concreteness, we consider only a single 
term in the complete expression for diagram B1a; the following 
arguments are general and also apply to the other terms and diagrams.
Consider therefore the  expression
\begin{align}
\label{GaugeWCHExample}
 \int\frac{d^4l}{(2\pi)^4}\;\int_{1/k}^{1/T}\!\!\!
\frac{dy}{(k y)^4} \int_{1/k}^{1/T}\!\!\!\frac{dw}{(k w)^4} 
\int_{1/k}^{1/T}\!\!\!\frac{dx}{(k x)^4}
\int_{(1-\delta)/T}^{1/T}\!\!\!dz \,\frac{T}{\delta} \,
\nonumber \\
\times \Delta^{\rm ZMS}_\perp(l,x,w)\, 
F_L^-(l,x,y)\; d^+F_L^-(l,y,z)\; d^+F_E^-(l,z,w)\,.
\end{align}
The wrong-chirality Higgs coupling appears at the vertex 
with bulk coordinate $z$, which is confined to  
$z\in (\frac{1-\delta}{T},\frac{1}{T})$ due to 
(\ref{higgsprofile}). Each of the propagators $d^+F_L^-(l,y,z)$ and 
$d^+F_E^-(l,z,w)$ vanishes if $z$ is 
exactly equal to $1/T$.

Let us start with the region of loop momentum much smaller than 
the inverse Higgs profile width $T/\delta$, in which case we can expand the 
wrong-chirality propagator functions for {\bf $l (1/T-z) \ll 1$} and $l/k
\ll 1$. An important point is that propagator functions such as 
(see (\ref{dpFmL}))
\begin{eqnarray}
d^+F^-_L(l, y, z) &=& 
- l \,\Theta(y - z)\,\frac{i  k^4 y^{5/2} z^{5/2} 
\tilde{S}_+(l, y,1/T, c_L)S_-(l,z, 1/k,c_L)}
{S_-(l,1/T, 1/k , c_L)} \nonumber \\
&& - \,l \,\Theta(z - y) \,
\frac{i k^4 y^{5/2} z^{5/2} S_-(l,z, 1/T,c_L) \tilde{S}_+(l, y, 1/k , c_L)}
{S_-(l,1/T, 1/k , c_L)}
\label{dpFmLexample}
\end{eqnarray}
are discontinuous at $y=z$.\footnote{This is clear from the fact that
  due to the equation of motion a derivative with respect to $z$ 
  acting on $d^+F_L^-(l,y,z)$ 
  generates factor of $\delta(y-z)$,} 
The second term is ${\cal O}(\delta)$, 
since $S_-(l,1/T, 1/T,c_L)=0$ and $z$ is within distance 
$\delta/T$ near $1/T$, while the first one is  
${\cal O}(1)$. But the first term is only operative, when 
$y>z$, which constrains $y$ to be within the narrow Higgs profile.
This immediately leads to the following counting for the 
four different orderings for the coordinates $z$,$w$ and $y$
(the arguments of the critical propagator functions): (a) $z>y,w$, 
(b)  $y>z>w$, (c)  $w>z>y$ and (d) $y,w>z$. 
\begin{itemize}
\item[(a)] For $z>y,w$ the product $d^+F_L^-(l,y,z) d^+F_E^-(l,z,w)$ 
is of order $\delta^2$. The $z$-integral provides a power of 
$\delta$ from the length of the integration interval and 
$T/\delta$ from the height of the Higgs profile, so the 
whole contribution from (a) vanishes as $\delta^2$ for $\delta\to 0$.
\item[(b)] 
In case of $y>z>w$,  $d^+F_E^-(l,z,w)$ still counts as ${\cal
  O}(\delta)$, but $d^+F_L^-(l,y,z)$ loses its suppression factor, 
since now the first line of (\ref{dpFmLexample}) is the relevant one.
Hence, $d^+F_L^-(l,y,z) d^+F_E^-(l,z,w)\sim \delta$.
Now note that the requirement  $y>z>w$ means that $y$ must be in 
      $(\frac{1-\delta}{T},\frac{1}{T})$, so the $y$ integration 
interval counts as $\mathcal{O}(\delta)$. Collecting all factors 
of $\delta$, we see that the loop momentum integrand 
again scales as $\delta^2$.
\item[(c)] For $w>z>y$ the counting is obviously analogous to
  (b).
\item[(d)] For $y,w>z$ both propagators lose their  $\delta$
  suppression. However, both, the $y$ and the $w$ integration, are now
  limited to the interval $(\frac{1-\delta}{T},\frac{1}{T})$.
  This restores the overall factor of $\delta^2$.
\end{itemize}
We see that irrespective of the ordering of bulk coordinates, the 
loop momentum integrand vanishes as $\delta^2$ for $\delta\to 0$ in 
the momentum region $l \ll T/\delta$. 
As this point we expand the propagator functions in  $l\gg T$
and $l/k \ll 1$ and find that after carrying out the angular
integration the loop momentum integrand  is nearly constant 
(that is, ${\cal O}(l^0)$) over the interval from $l\sim T$ to 
$l\sim T/\delta$. Hence the final integral over $dl$ provides a 
factor $1/\delta$ and the entire diagram with loop momentum 
cut off at $T/\delta$ vanishes as ${\cal O}(\delta)$ when the 
Higgs profile regularization is removed.

It remains to estimate the contribution from very large loop
momentum. When $l\gg T/\delta$ (but still $l\ll k$) we find that 
the propagators exhibit universal behaviour. Let $\Delta$ be the 
distance between starting and end point of the propagation
in the fifth dimension. Then the propagators all behave as 
$e^{-l\Delta}$. Hence, the coordinate integrals have support only  
if all coordinates are within a typical distance of order $1/l$ of
each other; otherwise the integrand is exponentially small. 
Changing the integration variables  from $(x,y,z,w)$ to 
$(z,z-w,w-x,x-y)$ we see that each integration over distance 
differences, e.g.~$z-w$, counts as $1/l$. Only 
the $z$ integral gives a factor of $\delta/T$ that is 
cancelled by the Higgs profile height $T/\delta$. 
Collecting all factors shows that after integration over the  
bulk coordinates the remaining 
loop-momentum integrand is independent of $\delta$ 
and scales as $dl/l^2$. Thus the integration over $l$ from 
$T/\delta$ to infinity scales as $\delta$. This proves that 
the entire expression (\ref{GaugeWCHExample}) vanishes in 
the limit $\delta\to 0$, when the regularization of the Higgs 
brane localization is removed. 

In a similar fashion all other terms with wrong-chirality Higgs couplings 
in gauge boson exchange diagrams can be shown to vanish in the 
limit $\delta \to 0$, so no correction needs to be applied to the 
calculation performed in the model with an exactly  brane-localized Higgs 
field.

\section{Muon anomalous magnetic moment}
\label{sec:result}

\subsection{Single-flavour approximation {\bf for gauge boson diagrams}}
\label{subsec:singleflavour}

We now discuss our result for the muon anomalous magnetic moment. 
We will mostly restrict ourselves to the single-flavour approximation, 
and discuss briefly below why it should be a good approximation to 
neglect lepton-flavour changing contributions 
for the gauge boson diagrams.

In this case the matching coefficients $\alpha\equiv \alpha_{22}$ 
(dipole operator) and $\beta\equiv \beta_{2222}$ (four-fermion 
operator) defined in (\ref{dim6broken}) are functions of 
$k$, $T$, the two bulk-mass parameters $c_L$, $c_E$, and the 
dimensionless 5D Yukawa coupling $Y_\mu \equiv  y^{\rm (5D)}_{22}k$.
We fix the Planck scale quantity $k$ to 
$2.44\times 10^{18}\,$GeV~\cite{Huber:2000ie}, and determine 
$c_E$ from the value of the muon mass $m_\mu=105.7\,$MeV through 
(\ref{4dyukawa}), for given $T$, $c_L$, and $Y_\mu$. The other parameters 
that enter the analysis are the Weinberg 
angle $s_W^2 =\sin^2 \theta_W= 0.231$, 
the electromagnetic coupling $\alpha_{\rm em}(M_Z)=1/128.94$, and the Higgs 
vacuum expectation value $v=246\,$GeV. 

The contribution to the anomalous magnetic moment from the four-fermion 
operator follows from (\ref{amures}) and (\ref{bsingflavour}):
\begin{equation}
[\Delta a_\mu]_{\rm 4f} = -\frac{m_\mu^2}{T^2}\,\frac{\beta}{4\pi^2} 
= \frac{\alpha_{\rm em}}{8\pi c_W^2}  
\,\frac{m_\mu^2}{T^2} \,\frac{1}{\ln\frac{1}{\epsilon}} \,
f(\epsilon,c_L,c_E)
\approx 1.2\cdot 10^{-13}\,\times (1\,\mbox{TeV}/T)^2.
\label{4fnum}
\end{equation}
In the last expression we adopted $T=1\,$TeV and symmetric bulk mass 
parameters $c_L=-c_E=0.578\ldots$, which leads to $f=0.95$, and 
made the dominant dependence on $T$ explicit. Thus, the four-fermion 
operator makes a negligible contribution to the anomalous magnetic 
moment compared to the dipole operator discussed below, 
mainly due to the $1/\ln(1/\epsilon)\approx 1/35$ suppression.
We find that $f\approx 1$ is always a good approximation, independent 
of the values of $T$ and $Y_\mu$, when the bulk mass parameters are 
near the symmetric value. This can be different when $c_L$ is larger 
than about 0.65, 
or smaller than 0.5, but (except for extreme choices of $c_L$ where 
$c_E$ becomes positive) $f$ is never large enough to compensate the 
$1/\ln(1/\epsilon)$ suppression. Therefore we drop the 
four-fermion operator contribution from the further discussion. 

The contribution from the electromagnetic dipole operator is 
determined by five-dimensional loop integrals. For their evaluation 
we have to rely on numerical integration and must add \eqref{IRTermInW8} 
to take the additional term from W8 into account. Before discussing the 
result, we provide an order-of-magnitude estimate of this contribution. 
Since the loop is dominated by momenta of order of the KK scale $T$, 
one might expect the dimensionless Wilson coefficient $\alpha_{22}$ 
in \eqref{dim6broken} to be of order $e\times \alpha_{\rm em}/(4\pi)$. 
According to (\ref{amures}) this would result in $\Delta a_\mu 
\sim  \alpha_{\rm em}/(4\pi) \times (m_\mu v)/T^2 \approx 
1.6 \cdot 10^{-8}\times (1\,\mbox{TeV}/T)^2$, which is very large.
This estimate is, however, too naive 
as it does not take into account the wave functions of the 
external 5D states.
The photon zero-mode profile \eqref{ZMProfilesBoson} provides
a factor of $1/\sqrt{\ln(k/T)}$. The conversion factors for 
the gauge couplings $g_5^{(\prime)} \to g_4^{(\prime)}$ 
amount to $\ln^{3/2}(k/T)$, since the one-loop diagrams involve 
three gauge couplings. Thus, we obtain an overall factor 
$\ln\left(1/\epsilon \right)$.
Furthermore, the external fermion zero-modes provide a factor 
of  $g_E^{(0)}(x)f_L^{(0)}(z)$. Since the integrals over the bulk 
coordinates are dominated by $x,z\sim 1/T$, we expect that 
$\alpha_{22}$ is roughly proportional to 
$g_E^{(0)}(1/T)f_L^{(0)}(1/T)$. Together with the vacuum expectation 
value $v/\sqrt{2}$ in (\ref{amures}) from the Higgs insertion, 
this counts as a factor of order $m_\mu$. The contribution to 
$\Delta a_\mu$ due to electromagnetic dipole operators is then estimated 
by\footnote{Note that one factor of $m_\mu$ arises from the loop 
diagram, the other from the definition of the  $F_2$ form factor 
as coefficient of $\sigma^{\mu\nu}q_\nu/2m_\mu$ rather 
than $\sigma^{\mu\nu}q_\nu$.}
\begin{align}
\label{DipoleEstimate}
\Delta a_\mu \approx  \frac{\alpha_{\rm em}}{4 \pi} \,
\frac{m_\mu^2}{T^2}\, 
\ln\left(1/\epsilon \right) \times 
{\rm Loop}(c_L,c_E,y^{\rm (5D)},T,\epsilon) \,,
\end{align}
where the remaining loop factor should be $\mathcal{O}(1)$. This estimate 
indicates that the contribution from the dipole operators is enhanced
by a factor of the order $\ln^2\left(1/\epsilon \right)\sim 10^3$ 
compared to the contribution from the four-fermion operator; this would 
lead to $\Delta a_\mu\approx 2.4 \cdot 10^{-10}$ for $T=1\,\rm TeV$, which 
is only a factor of three smaller than the present uncertainty of 
$a_\mu$. 

%%%%%%%%%%%%%%%%%%%%%%%%%%%%%%%%%%%%%%%%%%%%%%%%%%%%%%%%%%%%%%%%%%%%%%%
\begin{figure}[t]
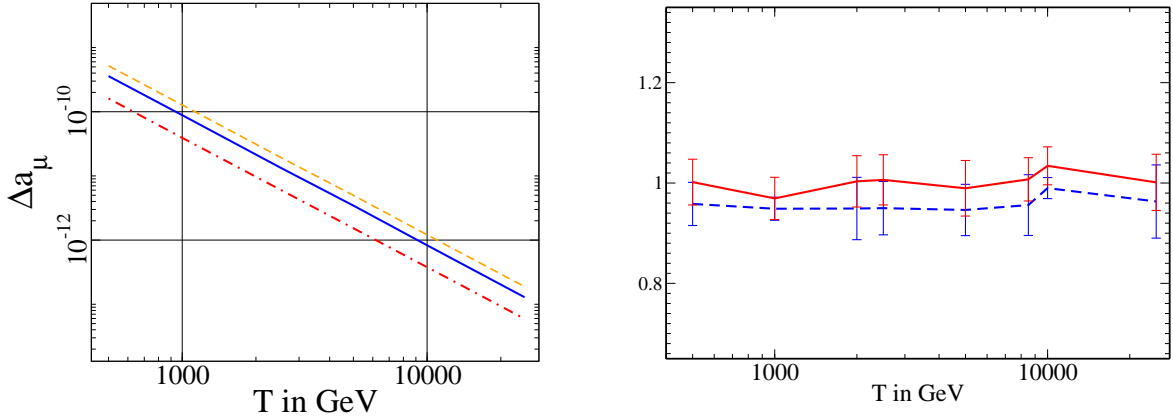

\begin{center}\hspace{-0.5cm}
\begin{minipage}{0.45\textwidth}
\includegraphics[width=1 \textwidth]{Sym1Plot.eps}
\end{minipage}
\hspace{1.0cm}
\begin{minipage}{0.45\textwidth}\vspace{-0.1cm}
\includegraphics[width=1 \textwidth]{Ratio.eps}
\end{minipage}
\end{center} 
\caption{ Left panel: Dependence of the different 
contributions to $\Delta a_\mu$ on the KK scale $T$:
$-[\Delta a_\mu]_B$ (dark grey/red, dot-dashed), 
$[\Delta a_\mu]_W$ (light-grey/orange, dashed) and the sum 
(black/blue, solid).
Right panel: $\Delta a_\mu(c_L,Y_\mu)/
\Delta a_\mu(c_L=-c_R=0.578425,Y_\mu=1)$ for 
$c_L=0.51, Y_\mu =1$ (dashed) and $c_L=-c_R, Y_\mu=10$ 
(solid) as a function of $T$.
\label{Tdependence:fig}}
\end{figure}

For our numerical study it is convenient to split  $\Delta a_\mu$ in 
contributions from an exchange of hypercharge $B$ bosons and $W$ 
bosons
\begin{align}
 \Delta a_\mu =[\Delta a_\mu]_B + [\Delta a_\mu]_W\;
\end{align}
and to study them separately. 
The left panel in figure~\ref{Tdependence:fig} 
shows the $T$ dependence of $-\left[\Delta a_\mu\right]_B$ (dot-dashed) and 
$\left[\Delta a_\mu\right]_W$ (dashed) as well as the sum (solid) for the 
parameter choice $c_L=-c_E$ and $Y_\mu=1$. Note that the contribution due to 
hypercharge bosons $B$ has opposite sign than the non-abelian contribution.
The dependence of $\Delta a_\mu$ on $T$ follows almost precisely  
the $1/T^2 \times \ln(k/T)$ scaling, see~\eqref{DipoleEstimate}.
Fitting the numerical $T$ dependence shown in figure~\ref{Tdependence:fig} 
to $1/T^{\,x} \times \ln(k/T)$ one finds a best fit for $x=1.997\pm 0.04$.
The contribution of \eqref{IRTermInW8} to $\left[\Delta a_\mu\right]_W$
is typically at the level of $2.5\%$. This is consistent with the absence 
of a $\ln(k/T)$ enhancement in \eqref{c3coeff}. 
The left panel shows the result for $\Delta a_\mu$ for different 
choices of $c_L$ and $Y_\mu$ normalized to  $\Delta a_\mu(c_L=-c_E,Y_\mu=1)$.
We see that the ratio is quite independent of the KK scale: the $T$ dependence 
is practically universal. Also different choices 
of $c_L$ and $Y_\mu$ change $\Delta a_\mu$ only mildly.

From figure~\ref{Tdependence:fig} we conclude not only that
the correction to $g_\mu-2$ is quite insensitive to the specific choice 
for $y^{(5D)}$ and the mass parameters $c_L$ and $c_E$, but also that 
the estimate \eqref{DipoleEstimate} agrees fairly well with the 
numerical results if not for some amount of 
cancellation between $[\Delta a_\mu]_W$ and $[\Delta a_\mu]_B$.
Explicit numerical values for a representative set of parameters 
are shown in table~\ref{TeVscaleTable} 
where a reference value of $T=1\,\rm TeV$ was used.   

%%%%%%%%%%%%%%%%%%%%%%%%%%%%%%%%%%%%%%%%%%%%%%%%%%%%%%%%%%%%%%%%%%%%%%
\begin{table}
\begin{center}
\begin{tabular}{cccccc}
   & $Y_\mu$ & $c_L$       &  $c_E$    & $\left[\Delta a_\mu\right]_B$ & $\left[\Delta a_\mu\right]_W$ \\ \hline \hline
   & $0.1$ & $0.53605$   &  $-0.53605$   & $-4.18(08)\cdot 10^{-11}$      & $1.27(2)\cdot 10^{-10}$     \\
   & $ 0.25$ & $ 0.553805$  &  $ -0.553805 $ & $ -4.31(19)\cdot 10^{-11} $   & $1.29(1)\cdot 10^{-10} $  \\
   & $ 1  $ &   $0.578425$  &  $-0.578425 $    & $-4.02(10)\cdot 10^{-11}   $   & $ 1.28(1)\cdot 10^{-10}  $ \\
   & $ 4 $ &   $ 0.601598$&  $ -0.601598$  & $ -4.05(15)\cdot 10^{-11} $   & $1.29(1)\cdot 10^{-10}$     \\
   & $ 10$ & $  0.616446$ &  $  -0.616446$ & $ -4.34(08)\cdot 10^{-11}$     & $ 1.29(2)\cdot 10^{-10} $   \\ \hline
   & $1 $  & $ 0.51$     &  $  -0.634898$ & $ -4.13(10)\cdot 10^{-11} $   & $ 1.28(2)\cdot 10^{-10} $   \\
   & $ 1   $ & $ 0.55$     &  $  -0.604918$& $  -4.20(15)\cdot 10^{-11} $  & $ 1.27(3)\cdot 10^{-10}  $\\
   &  $ 1  $ & $ 0.578425$   &  $ -0.578425$   & $  -4.02(10)\cdot 10^{-11}$    & $ 1.28(1)\cdot 10^{-10}  $\\
   &  $  1 $ &  $ 0.6 $    &  $ -0.555607 $ & $ -4.26(16) \cdot 10^{-11} $  & $ 1.29(2) \cdot 10^{-10} $\\
   &  $ 1  $ & $  0.64$    &  $  -0.501419 $& $ -4.06(10) \cdot 10^{-11} $  & $ 1.28(3) \cdot 10^{-10} $\\
\end{tabular}
\end{center}
\caption{Dependence of the contributions to $\Delta a_\mu$ on 
         mass parameter asymmetry and Yukawa coupling.
         $T$ has been fixed to $1\,\rm TeV$. The uncertainties include
         the numerical error from the integration, and the uncertainty 
         from the extrapolation and residual $\xi$ dependence. 
\label{TeVscaleTable}}
\end{table}
%%%%%%%%%%%%%%%%%%%%%%%%%%%%%%%%%%%%%%%%%%%%%%%%%%%%%%%%%%%%%%%%%%%%%%

The robustness of the result with respect to the 5D parameters
can be understood in the following way. Consider the contribution
from terms that do not have a fermion zero-mode propagator.
As already mentioned above each external (SM) fermion comes with
the corresponding zero-mode profile; in our case  $g_E^{(0)}$
and $f_L^{(0)}$ and we expect these mode factors together with the Yukawa 
to combine roughly into a factor of the lepton mass. The  
dependence on the 5D parameters will then arise from 
the zero-mode subtracted fermion propagators. Since the KK modes are 
quite insensitive to the 5D mass parameters the dependence is, 
as observed, mild.  

Contributions from terms containing an explicit fermion zero-mode propagator, 
such as the so-called ``off-shell'' contributions or the IR sensitive part of W8
have two additional zero-mode factors; for instance
\begin{align}
Y_\mu \,g_E^{(0)}(1/T) f_L^{(0)}(1/T) \cdot f_L^{(0)}(x) f_L^{(0)}(z)
\sim \frac{m_\ell}{v}  \cdot f_L^{(0)}(x) f_L^{(0)}(z).
\end{align}
The product $f_L^{(0)}(x) f_L^{(0)}(z)$ is very sensitive to the choice of the 
5D mass parameters and one expects a much more pronounced 
dependence on 5D masses and Yukawa couplings. However, as already argued
in Sec.~\ref{subsec:external} these terms will typically be suppressed 
and, hence, this effect is not seen in the total.  

We will illustrate this using the non-abelian diagrams as an example.  
Consider the two special cases: a fixed doublet 5D mass such that $c_L=0.5$ 
and the symmetric choice $c_L=-c_E$. 
In all non-abelian diagrams contributing to 
the $\bar L \sigma _{\mu\nu} E W^{\mu\nu}$
structure the actual loop is independent of $c_E$; thus after fixing $c_L=0.5$
the remaining dependence on $c_E$ from the external lines and on $Y_\ell$ will always combine to 
a factor of the lepton mass $m_\ell$ -- irrespective whether a 
zero-mode propagator is involved or not.
For the symmetric choice $c_L=-c_E$
the identity $f_L^{(0)}(x)=g_E^{(0)}(x)$ suggests that the explicit factor
$f_L^{(0)}(x) f_L^{(0)}(z)$ for zero-mode propagators should count 
as an additional power of $m_\ell$ if the loop is dominated by the 
region very close to the IR brane. In this case the term with a 
zero-mode propagator will scale like $m_\ell^2$ rather than 
$m_\ell$ as for the remaining contributions. In particular, the 
``off-shell'' diagrams which by definition have an explicit zero-mode 
propagator are sensitive to the 5D mass parameters and not
necessarily proportional to the lepton mass. 

%%%%%%%%%%%%%%%%%%%%%%%%%%%%%%%%%%%%%%%%%%%%%%%%%%%%%%%%%%%%%%%%%%%%%%%%
\begin{figure}[t]
% Add picture
\begin{center}
%\vspace*{-2cm}
\hspace*{-1.5cm}
 \includegraphics[width=0.7\textwidth]{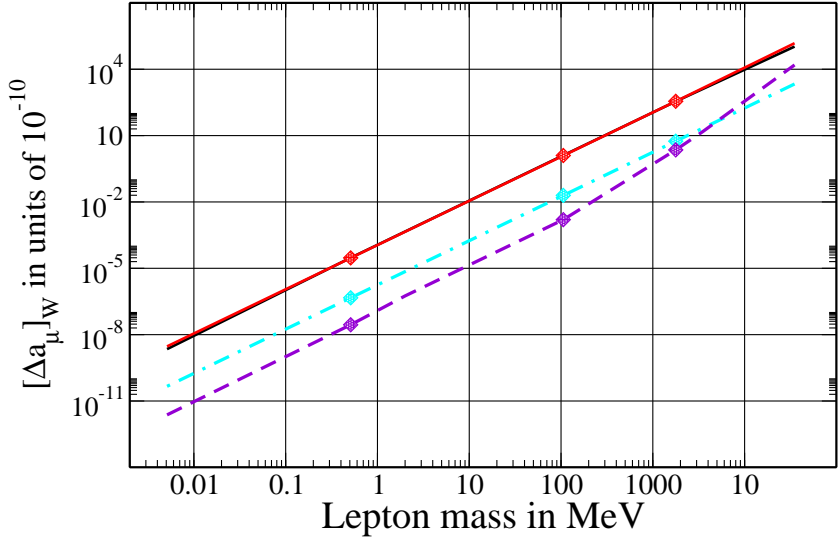}
%\vspace*{-1.5cm}
\end{center}
\caption{Lepton mass dependence of total non-abelian (solid) and 
non-abelian ``off-shell'' contributions to $a_\mu$; the dashed line 
refers to fixed $c_L=0.5001$, the dot-dashed one to the 
symmetric choice $c_L=-c_R$.
Diamonds correspond to electron, muon and tau mass, respectively. 
\label{MassDep:fig}} 
\end{figure}
%%%%%%%%%%%%%%%%%%%%%%%%%%%%%%%%%%%%%%%%%%%%%%%%%%%%%%%%%%%%%%%%%%%%%%%%
%%%%%%%%%%%%%%%%%%%%%%%%%%%%%%%%%%%%%%%%%%%%%%%%%%%%%%%%%%%%%%%%%%%%%%%%
\begin{figure}[t]
\begin{center}
\hspace*{-1.5cm}
\includegraphics[width=0.7\textwidth]{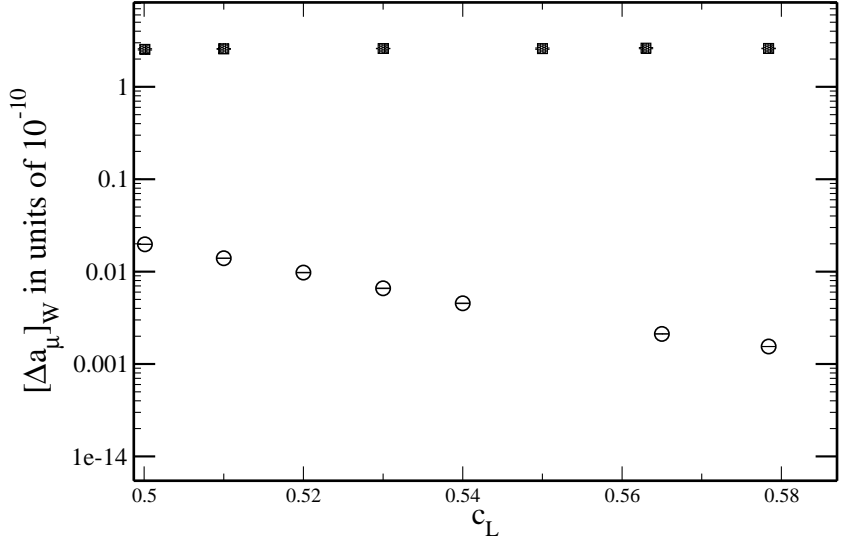}
\end{center}
\caption{Dependence of $[a_\mu]_W$ on $c_L$. Squares correspond to the 
sum over all diagrams, circles to the ``off-shell'' terms only. 
\label{zeromodesVSfull:fig}} 
\end{figure}
%%%%%%%%%%%%%%%%%%%%%%%%%%%%%%%%%%%%%%%%%%%%%%%%%%%%%%%%%%%%%%%%%%%%%%%

To verify whether this intuitive picture is justified we study 
the lepton mass dependence of $[\Delta a_\mu]_W$ numerically.     
Figure~\ref{MassDep:fig} shows the full contribution 
of all $W$ diagrams compared to the contribution of ``off-shell'' 
terms as a function of the lepton mass.\footnote{Only the sum over 
all diagrams is gauge independent; 
the pure off-shell terms are gauge dependent. For figures~\ref{MassDep:fig} 
and \ref{zeromodesVSfull:fig} Feynman gauge was used. Furthermore, 
recall from the previous footnote that $\Delta a_\mu$ has one 
additional factor of $m_\ell$ relative to the coefficient function 
$\alpha_{22}$.}
Both $T=1\;\rm TeV$ and $Y_\ell=1$ are kept fixed.  
The total non-abelian contributions (solid lines) for both, the asymmetric and
symmetric case are on top of each other and cannot be distinguished 
on the logarithmic scale; both are proportional to $m_\ell^2$.   
The ``off-shell'' terms show exactly the same scaling for fixed doublet 
modes with $c_L=1/2$  
(dot-dashed line) and are a factor $60$ smaller than the total 
contribution in agreement with our previous estimate. 
For $c_L = -c_E$ (dashed) the lepton-mass dependence is more involved.
Following our argument above we 
expect $m_\ell^3$ behaviour. This is, however, only realized for lepton
masses above $2\,\rm GeV$. For leptons
lighter than the muon we find same $m_\ell^2$ dependence as in the
$c_L=1/2$ case. The source of 
this unexpected behaviour can be traced to two diagrams: W1 and W4. Both W1
and W4 contain a contribution where only fermion zero-modes propagate 
in the loop. Our argument for the expected 
suppression of the off-shell terms by one additional power of the lepton
mass relies on IR-brane dominance of the coordinate integrals due to the
localization of the KK modes (see Sec.~\ref{subsec:external}). For the
pure fermion zero-mode contribution this dominance is not present and 
these terms are only suppressed by a mass-independent factor instead of 
$m_\ell/v$. The zero-mode terms are thus less suppressed for very light
leptons and give rise
to the $m^2_\ell$ behaviour in the low mass region of
figure~\ref{MassDep:fig}. 
The $c_L$  dependence of the different contributions to $[a_\mu]_W$ 
is illustrated in figure~\ref{zeromodesVSfull:fig}. Again 
sum over all diagrams shows almost no sensitivity whereas the small
off-shell terms are quite sensitive to the specific choice for $c_L$.

Using the overall dependence on $T^2$ and the 
relative insensitivity to mass parameters and 
Yukawa couplings we can condense our result for the correction to $(g-2)_\mu$
in the minimal RS model into 
\begin{equation}
\label{eq:ResultForGMinus2}
\Delta a_\mu \approx 8.8\cdot 10^{-11}\,\times (1\,\mbox{TeV}/T)^2 \,,
\end{equation}
which is our main result. 
The exact results differ from this approximation by less than 10\% for 
$Y_\mu \in (0.1,10)$, $T\in (0.5\,{\rm TeV}, 50\,{\rm TeV})$ and 
$c_L$ between $0.5$ and the symmetric choice. 
The difference between the current experimental average 
of the anomalous magnetic moments of $\mu^+$ and $\mu^-$ and the
Standard Model prediction is given by \cite{pdg2012} 
\begin{equation}
\label{SMdeltaAmu}
 a_\mu^{\rm exp}-a^{\rm SM}_\mu=287(63)(49)\times 10^{-11}
\end{equation}
where the two errors correspond to the combined experimental
and theoretical uncertainties, respectively. 
Since we find $\Delta a_\mu$ to be positive in the RS model,
$a_\mu^{\rm exp}-a^{\rm RS}_\mu$ will be smaller than 
$a_\mu^{\rm exp}-a^{\rm SM}_\mu$, decreasing 
the discrepancy between theory and experiment. 
However, the estimate \eqref{eq:ResultForGMinus2} shows 
that for $T>1\;{\rm TeV}$ the correction to $a^{\rm SM}_\mu$ is 
too small. Even for a low KK scale of $T\approx 500\;{\rm GeV}$ 
which corresponds to lowest KK excitations presently being 
excluded by direct searches, 
$\Delta a_\mu$ only reaches the level of $40 \times 10^{-11}$
--- of order of the error on $a_\mu$, but 
too small to reconcile theory and experiment 
at the $2\sigma$ level. 

The above discussion applies to the gauge boson contribution 
to $a_\mu$. In Sec.~\ref{subsec:Higgs} we already showed  
that in models with anarchic Yukawa matrices the 
Higgs-exchange contribution is limited to a few 
times $10^{-12}$, barring accidental cancellations that 
invalidate the $\mu\to e\gamma$ constraint. Abandoning the 
anarchy assumption, we may turn the argument around and try 
to glean some information on the specific product of three Yukawa matrices 
that enters the Higgs contribution from  $g_\mu-2$. To this end, we 
define the dimensionless quantity 
\begin{align} 
\label{eqn_Ytwo}
\langle YY^\dagger \rangle_{\mu}=
\frac{\operatorname{Re}\left[\sum_{lm}[U^\dagger]_{2l}f_{L_l}^{(0)}(1/T) [Y Y^\dagger Y]_{lm} 
g_{E_m}^{(0)}(1/T)V_{m2}\right]}{\sum_{lm}[U^\dagger]_{2l}
f_{L_l}^{(0)}(1/T) Y_{lm} g_{E_m}^{(0)}(1/T)V_{m2}}\,,
\end{align}
which parametrises the flavour dependence of the contribution 
from \eqref{aijWCHC} to $g_\mu-2$ 
compared to a term with only a single Yukawa matrix. 
Note that the denominator of \eqref{eqn_Ytwo} is 
simply the 4D muon Yukawa coupling $y_\mu$. From 
(\ref{amures}) and \eqref{aijWCHC} we obtain 
\begin{equation}
\Delta a_{\,u}^{\rm WCHC} = \frac{1}{48\pi^2}\, 
\frac{m_\mu^2}{T^2}
\,\langle YY^\dagger \rangle_{\mu}\,.
\end{equation}
Adding the gauge-boson contribution this can be written as 
\begin{equation}
\Delta a_\mu = \left[8.8 + 2.4  \, \langle YY^\dagger \rangle_{\mu} 
\right]\cdot 10^{-11}\,\times (1\,\mbox{TeV}/T)^2
\end{equation}
With experimental and theoretical errors added linearly, the 
measurement (\ref{SMdeltaAmu}) can then translated into the 
bound 
\begin{align}
\label{eq_HiggsBoundOnY}
-25 < \langle YY^\dagger \rangle_{\mu} \times (1\,\mbox{TeV}/T)^2 < 260\,,
\end{align}
where we required that $a_\mu^{\rm exp}-a_\mu^{\rm RS}$ stays 
compatible with zero at the $3 \sigma$ level.  
The pronounced asymmetry of upper and lower bound arises from the 
non-zero value of (\ref{SMdeltaAmu}). The limit is rather loose, 
reflecting the small coefficient of the Higgs contribution, 
constraining the Yukawa entries to ``average values'' of around 
$5$ to $15$ for $T=1\,$TeV. For $\langle YY^\dagger \rangle_{\mu} 
\approx 20$ the Higgs contribution is of the same order as the
present theoretical uncertainty on $g_\mu-2$, again for $T=1\,$TeV. 
The bound weakens for larger KK scales.

%%%%%%%%%%%%%%%%%%%%%%%%%%%%%%%%%%%%%%%%%%%%%%%%%%%%%%%%%%%%%%%%%%%%%%
\subsection{Beyond single-flavour}

The generalization to the case with three 
lepton flavours is in principle straightforward:
the Yukawa factors have to be promoted to the $3\times 3$ Yukawa matrix
and the external muon states have to be represented in the mass basis,
see (\ref{eq:changeBasis}). In practice, this complicates the 
calculation, since it now depends on six bulk mass parameters 
and the full 5D Yukawa matrix, and only three of these parameters 
can be eliminated by the known lepton masses. We now argue that 
this complication is not relevant for the gauge-boson contribution to  
the anomalous magnetic moment. 

In the interaction basis the only source of 
flavour change is the Yukawa matrix itself. Thus, independent of its 
topology each Feynman diagram can schematically be written as
\begin{align}
\label{eq:ThreeFlavourContrib}
 \sum_{i,j}  U^\dagger_{ni}M^{(1)}_{ii} 
y_{ij}^{\rm (5D)}M^{(2)}_{jj} V_{jm},
\end{align}
where $n=m=2$ for external muon states.
Here $M^{(1)}$ and $M^{(2)}$ collect all 
factors depending on only flavour indices $i$ and $j$, respectively.
For the single-flavour approximation we observed that the 
dominant contribution to the dipole-operator 
Wilson coefficient $\alpha_{ij}$ (and thus to
$g_\mu-2$) is largely insensitive to the specific choice for the 5D masses 
or Yukawa couplings but depends linearly on the lepton mass.
For the three-flavour case this means that the dominant 
contribution to $g_\mu-2$ is roughly proportional to the mass matrix:  
\begin{align}
\label{eq:domContrib}
         \left[M^{(1)}_{ii} y_{ij}^{5D}M^{(2)}_{jj}\right]_{\rm dom}  
\propto  f^{(0)}_{L_i}(1/T)y_{ij}^{5D}g^{(0)}_{E_j}(1/T)\,. 
\end{align}  
As the matrices $U$ and $V$ diagonalize the mass matrix
\begin{align}
\frac{T^3}{k^3} \frac{v}{\sqrt{2}} 
\sum_{i,j}\,U^\dagger_{ni} f^{(0)}_{L_i}(1/T)  
               y_{ij}^{\rm (5D)}g^{(0)}_{E_j}(1/T) V_{im}
=  {\rm diag}\left\lbrace m_e, m_\mu, m_\tau \right\rbrace_{nm}\,,
\end{align}
the flavour dependence of the diagrams approximately combines into a 
lepton mass factor, irrespective of the specific choices of the 
bulk mass parameters $c_{L_i}$, $c_{E_j}$ and 5D Yukawa matrix. 
Hence we can adopt the simplest choice for the calculation by 
taking the Yukawa matrix diagonal, which corresponds to the 
single-flavour approximation. The anomalous magnetic moment 
of the muon is therefore reliably determined in this approximation, 
up to small corrections to (\ref{eq:domContrib}), which are 
no larger than, e.g., the effect of unaccounted
higher dimensional operators. 

The fact that \eqref{eq:ThreeFlavourContrib} is, up to small 
corrections, already diagonal, implies that lepton flavour-violating 
transitions are suppressed. Flavour-changing processes -- such as 
$\mu \to e\gamma$ in the lepton sector or $b\to s \gamma$ in the 
quark sector -- are therefore sensitive to the sub-dominant contributions 
that depend strongly on the 5D mass parameters. Such terms arise, 
for example, when the internal fermion lines of a diagram 
propagate zero modes and the gauge boson is a KK mode 
(see \cite{Csaki:2010aj}), or via the contribution of
the ``off-shell'' part of the $\bar L_iL_i\gamma$ vertex with a subsequent 
mass insertion as discussed in the previous section. 
In this situation the suppression of ``off-shell'' 
diagrams is lifted and the terms have to be taken into account.
Also Higgs exchange, though related to higher-dimensional operators,
is relevant to flavour violation, since its different Yukawa 
matrix dependence produces terms that are not aligned with 
the mass matrix \cite{Agashe:2006iy,Csaki:2010aj}.

%%%%%%%%%%%%%%%%%%%%%%%%%%%%%%%%%%%%%%%%%%%%%%%%%%%%%%%%%%%%%%%%%%%%%%%%   
\subsection{Comparison to previous work}
\label{sec:compare}

The first calculation of the muon anomalous magnetic moment in the 
RS model was done very soon \cite{Davoudiasl:2000my} after the  
invention of the model. The authors used the 4D formalism and computed 
multiple KK sums. They (incorrectly) concluded that external mass 
insertion diagrams are suppressed by a factor $(m_\mu/M_{\rm KK})^2$, 
which eliminates all diagrams except B1a and B1b in 
figure~\ref{fig:diagsall}. In these diagrams a misplaced chiral 
projector at the photon vertex eliminates one of the two terms in the 
integrand (\ref{diag1}), (\ref{diag2}). The final result for 
the gauge-boson contribution to $\Delta a_\mu$ 
in \cite{Davoudiasl:2000my} is opposite in sign to ours, 
which is consistent with the fact that the abelian 
hypercharge contribution $[\Delta a_\mu]_B$ is negative. The final result 
(\ref{eq:ResultForGMinus2}) is, however, dominated by $W$-exchange, which 
was neglected in~\cite{Davoudiasl:2000my}. Ref.~\cite{Davoudiasl:2000my} 
also considered the Higgs contribution and found that it 
puts important constraints on the bulk mass parameters. However, 
their Higgs contribution does not seem to contain the 
$(v/T)^2$ suppression relative to the gauge-boson 
contribution that is expected on general grounds, see 
Sec.~\ref{subsec:Higgs}, when the wrong-chirality Higgs couplings 
and ``off-shell'' terms are neglected as was done~\cite{Davoudiasl:2000my}.

The only other estimate of $\Delta a_\mu$ in the RS model with SM 
model fields in the bulk we 
are aware of is contained in \cite{Goertz:2011gk}. The calculation 
is restricted to a one-loop vertex diagram with KK photon exchange 
but only zero-mode fermions in the loop. This subset of 
contributions significantly underestimates the true result, 
since it is suppressed relative to (\ref{DipoleEstimate}) by a 
factor $\ln^2(1/\epsilon) \sim 10^3$ due to a cancellation in 
the photon KK sum.

The calculations of lepton-flavour violating $\ell_i\to \ell_j \gamma$ 
processes \cite{Agashe:2006iy,Csaki:2010aj} involve the same 
diagram topologies as $(g-2)_\mu$, but the restriction to flavour violation 
allows to make some simplifying assumptions. Ref.~\cite{Agashe:2006iy} 
works in the KK picture and includes only the first KK excitation, 
while \cite{Csaki:2010aj} adopts the 5D formalism as we have done 
in this paper. These works do not 
consider external insertion diagrams, though this has been 
corrected in \cite{Csaki:2010ajv3}. The Higgs exchange diagrams 
that can be neglected for $(g-2)_\mu$ provide the dominant source 
of flavour violation and are found to be of order $1/T^4$ (again, 
in the absence of wrong-chirality Higgs couplings and ``off-shell'' 
terms) consistent with 
(\ref{dim8Higgs}). Ref.~\cite{Csaki:2010aj} 
also considers the gauge-boson contributions, but assumes 
that only internal zero-mode fermions are relevant to flavour 
violation.  Since we have seen that the 
KK fermions in the loop are by far the dominant contribution 
to flavour-conserving quantities like $(g-2)_\mu$, it would be 
interesting (but numerically challenging) 
to investigate how the small non-alignment 
of these terms feeds into flavour violation.

%%%%%%%%%%%%%%%%%%%%%%%%%%%%%%%%%%%%%%%%%%%%%%%%%%%%%%%%%%%%%%%%%%%%%%%%%
\section{Conclusion}
\label{sec:conclude}

In this paper we performed the first complete calculation of the 
gauge boson exchange contribution to the muon anomalous magnetic moment 
in the minimal Randall-Sundrum model. 
We find that the additional contribution to $a_\mu$ is enhanced 
by a factor $\ln(k/T)$ relative to the naive estimate 
$\alpha_{\rm em}/(4 \pi)\times (m_\mu/T)^2$, resulting in 
\begin{equation}
\Delta a_\mu \approx 8.8\cdot 10^{-11}\,\times (1\,\mbox{TeV}/T)^2 \,,
\end{equation}
nearly independent of the bulk mass and Yukawa coupling parameters 
of the model. For $T\approx 500\,$GeV, which corresponds to first KK 
excitations around $1.3\,$TeV, this is of order of the theoretical 
uncertainty on $a_\mu$. The minimal RS model is, however, already 
excluded by electroweak precision tests unless $T$ is much larger, 
which makes the correction to $g_\mu-2$ unobservable. $T$ may be lower 
in models with custodial symmetry, but in this case the contribution 
from the extra states remains to be computed. The number given above 
refers to the model-independent contribution from gauge-boson 
exchange. We also determined the Higgs-exchange mediated contributions to 
$(g-2)_\mu$; they depend on an unconstrained
combination of 5D Yukawa couplings. In anarchic scenarios this 
combination is constrained by $\mu\to e\gamma$, such that the 
Higgs contributions to $\Delta a_\mu$ are below $2\cdot 10^{-12}$.
In the general scenario the Higgs contribution depends
strongly on the model parameters. We used this to derive 
a rough limit on the product of Yukawa factors and KK scale $T$.

The 5D framework developed here can be extended to an investigation of 
lepton and quark flavour-violating radiative processes in warped 
geometries that goes beyond the approximations adopted 
in~\cite{Agashe:2006iy,Csaki:2010aj,Blanke:2012tv}. We hope to 
report on such results in the future. On the more formal side, it would 
be interesting to investigate the proper gauge-invariant regularization and 
renormalization of the RS model as an effective field theory, and the 
5D formalism seems to be the suitable framework.

\vspace*{0.5em}
\noindent
\subsubsection*{Acknowledgement}
We thank A.~M\"uck for many helpful discussions. We also thank 
K.~Agashe, C.~Delaunay, and R. Sundrum for their comments on 
wrong-chirality Higgs couplings. 
MB and JR thank the Kavli Institute for Theoretical Physics 
at UC Santa Barbara for hospitality, while part of this work 
was done. This work is supported in part by the 
Sonder\-for\-schungs\-bereich/Trans\-regio~9 ``Computergest\"utzte
Theoreti\-sche Teilchenphysik'' and the Gottfried Wilhelm 
Leibniz programme of the Deutsche Forschungsgemeinschaft (DFG), and
by the National Science Foundation under Grant No. NSF PHY05-51164.

%%%%%%%%%%%%%%%%%%%%%%%%%%%%%%%%%%%%%%%%%%%%%%%%%%%%%%%%%%%%%%%%%%%%%%%%%%%%%%%%%
\appendix

\section{5D Feynman rules}
\label{ap:5Drules}

In the following we provide the Feynman rules for the five-dimensional 
electroweak Standard model (\ref{5Daction}) in the Anti-de-Sitter 
space time (\ref{metric}). The rules consist of the bulk mode functions, 
the vertices and the 5D propagators.

\subsection{Equations of motion and mode profiles}
\label{App:EOMs}
\subsubsection{Fermions}
\label{App:EOMsFermions}
The fermionic part of the action was given in \eqref{5Daction}.
Upon inserting the explicit expressions for the metric and the vielbein
we obtain for a generic fermion field $\psi$
\begin{align}
\label{RS_FermionAction:Eq}
S_F = & \int \rd ^4x \int \rd z\;\left[\frac{1}{kz}\right]^4 
 \bigg\lbrace \bar \psi \left(  
 i\slashed \partial +i\Gamma^5(\partial_z-\frac{2}{z})-\frac{c}{z}\right)
 \psi
 \nonumber \\
&\qquad\qquad\quad +  \bar \psi\left(
  g'_5 \frac{Y}{2} \slashed{B}   + g_5 \frac{\tau^a}{2} \slashed{W}^a +
  g'_5 \frac{Y}{2} \Gamma^5{B_5} + g_5 \frac{\tau^a}{2}\Gamma^5{W^a_5} 
  \right) \psi\bigg\rbrace\,,
\end{align}
where $Y$ is the hypercharge of the fermion $\psi$, and $\tau^a\to 0$ in 
case of a SU(2) singlet fermion.
The dimensionless parameter $c$ is related to the 5D mass $M$ via $c=M/k$.
The fifth Dirac matrix is chosen such that  
\begin{align}
\Gamma^5=-i \gamma_5=i
   \begin{pmatrix}
     \mathds{1} &    0      \\
             0   & -\mathds{1}  
   \end{pmatrix}
\end{align}
in the chiral representation. 
Splitting the fermion field into left- and right-handed components we obtain
\begin{align}\label{RS_FermionAction2:Eq}
 S_F =& \int \rd ^4x \int \rd z\;\left[ \frac{1}{kz}\right]^4 
 \left(\bar\psi_R  i\slashed \partial \psi_R + \bar\psi_L  i\slashed 
  \partial \psi_L
  +\bar \psi_L(\partial_z-\frac{2}{z})\psi_R \right. \nn \\
&\qquad\qquad\left.- \bar \psi_R(\partial_z-\frac{2}{z})\psi_L
  -\frac{c}{z}\bar \psi_R\psi_L - \frac{c}{z}\bar \psi_L\psi_R \right) + 
  \ldots\,.
\end{align}
where $\psi_L=P_L\psi$ and  $\psi_R=P_R\psi$ with
\begin{align}
 \label{chiralProjectors}
P_{R/L}=\frac{1 \pm \gamma_5}{2}\,,
\end{align}
and only the bilinear terms are given explicitly.

Substituting the factorization ansatz $\psi_{L/R}(x,z)=
\sum_n f_{L/R}^{(n)}(z) \psi_{L/R}^{(n)}(x)$ into the action, we obtain
the mode equations \cite{Grossman:1999ra}
\begin{subequations}
\label{EoMFermion}
\begin{equation}
  -\frac{c}{z}f_R^{(n)} + \left( \partial_z - \frac{2}{z} \right) f_R^{(n)} 
  = - m_n f_L^{(n)}\,,
\end{equation}
\begin{equation}
 -\frac{c}{z}f_L^{(n)} - \left( \partial_z - \frac{2}{z} \right) f_L^{(n)} = - m_n f_R^{(n)}\,.
\end{equation}
\end{subequations}
The orthonormality conditions
\begin{align}\label{OrthonormFermion}
 \int_{1/k}^{1/T}\rd z\,\left[\frac{1}{kz}\right]^4 f_L^{(n)}(z)f_L^{(m)}(z)= 
\int_{1/k}^{1/T}\rd z\,\left[\frac{1}{kz}\right]^4f_R^{(n)}(z)f_R^{(m)}(z)
=\delta_{nm}
\end{align}
fix the normalization of the modes. The solutions for the zero mass eigenvalue 
are particularly simple:
\begin{equation}
f_R^{(0)}(z) \sim  z^{2+c}\,,\quad \qquad 
f_L^{(0)}(z) \sim  z^{2-c}\,.
\end{equation} 

However, to obtain the SM in the low-energy limit 
we have to prescribe boundary conditions\footnote{This is equivalent to 
assigning different parities under orbifolding to left- and
 right-handed fields.}
to exclude one of the zero mode of definite handedness. If $\psi$ describes
a lepton SU(2) singlet, i.e.~the right-handed massless field $E$ in the SM, 
one needs to impose the boundary condition 
\begin{equation}
\label{FermionBCforE}
\left.f^{(n)}_L(z)\right|_{z=1/k,1/T}=0 
\end{equation}
which removes the left-handed solution for the zero-mode and 
fixes the boundary condition for the right-handed modes via
\eqref{EoMFermion}. Similarly, for left-handed SM fields
 the correct massless mode for the doublet lepton field $L$ is
 obtained by imposing the condition
\begin{equation}
\label{FermionBCforL}
\left.f^{(n)}_R(z)\right|_{z=1/k,1/T}=0 
\end{equation}
on the right-handed modes. 

In the following the labels on the fermion mode functions
 refer to the doublet ($L$) and singlet ($E$) field solution; in particular,
the subscript $L$ on a fermion mode or field will henceforth always refer to
the lepton doublet $L$ and not to its handedness. The left-handed 
(right-handed) mode functions will instead be denoted by $f_\psi(z)$ 
($g_\psi(z)$). The explicit expressions for the (massless) zero-modes are
\begin{subequations}
\begin{align}
\label{FermionZeroModesSinglet}
g_E^{(0)}(z)& = 
\sqrt{\frac{1+2c_E}{1-(T/k)^{1+2c_E}}}\sqrt{T}\,(kz)^2(Tz)^{c_E}\,, \\
\label{FermionZeroModesDoublet}
f_L^{(0)}(z)& = 
\sqrt{\frac{1-2c_L}{1-(T/k)^{1-2c_L}}}\sqrt{T}\,(kz)^2(Tz)^{-c_L}\,.
\end{align}
\end{subequations}
The mode functions for non-zero eigenvalue are given by
\begin{subequations}
\begin{align}
 f^{(n)}_E(z)=C_E\,z^{5/2} \left(Y_{c_E+1/2}(m_n z
   )-\frac{Y_{c_E+1/2}\left(\frac{m_n }{T}\right) J_{c_E+1/2}(m_n z
  )}{J_{c_E+1/2}\left(\frac{m_n }{T}\right)}\right)\,,
\\
 g^{(n)}_E(z)=C_E\,z^{5/2} 
\left(Y_{c_E-1/2}(m_n z)-\frac{Y_{c_E+1/2}\left(\frac{m_n}{T}\right)
 J_{c_E-1/2}(m_n z)}{J_{c_E+1/2}\left(\frac{m_n}{T}\right)}\right)
\end{align}
and
 \begin{align}
f^{(n)}_L(z)= C_L\,z^{5/2}\left( Y_{c_L+1/2}(m_n z)-\frac{
   Y_{c_L-1/2}\left(\frac{m_n}{T}\right) J_{c_L+1/2}(
   m_n z)}{J_{c_L-1/2}\left(\frac{m_n}{T} \right) }\right)\,,
\\
 g^{(n)}_L(z)=C_L 
\,z^{5/2} \left(Y_{c_L-1/2}( m_n z)
     -\frac{Y_{c_L-1/2}\left(\frac{m_n}{T}\right) J_{c_L-1/2}(
  m_n z)}{J_{c_L-1/2}\left(\frac{m_n}{T}\right)}\right)\,.
\end{align}
\end{subequations}
Applying the boundary conditions~(\ref{FermionBCforE}), 
(\ref{FermionBCforL}), adjusted to the present notation, the mass 
eigenvalues $m_n$ are determined from
\begin{align}
Y_{c_L-1/2}\left(\frac{m_n}{k}\right) 
J_{c_L-1/2}\left(\frac{m_n}{T}\right)-J_{c_L-1/2}\left(\frac{m_n}{k}\right)
   Y_{c_L-1/2}\left(\frac{m_n}{T}\right)=0
\end{align}
for the doublet and from
\begin{align}
Y_{c_E+1/2}\left(\frac{m_n}{k}\right) 
J_{c_E+1/2}\left(\frac{m_n}{T}\right)-J_{c_E+1/2}\left(\frac{m_n}{k}\right)
   Y_{c_E+1/2}\left(\frac{m_n}{T}\right)=0
\end{align}
for the singlet field.
The normalization factors $C_E$ and $C_L$ can be computed via 
the orthogonality relation \eqref{OrthonormFermion}.

\subsubsection{Gauge bosons}
\label{App:EOMsGaugeBosons}
Let us start by repeating the expression for the different pieces 
of the action. The gauge part of the action for the non-abelian field 
can be written as
\begin{align}\label{GaugeAction}
 S_{\rm G}=\int& d^4x \int \rd z\; 
\frac{1}{2kz}\left[ W_{\mu}\left(\eta^{\mu\nu} \partial^2 -\partial^{\mu} 
\partial^{\nu } - \eta^{\mu\nu} z\partial_z 
\left( \frac{1}{z}\,\partial_z\right)  \right) W_{\nu} +\right.
\nn \\
& - W_{5}\partial^2W_{5} + 2 W_5\partial_z \partial^{\mu}W_{\mu} 
- 2 g_5 \epsilon^{abc} \partial_{\mu} W_{\nu}^a W^{\mu,b} W^{\nu,c} \nn \\ 
&  - \frac{g_5^2}{2}\epsilon^{abc}\epsilon^{ade}
W^b_\mu W^c_\nu W^{\mu,d}W^{\nu,e}
  +  g_5 \epsilon^{abc}\partial_\nu W^a_5 W^{\nu,b}W^c_5 \nn \\
&  -  g_5 \epsilon^{abc} \partial_z W^a_\nu W^{\nu,b}W^c_5 
 + \frac{g_5^2}{2}\epsilon^{abc}\epsilon^{ade}W^b_\mu W^c_5 W^{\mu,d}W^{e}_5  
\bigg] 
\end{align}
where $\partial_5\equiv \partial_z$, $\partial^2 = \partial_\mu\partial^\mu$ 
and $\eta^{55}=-1$ were already used and $\mu, \nu =0,1,2,3$. 
The action for the abelian field $B$ follows readily via the replacement 
$W\to B$ and $\epsilon^{abc}\to 0$.

The gauge fixing part of the action is chosen such that the mixed 
$W_5 W_\mu$ term  in the second line of \eqref{GaugeAction} is 
removed and no mixing of the 4D vector
boson $W^\mu$ and the Goldstone $W_5$ can occur \cite{Randall:2001gb}:
\begin{align}
 S_{\rm GF}&=\int\rd^4x \int \rd z\; 
\frac{-1}{2\xi kz }\left[ \partial_{\mu} W^\mu
-\xi z\partial_z\left(\frac{1}{z}\,W_5\right) \right]^2\;.
\end{align}
The bilinear part of the gauge action then takes the form
\begin{align}
 S_{\rm G+GF}=\int\rd^4x \int \rd z\; \frac{1}{2kz}&
\left[ W_{\mu}\left(\partial^2\eta^{\mu\nu}
        -\left(1-\frac{1}{\xi}\right)\partial^{\mu} \partial^{\nu } 
        - z\partial_z \left( \frac{1}{z}\partial_z\right)\eta^{\mu\nu}  
\right) W_{\nu} +\right.
\nn \\
& - W_{5}\partial^2 W_5 + \xi W_5 \partial_z\left(z\partial_z 
\left(\frac{1}{z}\,W_5\right)\right) \bigg]+\ldots\,,
\end{align}
where the boundary terms left after integration by parts vanish as
the $W_5$ field vanishes on the branes. In the orbifold picture this 
follows directly if $W_5$ is assigned odd parity under orbifolding;
this is required to reproduce the observed low-energy particle spectrum.
Covariant gauge fixing implies the existence of ghosts which, however,
are not needed in this work, since ghosts do not couple to the 
$W_5$ field.

Using the factorization ansatz 
$ W^\mu(x,z) = \sum_n W^{(n)\mu}(x) f_W^{(n)}(z) $ for the gauge field,  
the profiles $ f_W^{(n)}(z)$ can be determined via the equation of motion  
\begin{align}
z\partial_z \left(\frac{1}{z}\,\partial_z f_W^{(n)}(z)\right) 
= - m_n^2 f_W^{(n)}(z)\,,
\end{align}
the boundary conditions on the branes
\begin{align}
\left.\partial_z f_W^{(n)}(z)\right|_{z=1/T,1/k}=0\,,
\end{align}
and the normalization
\begin{align}
\label{BosonModeNormalization}
 \int_{1/k}^{1/T} \frac{dz}{kz} \,f_W^{(n)}(z)f_W^{(m)}(z)=\delta_{nm}\;.
\end{align}
Note that the choice of the boundary condition follows from
the need for a massless gauge boson mode to be present after
integrating out the fifth dimension.

The solution for zero mass eigenvalue has no dependence on the 
5D coordinate and is given by
\begin{align}
 \label{ZMProfilesBoson}
f^{(0)}_W(z)& =\sqrt{\frac{k}{\ln\frac{k}{T}}}\,.
\end{align}
The mode functions for non-zero mass eigenvalue
are given by 
\begin{align}
 \label{KKProfilesBoson}
f^{(n)}_W(z)& = C_W  z \left( J_1(m_n z) -
\frac{J_0(m_n/k)}{Y_0(m_n/k)} Y_1(m_n z)\right)\,,
\end{align}
where the eigenvalues $m_n$ are given by the $n$-th zero of 
\begin{align}
 \label{KKmass:eq}
 J_0(m_n/T) -\frac{J_0(m_n/k)Y_0(m_n/T)}{Y_0(m_n/k)}=0
\end{align}
and $C_W$ is determined by \eqref{BosonModeNormalization}.
Similarly, the mode equation for the Goldstone modes $W_5$ 
is given by 
 \begin{align}
\xi \partial_z \left(z\partial_z \frac{1}{z}f_{W_5}^{(n)}(z)\right)
= - \xi m_n^2 f_{W_5}^{(n)}(z)\,.
\end{align}
The boundary conditions
\begin{align}
\left. f_{W_5}^{(n)}(z)\right|_{z=1/T,1/k}=0
\end{align}
remove the zero-eigenvalue solution from the 
spectrum. One finds 
\begin{align}\label{FiveModeProfile:eq}
f^{(n)}_{W_5}(z) = 
C_{W_5}z\left(J_0(m_n z) -\frac{J_0(m_n/k)}{Y_0(m_n/k)}Y_0(m_n z)\right)
\end{align}
and the mass eigenvalues $m_n$ are the same as those for the 4D vector 
boson $W_\mu$.

Using the equations of motion yields the useful identities:
\begin{eqnarray}
\partial_z f_W^{(n)}(z) &=&  m_n f_{W_5}^{(n)}(z)
\label{eq:gaugemoderelation}\\
\partial_z f_{W_5}^{(n)}(z) &=& 
\frac{1}{z} f_{W_5}^{(n)}(z) - m_n f_{W}^{(n)}(z)\,.
\end{eqnarray}

\subsection{Vertices}
\label{App:vertices}

The Feynman rules for the vertices can be read off directly from the 
action. We denote non-abelian gauge fields $W^\mu$ by curly lines and the 
abelian field $B^\mu$ by wavy lines. The Goldstones, $W_5$ and $B_5$, 
are indicated by a superimposed solid line.

\subsubsection*{Fermion-gauge field vertices}
In the unbroken theory all gauge interactions are 
flavour diagonal. Therefore, the flavour index $i$ is
conserved in all boson-fermion vertices. Below,
$a$ denotes the SU(2) index of the gauge boson of 
weak isospin and $Y$ is the hypercharge for the fermion. 
For every vertex, there is an integral $\int_{1/k}^{1/T}dz$ over 
the bulk coordinate of the vertex.
\vspace{0.7cm}
\\
\begin{minipage}{0.27\textwidth}
%%\JaxoComment:
%%\JaxoScale{0.3}
\begin{center}
 \hspace*{0.5cm}
 \includegraphics[width=0.8\textwidth]{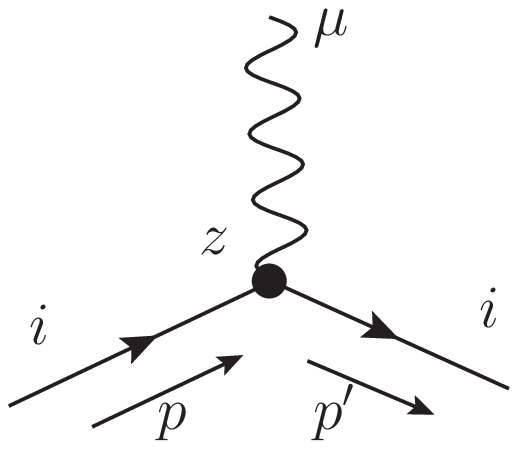}\\[0.5cm]
 \hspace*{0.5cm}
 \includegraphics[width=0.8\textwidth]{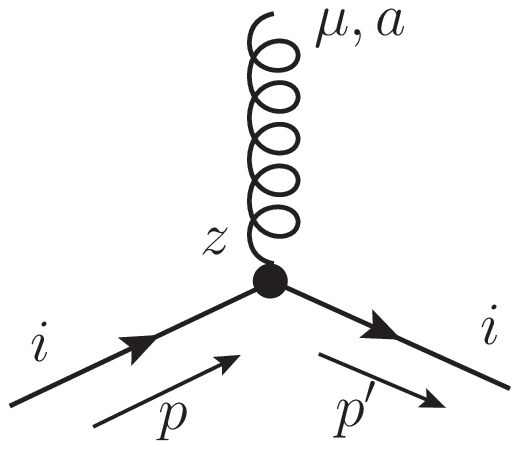}
%rs_VertexFermionAbel.eps: 0x0 pixel, 300dpi, 0.00x0.00 cm, bb=197 576 398 721
\end{center}
\end{minipage}
\begin{minipage}{0.18\textwidth}
\begin{eqnarray*}
&&\\
&& \frac{1}{(kz)^4}\,ig_5' \frac{Y}{2} \gamma^\mu 
\\[1.7cm]
&& \frac{1}{(kz)^4}\,ig_5  \frac{\tau^a}{2} \gamma^\mu
\end{eqnarray*}
\end{minipage}$\;\;\;\;\;\;\;\;$
\begin{minipage}{0.27\textwidth}
\begin{center}
 \includegraphics[width=0.8\textwidth]{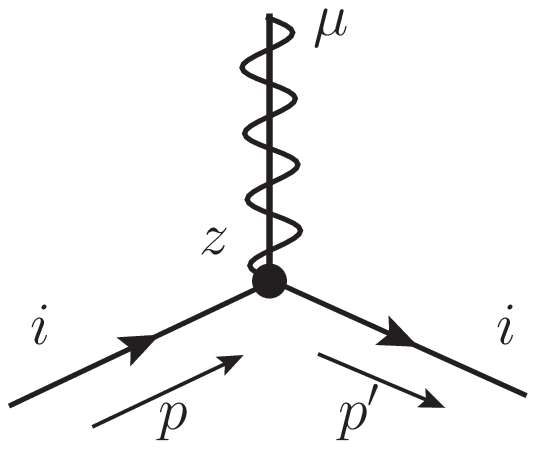}\\[0.5cm]
 \includegraphics[width=0.8\textwidth]{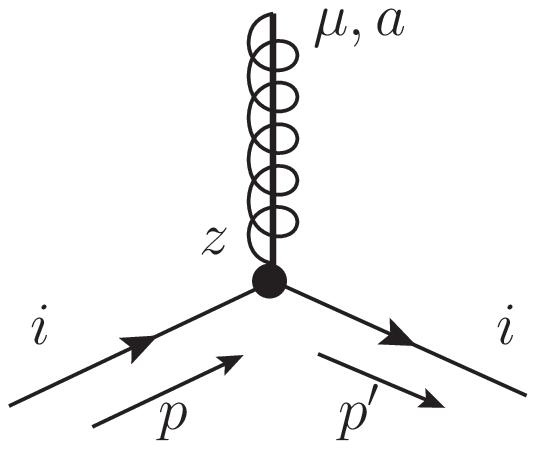}
%rs_VertexFermionAbel.eps: 0x0 pixel, 300dpi, 0.00x0.00 cm, bb=197 576 398 721
\end{center}
\end{minipage}
\begin{minipage}{0.18\textwidth}
\begin{eqnarray*}
&&\\
&& \hspace*{-0.5cm} \frac{1}{(kz)^4}\,g_5' \frac{Y}{2} \gamma_5
\\[1.7cm]
&&  \hspace*{-0.5cm} \frac{1}{(kz)^4}\,g_5 \frac{\tau^a}{2}\gamma_5
\end{eqnarray*}
\end{minipage}
\vspace{0.3cm}
   
\subsubsection*{Gauge boson self-interactions}

Here we only give the rules for trilinear gauge boson self-interactions.
$\lambda, \nu\text{ and }\mu$ are Lorentz indices and $a,b,c$ 
are SU(2) indices. The coordinate space derivatives always act on
on the $5$-coordinate of the vertex and only on the propagator/external state
with the indicated SU(2) index. 
For every vertex, there is an integral $\int_{1/k}^{1/T}dz$ over 
the bulk coordinate of the vertex.\\[0.7cm]
\noindent
\begin{minipage}{0.23\textwidth}
\begin{center}
 \hspace*{0.7cm}
 \includegraphics[width=0.85\textwidth]{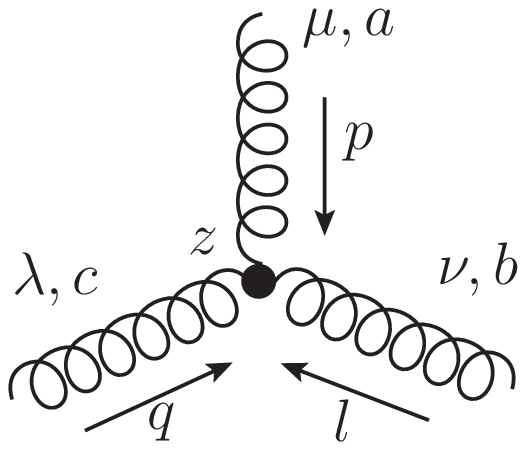}
\end{center}
\end{minipage}
\begin{minipage}{0.22\textwidth}
\vspace{-0.2cm}
\begin{align*}  \hspace*{0.8cm}
                - \frac{1}{kz}g_5\, \epsilon^{abc} 
                  \left( \left(p-q\right)_\nu \eta_{\mu\lambda} 
                + \left(q-l\right)_\mu \eta_{\lambda\nu} +
                  \left(l-p\right)_\lambda \eta_{\nu\mu} \right) 
\end{align*}
\end{minipage}\\[0.5cm]
\begin{minipage}{0.23\textwidth}
\begin{center}
 \hspace*{0.7cm}
 \includegraphics[width=0.85\textwidth]{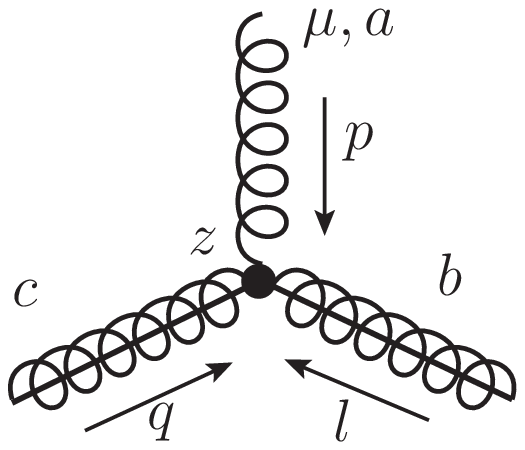}
\end{center}
\end{minipage}
\begin{minipage}{0.22\textwidth}
\begin{center}
\begin{align*}
 \hspace*{0.8cm}
+\frac{1}{kz}g_5 \,\epsilon^{abc} \left(q-l \right)_\mu\end{align*}
\end{center}
\end{minipage}\\[0.5cm]
\begin{minipage}{0.23\textwidth}
\begin{center}
 \hspace*{0.7cm}
 \includegraphics[width=0.85\textwidth]{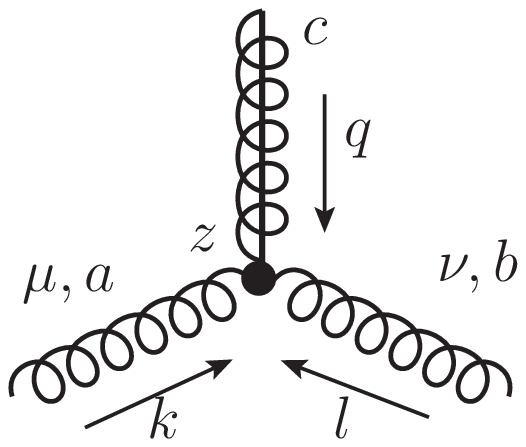}
\end{center}
\end{minipage}
\begin{minipage}{0.22\textwidth}
\begin{align*}
 \hspace*{0.8cm}
-i\frac{1}{kz}g_5 \,\epsilon^{abc} \eta_{\mu\nu} 
\left(\left.\partial_z\right|_{\text{on `a'}}
-\left.\partial_z\right|_{\text{on `b'}} \right)
\end{align*}
\end{minipage}
\vspace{0.3cm}
   
\subsubsection*{Higgs interactions}

As the Higgs field is IR brane localized all interactions occur at $1/T$ 
in the 5th dimension; therefore, we omit the $5$-coordinate of the vertices. 
The only flavour changing interaction of the theory (with our basis choice) 
is introduced by the Yukawa coupling of the Higgs to
fermions. Below $m$ and $n$ are SU(2) indices of the Higgs doublet, 
$i$ and $j$ are flavour indices. The Higgs field is represented by dashed 
lines.

\vspace{0.45cm}

\begin{minipage}{0.23\textwidth}
\begin{center}
 \hspace*{0.0cm}
 \includegraphics[width=1.05\textwidth]{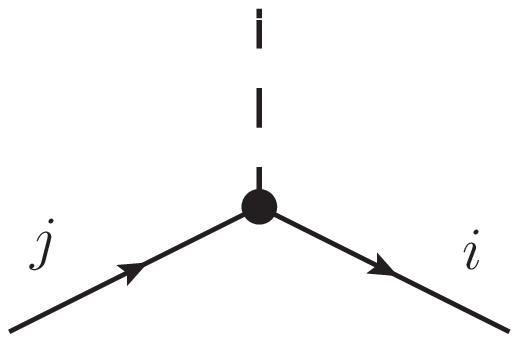}\\[0.7cm]
 \includegraphics[width=0.9\textwidth]{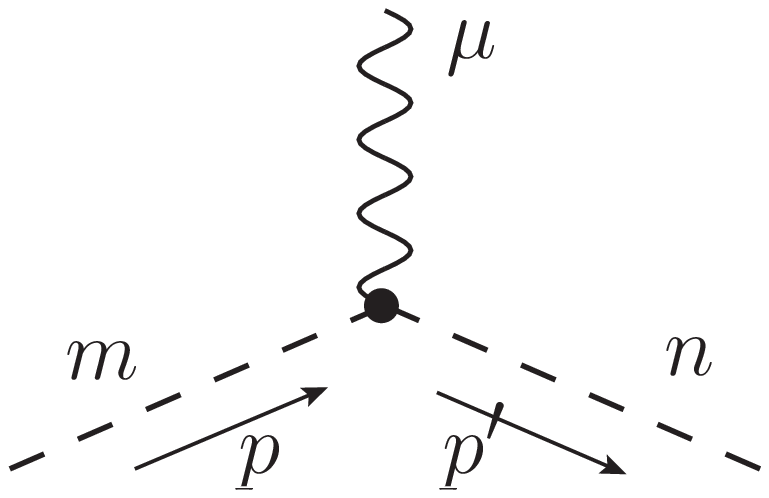}\\[0.8cm]
 \includegraphics[width=0.9\textwidth]{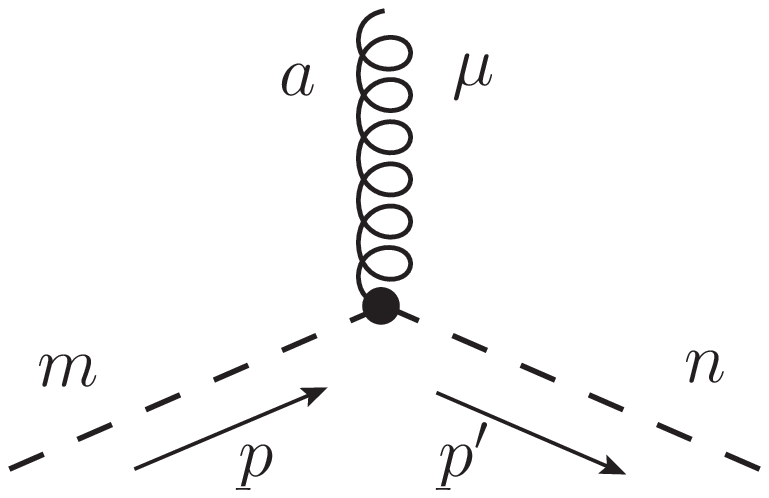}
\end{center}
\end{minipage}
\begin{minipage}{0.22\textwidth}
\vspace{-0.0cm}
\begin{eqnarray*}
&& -i \frac{T^3}{k^3} \,y^{(5D)}_{ij} 
\\[1.6cm]
&& i g_5' \frac{Y}{2} \left( p + p' \right)^\mu 
\\[1.6cm]
&&i g_5  \frac{\tau^a}{2}  \left( p + p' \right)^\mu
\end{eqnarray*}
\end{minipage}
$\qquad $
\begin{minipage}{0.23\textwidth}
\begin{center}
 \vspace*{0.4cm}
 \includegraphics[width=0.9\textwidth]{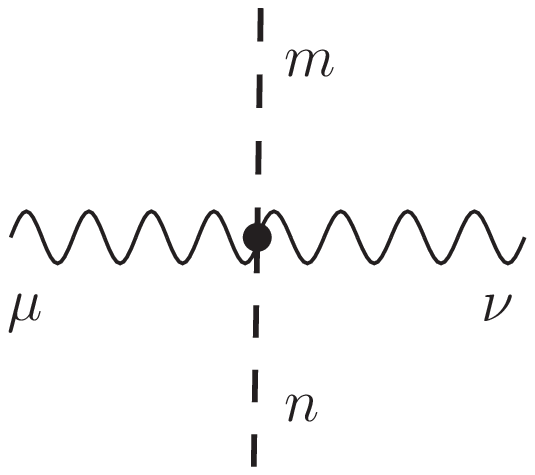}\\[0.5cm]
 \includegraphics[width=0.9\textwidth]{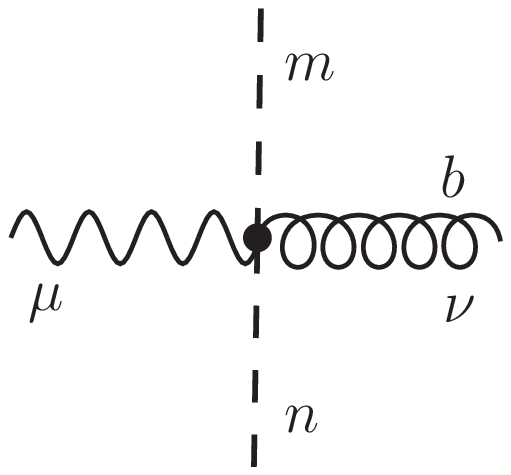}\\[0.4cm]
 \includegraphics[width=0.9\textwidth]{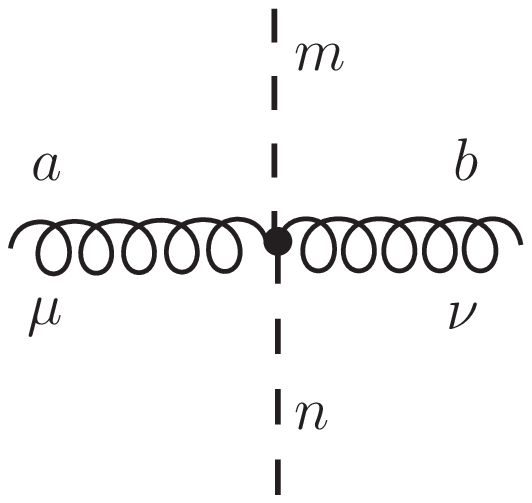}
\end{center}
\end{minipage}
\begin{minipage}{0.22\textwidth}
\vspace{-0.0cm}
\begin{eqnarray*}
&& \hspace*{-0.5cm}\frac{i}{2}\,{g^\prime}_5^2 Y^2 \eta_{\mu\nu}  \mathds{1}
\\[1.6cm]
&&  \hspace*{-0.5cm}i g_5g_5'  \,\frac{Y \tau^a}{2} \eta_{\mu\nu}  
\\[1.6cm]
&&  \hspace*{-0.5cm}\frac{i}{2} \,g_5^2 \eta_{\mu\nu} \delta^{ab} \mathds{1}
\end{eqnarray*}
\end{minipage}
\vspace*{0.3cm}

\subsection{Propagators}
\label{App:Propagators}

The 5D propagators for bosons and fermions have been 
derived in the literature, see \cite{Randall:2001gb} and 
\cite{Csaki:2010aj} respectively. However, we use 
slightly different conventions, e.g.~our choice for the 
sign of the fermion mass term in the 5D Lagrangian differs  
from \cite{Csaki:2010aj}. For completeness, we will therefore 
briefly sketch the strategy used to derive the propagators.
Note that we try to use the same notation as the existing 
literature in order to allow for an easier comparison
with previous results.  

\subsubsection{Fermion propagators}

The fermion propagator is given by the 
inverse of the 5D Dirac operator 
\begin{align}
\label{InverseDiracOp}
\left[\frac{1}{kz}\right]^4\mathcal{D}\Delta(p,z,z')=i\delta(z-z')
\mathds{1}\,. 
\end{align}
In the mixed 4D momentum/bulk coordinate space representation we can replace
$i\slashed{\partial}\to\slashed{p}$ and obtain
\begin{align}
\label{defDiracOp}
\mathcal{D}= \slashed p +i\Gamma^5(\partial_z-\frac{2}{z})-\frac{c}{z}\;.
\end{align}
Following \cite{Csaki:2010aj} we define a new function $F$ via
\begin{align}
\label{DefinitionReducedPropagator}
 \Delta(p,z,z')= \left[-\slashed{p} -i\Gamma^5(\partial_z-\frac{2}{z})-\frac{c}{z}\right]F(p,z,z')\;.
\end{align}
$F$ is also a matrix in Dirac space and once we know $F$ we also know the 
full fermion propagator. 
Inserting \eqref{DefinitionReducedPropagator} into \eqref{InverseDiracOp} 
and using the explicit representation for $i\Gamma^5$ gives
\begin{align}
\label{matrixequationForF}
 \left(\begin{array}{cc}
         -p^2-\partial_z^2+\frac{c^2-c-6}{z^2} +\frac{4}{z}\partial_z&0\\
0& -p^2-\partial_z^2+\frac{c^2+c-6}{z^2} +\frac{4}{z}\partial_z
        \end{array}
\right) F(p,z,z') =i(kz)^4 \delta(z-z')\mathds{1}_{4\times 4}\,,
\end{align}
where each entry of the matrix is itself a $2\times 2$ matrix.
The structure of \eqref{matrixequationForF} suggests a solution of the form
\begin{align}
 F(p,z,z')= P_L F^-(p,z,z') +  P_R F^+(p,z,z')\;,
\end{align}
where $F^\pm$ are now scalar-valued functions.
The differential equation then can be solved in two separate regions: 
$z< z'$ and $z>z'$.
In each region one finds two equations
\begin{subequations}
\begin{align}
\left[-p^2-\partial_z^2+\frac{c^2-c-6}{z^2}
   +\frac{4}{z}\partial_z\right] F^-(p,z,z')= 0\\
\left[-p^2-\partial_z^2+\frac{c^2+c-6}{z^2}
   +\frac{4}{z}\partial_z\right] F^+(p,z,z') = 0
\end{align}
\end{subequations}
which can be solved: 
\begin{align}
 F^{\pm}(p,z,z')=\left\lbrace
 \begin{array}{c}
 z^{5/2} \left( A_\pm^> {J}_{c\pm 1/2}(pz)+B_\pm^> {Y}_{c\pm 1/2}(pz)\right)
 \qquad \text{ for } z > z'\\[0.2cm]
 z^{5/2} \left( A_\pm^< {J}_{c\pm 1/2}(pz)+B_\pm^< {Y}_{c\pm 1/2}(pz)\right)
 \qquad \text{ for } z' > z  
 \end{array} \right.
\end{align}
Here, $J$ and $Y$ are Bessel functions and the coefficients 
$A_\pm^{>/<}$ and $B_\pm^{>/<}$ are 
functions of $z'$ and $p$ that are fixed by the boundary conditions. 
Continuity at $z=z'$, 
\begin{align} 
F_>^\pm(p,z'+\epsilon,z')-F_<^\pm(p,z'-\epsilon,z') \to 0 \quad 
{\rm for}\quad \epsilon\to 0\,,
\end{align}
and the ``cusp condition'' for the delta-function,
\begin{align} 
\lim_{\epsilon \to 0} \;\int_{z'-\epsilon}^{z'+\epsilon}d z\;
\left[-p^2-\partial_z^2+\frac{c^2\pm c-6}{z^2} +\frac{4}{z}\partial_z\right]
F^\pm(p,z,z')=i(kz')^4
\end{align}
provide two relations. Two additional conditions
that are required for an unambiguous determination of the 
propagator come from the boundary conditions on the branes.
One has to enforce that the fermion propagator only contains 
the zero-mode of the correct handedness.
These conditions are most conveniently found by explicitly 
writing~\eqref{DefinitionReducedPropagator} in matrix form. 
For the propagator of the SU(2) singlet field $E$ one finds 
\begin{align}
 \label{Eq:FermionPropForBCs}
\Delta(p,z,z')=\left(
  \begin{array}{cc}
  (\partial_z -\frac{2}{z}-\frac{c}{z}) F^-_E(p,z,z') \mathds{1} & 
  -\sigma^\mu p_\mu F^+_E(p,z,z')\\[0.2cm]
  -\sigma^\mu p_{\mu} F^-_E(p,z,z')  &   
 (-\partial_z + \frac{2}{z}-\frac{c}{z}) F^+_E(p,z,z')\mathds{1} \\
  \end{array}
\right)\,.
\end{align}
The off-diagonal elements of \eqref{Eq:FermionPropForBCs} conserve the
handedness during propagation; the diagonal elements lead to mixing of
modes of different handedness. Since in the mode representation (see 
(\ref{KKdecFermionPropagator}) below) the upper-right 
submatrix corresponds to the pure ``left'' case, its entries must 
vanish on the branes for all values of $p^\mu$, as follows 
from~\eqref{FermionBCforE}; thus
\begin{align}
\label{boundCondRHZMpart1}
          \left. F^+_E(p,z,z')\right|_{z=1/k,1/T}=0\,.
\end{align}
Similarly, the upper-left submatrix must vanish for $z=1/T,1/k$ as 
the left-handed components of the singlet field must vanish.{\footnote{Note 
that this is not true for $z'=1/T,1/k$.}} This gives the second  
boundary condition:
\begin{align}
\label{boundCondRHZMpart2}
\left. [\partial_z -\frac{2}{z}-\frac{c}{z}]F^-_E(p,z,z') 
\right|_{z=1/k,1/T}=0\,.
\end{align}
Analogously, the boundary conditions for the doublet fermion
propagators are given by
\begin{subequations}
\begin{align}
\label{boundCondLHZM}
\left. F^-_L(p,z,z')\right|_{z=1/k,1/T}=0\,,\\
\left. [-\partial_z +\frac{2}{z}-\frac{c}{z}]F^+_L (p,z,z')
\right|_{z=1/k,1/T}=0\,.
\end{align}
\end{subequations}
It is now straightforward but tedious to determine the coefficient 
functions $A$ and $B$ which again can be expressed in terms of Bessel 
functions $J$ and $Y$. We give the results 
in a more concise form below.
 
For our calculations it is most convenient to decompose the fermion 
propagators into their RL, LR, LL and RR chiral components:
\begin{align}
\label{FermionPropagator}
 \Delta^{X}(p,x,y) =& 
-F^-_{X}(p,x,y) \slashed{p} P_L-F^+_{X}(p,x,y)
 \slashed{p} P_R
\nonumber\\  & 
+ d^+F^-_{X}(p,x,y)P_L + d^-F^+_{X}(p,x,y)P_R\,,
\end{align}
where the  $X$ can be either $E$ or $L$. 
The differential operators
\begin{align}
d^+= \partial_x -\frac{2}{x} - \frac{c}{x} && d^-= -\partial_x + 
\frac{2}{x} - \frac{c}{x}
\end{align}
always act on the first bulk position argument 
($x$ in \eqref{FermionPropagator}).
The decomposition \eqref{FermionPropagator} also allows for a simple 
representation in terms of Kaluza-Klein mode sums: 
\begin{subequations}
\begin{eqnarray}
\label{KKdecFermionPropagator}
F^+_{L}(p,x,y) &=&\sum_{n} f^{(n)}_L(x)\frac{-i}{p^2-m_n^2}f^{(n)}_L(y) \\ 
F^-_{L}(p,x,y)&=&\sum_{n} g^{(n)}_L(x)\frac{-i}{p^2-m_n^2}g^{(n)}_L(y) \\
d^-F^+_{L}(p,x,y)&=&\sum_{n} g^{(n)}_L(x)\frac{i m_n}{p^2-m_n^2}f^{(n)}_L(y)
\\
d^+F^-_{L}(p,x,y)&=&\sum_{n} f^{(n)}_L(x)\frac{i m_n }{p^2-m_n^2}g^{(n)}_L(y)\,.
\end{eqnarray}
\end{subequations}
Analogous expressions hold for the singlet field $E$.

The chiral components $d^\pm F^\mp$ and $F^{\pm}$ are scalar
functions; the Dirac and Lorentz structures 
of the propagator are contained only in the projectors and $\slashed{p}$.
Since the final evaluation of the loop integrals is performed 
numerically we need the analytic expressions for these scalar functions
after a Wick rotation has been performed.
In our matching calculation this rotation must be performed 
after the (Minkowski) expressions for the Wilson coefficients have 
been extracted and the Dirac and Lorentz algebra are already done. 
The Wilson coefficients 
are then given as integrals over scalar functions. 
The Wick rotation effectively amounts to the replacement $p^2\to -p^2$ or 
$p\to i p$ with $p=\sqrt{p^2}$, where $p$ is the loop momentum. 
The Bessel functions $J$ and $Y$ are replaced by 
the modified Bessel functions $I$ and $K$  of the first and second
kind, respectively, via
\begin{align} 
 J_\nu(ix) = i^\nu I_\nu (x) &&
 Y_\nu (ix) = i^{\nu+1}I_\nu (x) - \frac{2}{\pi} \,i^{-\nu} K_\nu(x)\,.
\end{align}

To write the chiral components of the fermion propagator 
in a compact way we make use of the following functions which were
first introduced in \cite{Csaki:2010aj}:
\begin{subequations}
\begin{eqnarray}
S_+(p,x, y,  c) &=& I_{c + 1/2}(px) K_{c + 1/2}(py) - 
                 K_{c + 1/2}(px) I_{c + 1/2}(py)\\
S_-(p,x, y,  c) &=& I_{c - 1/2}(px) K_{c - 1/2}(py) - 
                 K_{c - 1/2}(px) I_{c - 1/2}(py)\\
\tilde{S}_+(p,x, y,  c) &=& I_{c + 1/2}(px) K_{c - 1/2}(py) + 
                 K_{c + 1/2}(px) I_{c - 1/2}(py)\\
\tilde{S}_-(p,x, y,  c) &=& I_{c - 1/2}(px) K_{c + 1/2}(py) + 
                 K_{c - 1/2}(px) I_{c + 1/2}(py)\,.
\end{eqnarray}
\end{subequations}
Additionally, the following functions are useful when Taylor expansions 
of the propagators are needed 
\begin{subequations}
\begin{align}
S'_+(p, x, y,c) =&  \frac{1 + 2c}{2p^2} S_+(p,x,y, c) - 
   \frac{x}{2p} \tilde{S}_-(p,x,y, c) + \frac{y}{2 p}\tilde{S}_+(p,x,y, c)\\
S'_-(p, x, y,  c) =&  \frac{1 - 2 c}{2p^2} S_-(p, x, y, c) 
   -\frac{x}{2 p}\tilde{S}_+(p,x, y, c) + 
\frac{y}{2 p}\tilde{S}_-(p, x, y, c)\\
\tilde{S}'_+(p, x, y,  c) =&  \frac{1}{2 p^2}\tilde{S}_+(p,x,y, c) - 
\frac{x}{2 p}S_-(p, x,  y, c) + 
  \frac{y}{2 p} S_+(p, x,  y, c)\\
\tilde{S}'_-(p,x, y,  c) =& \frac{1}{2p^2} \tilde{S}_-(p,x,y, c) - 
\frac{x}{2 p} S_+(p, x,y, c) + \frac{y}{2p} S_-(p, x,  y, c)\,.
\end{align}
\end{subequations}
The corresponding expressions in Minkowski space can be recovered with help of 
the following substitution rules:
\begin{subequations}
\label{EuclidToMinkowski}
\begin{eqnarray}
S_+[I,K]\to -\frac{\pi}{2} S_+[J,Y] && 
\qquad \tilde S_+[I_\nu,K_\mu]\to 
i \frac{\pi}{2} \tilde S_+[J_\nu,(-1)^{\mu-c+1/2}Y_\mu]\\
S_-[I,K]\to -\frac{\pi}{2} S_-[J,Y] && 
\qquad \tilde S_-[I_\nu,K_\mu]\to 
-i\frac{\pi}{2} \tilde S_-[J_\nu,(-1)^{\mu-c-1/2}Y_\mu]
\end{eqnarray}
\end{subequations}
where the brackets $[\cdot,\cdot]$ show how $I$ and $K$ have to be 
replaced with the Bessel functions $J$ and $Y$, respectively. 

For the right-handed singlet fermion field the scalar functions 
are given by
\begin{subequations}
\label{SingletPropagators}
\begin{eqnarray}
 F^+_E(p, x, y) &=& 
 -\Theta(x - y) \frac{i k^4x^{5/2}y^{5/2} S_+(p,x, 1/T , c_E)S_+(p,y, 1/k , c_E)}
                      {S_+(p,1/T, 1/k , c_E)} \nonumber \\
 && - \, \Theta(y - x) \,\frac{ik^4 x^{5/2}y^{5/2} S_+(p,y, 1/T , c_E) 
 S_+(p,x, 1/k ,c_E)}{S_+(p,1/T,1/k , c_E)}\\
%%%%%%%%%%%%%%%%%
F^-_E(p,x,y) &=&
 \Theta(x - y)\,\frac{i k^4 x^{5/2} y^{5/2} \tilde{S}_-(p,x,1/T,c_E) 
 \tilde{S}_-(p,y,1/k,c_E)} {S_+(p,1/T,1/k,c_E)}\nonumber \\
 && + \,\Theta(y - x)\,\frac{i k^4 x^{5/2} y^{5/2}  
 \tilde{S}_-(p,y,1/T,c_E) \tilde{S}_-(p,x,1/k,c_E)} {S_+(p,1/T,1/k,c_E)}\\
%%%%%%%%%%%%%%%%
d^+F^-_E(p, x, y) &=& 
  p \,\Theta(x - y)\,\frac{i  k^4 x^{5/2} y^{5/2} {S}_+(p, x,1/T, c_E)
  \tilde{S}_-(p,y, 1/k,c_E)} {S_+(p,1/T, 1/k , c_E)} \nonumber \\
  && + \, p \,\Theta(y - x) \,\frac{i k^4 x^{5/2} y^{5/2} 
  \tilde{S}_-(p,y, 1/T,c_E) {S}_+(p, x, 1/k , c_E)} {S_-(p,1/T, 1/k , c_E)}\\
%%%%%%%%%%%%%%%%
d^-F^+_E(p,x,y) &=& 
 p \,\Theta(x - y)\,\frac{i k^4 x^{5/2} y^{5/2} 
 \tilde{S}_-(p,x,1/T,c_E){S}_+ (p,y,1/k,c_E)}{S_+(p,1/T,1/k,c_E)}\nonumber \\
 && +\,p \,\Theta(y - x)\,\frac{i k^4 x^{5/2} y^{5/2} 
 {S}_+(p,y,1/T,c_E) \tilde{S}_-(p,x,1/k,c_E)}{S_+(p,1/T,1/k,c_E)}\,.
\end{eqnarray}
\end{subequations}
For the left-handed doublet field $L$ we find 
\begin{subequations}
\label{DoubletPropagators}
\begin{align}
 F^+_L(p, x, y)=& \,
\Theta(x - y) \,\frac{i k^4x^{5/2} y^{5/2} 
    \tilde{S}_+(p,x, 1/T,c_L)\tilde{S}_+(p,y, 1/k ,c_L)}
   {S_-(p,1/T, 1/k,c_L)}\nonumber \\
& +  \Theta(y - x) \,\frac{i k^4 x^{5/2} y^{5/2} 
   \tilde{S}_+(p,y,1/T,c_L)\tilde{S}_+(p,x,1/k,c_L)}
   {S_-(p,1/T,1/k,c_L)} \\
%%%%%%%%%%%%%%%%
F^-_L(p,x,y) =& - \Theta(x - y) \,\frac{i k^4 x^{5/2} y^{5/2} S_-(p,x,1/T,c_L) S_-(p,y,1/k,c_L)}
                                            {S_-(p,1/T,1/k,c_L)}\nonumber\\
                 &-\Theta(y - x) \,\frac{i k^4 x^{5/2} y^{5/2} S_-(p,y,1/T,c_L) S_-(p,x,1/k,c_L)}
                                      {S_-(p,1/T,1/k,c_L)}\\
%%%%%%%%%%%%%%%%
d^+F^-_L(p, x, y) = 
  & - p \,\Theta(x - y)\,\frac{i  k^4 x^{5/2} y^{5/2} \tilde{S}_+(p, x,1/T, c_L)S_-(p,y, 1/k,c_L)}
                            {S_-(p,1/T, 1/k , c_L)} \nonumber \\
  & - p \,
   \Theta(y - x) \,\frac{i k^4 x^{5/2} y^{5/2} S_-(p,y, 1/T,c_L) \tilde{S}_+(p, x, 1/k , c_L)}
                      {S_-(p,1/T, 1/k , c_L)}
\label{dpFmL}\\
%%%%%%%%%%%%%%%%
d^-F^+_L(p,x,y) = 
 & -p \,\Theta(x - y)\,\frac{i k^4 x^{5/2} y^{5/2} S_-(p,x,1/T,c_L) \tilde{S}_+(p,y,1/k,c_L)}
                          {S_-(p,1/T,1/k,c_L)}\nonumber \\
 & -p \,\Theta(y - x)\,\frac{i k^4 x^{5/2} y^{5/2} \tilde{S}_+(p,y,1/T,c_L) S_-(p,x,1/k,c_L)}
                          {S_-(p,1/T,1/k,c_L)}\,.
\end{align}
\end{subequations}
Note that here $p=\sqrt{p^2}=\sqrt{p_0^2+p_1^2+p_2^2+p_3^2}\,$ is the 
Euclidean momentum.

\subsubsection{Gauge boson propagators}

The derivation of the propagators for the gauge sector follows
\cite{Randall:2001gb}.
The $R_\xi$ gauge is used to disentangle the orbifolding-even
vector components of the 5D gauge field from the  orbifolding-odd
 fifth component. The propagator for the ``vector part''
is then determined by the equation
\begin{align}
\label{EquationForVectorPropagator}
 \left[-p^2\eta^{\mu\rho}+p^{\mu} 
  p^{\rho }\left(1-\frac{1}{\xi}\right) - 
\eta^{\mu\rho} z\partial_z \left( \frac{1}{z}\partial_z\right)
\right]\Delta_{\rho\nu}(p,z,z')=ikz\,\delta(z-z') \delta^{\mu}_{\nu}\,.
\end{align}
Substituting the ansatz
\begin{align}
 \Delta^{\mu\nu}(p,x,y,\xi)= \Delta_\perp (p,x,y) \left(\eta^{\mu\nu}
   - \frac{p^\mu p^\nu}{p^2}\right)
+\frac{p^\mu p^\nu}{p^2}\Delta_\parallel(p,x,y,\xi)
\label{gaugepropansatz}
\end{align}
into \eqref{EquationForVectorPropagator} gives two equations
\begin{subequations}
\begin{align}
\label{1stEqGaugeProp}
 &(p^2+z\partial_z\frac{1}{z} \partial_z) \Delta_\perp(p,z,z')= 
-i kz \,\delta(z-z')\,,\\
\label{2ndEqGaugeProp}
 &(-p^2-z\partial_z\frac{1}{z} \partial_z) \Delta_\perp(p,z,z')
-\frac{1}{\xi}\left( -p^2-\xi z\partial_z\frac{1}{z} \partial_z
\right) \Delta_\parallel(p,z,z',\xi)=0\,.
\end{align}
\end{subequations}
Putting $\Delta_\parallel(p,z,z',\xi)=\Delta_\perp(p/\sqrt{\xi},z,z')$
the second equation is automatically satisfied, if $\Delta_\perp(p,z,z')$
solves the first one.
Eq.~\eqref{1stEqGaugeProp} can be solved with the same strategy 
employed before for the fermion case. Here the boundary conditions read
\begin{align}
 \partial_z \!\left.\Delta_\perp(p,z,z')\right|_{z=1/k,1/T}=0\,.
\end{align}

The propagator of the fifth component of the gauge field 
is denoted by $ \Delta_{5}(p,x,y,\xi)$ and follows readily 
from 
\begin{align}
 \left[p^2 + \xi\partial_z z \partial_z \frac{1}{z}\right]
\Delta_5(p,z,z')= i kz\,\delta(z-z')\,,
\label{gaugeprop5def}
\end{align}
and the boundary condition 
\begin{align}
 \left.\Delta_5(p,z,z')\right|_{z=1/k,1/T}=0\,.
\end{align}

Again we only give the Euclidean propagators:
\begin{eqnarray}
\label{gaugeprop1}
\Delta_\perp(p, x, y) &=& 
\Theta(x-y) \,\frac{i k x y \tilde{S}_+(p,x,1/T,1/2) \tilde{S}_+(p,y,1/k,1/2)}
{S_-(\tilde p,1/T,1/k,1/2)}
\nonumber \\
&& + \,\Theta(y-x)\,\frac{i k x y \tilde{S}_+(p,y,1/T,1/2) 
\tilde{S}_+(p,x,1/k,1/2)}{S_-(\tilde p,1/T,1/k,1/2)}
\\
%%%%%%%%%%%%%%%%%%%%%%%%
\label{gaugeprop2}
\Delta_\parallel(p, x, y,\xi) &=& \Delta_\perp(p/\sqrt{\xi}, x, y) 
\\[0.1cm]
%%%%%%%%%%%%%%%%%%%%%%%%
\xi \Delta_5(p, x, y,\xi) &=& 
\Theta(x-y) \,\frac{i k x y S_-(\tilde p,x,1/T,1/2)S_-(\tilde p,y,1/k,1/2)}
{S_-(\tilde p,1/T,1/k,1/2)}
\nonumber \\
&& + \,\Theta(y-x)  \,\frac{i k x y S_-(\tilde p,y,1/T,1/2)
S_-(\tilde p,x,1/k,1/2)}{S_-(\tilde p,1/T,1/k,1/2)}
\label{gaugeprop5}
\end{eqnarray}
where $\tilde p =p/\sqrt{\xi}$. 
The Minkowski space expressions can be recovered 
via \eqref{EuclidToMinkowski}.
The SM contribution to $(g-2)_\mu$ is, from point of view of the 
matching calculation at the KK scale, a long-distance contribution.
As described in the main text, this contribution is most effectively 
removed by subtracting the contribution of the SM gauge boson 
from each 5D boson propagator, which is equivalent to subtracting 
the zero mode. We therefore 
define the zero-mode subtracted (ZMS) propagators
\begin{eqnarray}
\label{ZMSgaugeprop}
\Delta^{\rm ZMS}_\perp(p, x, y)&=&
\Delta_\perp(p, x, y) - \frac{i}{p^2}\,f_\gamma^{(0)}(x)f_\gamma^{(0)}(y)\\
\Delta^{\rm ZMS}_\parallel(p, x, y,\xi) &=& 
\Delta^{\rm ZMS}_\perp(p/\sqrt{\xi}, x, y)\;. 
\end{eqnarray}
The boundary conditions for the fifth component of the gauge fields  
prevent the existence of a massless mode; no subtraction 
is necessary for $\Delta_5(p, x, y,\xi)$.

\subsubsection{Higgs propagator}
The Higgs field is IR brane localized. In the unbroken phase 
the Higgs propagator has a tachyonic mass $\mu^2$. Since $\mu^2\sim
v^2 \ll T^2$ we can treat the mass term as a perturbation of order 
$v^2$ in the calculation of the matching coefficients, 
and work with a standard massless scalar 4D propagator for the 
Higgs,
\begin{align}
 \Delta_H(x,y)_{mn}=   \int \frac{\rd^4p}{(2\pi)^4} \,e^{-ip(x-y)}\,
\frac{i \delta_{mn}}{p^2}\,,
\end{align}
where $m,n=1,2$ are SU(2) indices.

\section{Expanded propagators}
\label{App:Expansions}
If the virtuality of the loop momentum is of the 
order of the KK scale $T$ one can expand the propagators 
in powers of the small external momenta.
We effectively perform a Taylor expansion of the form
\begin{align}
 \Delta(p-k,x,y)\approx  \Delta(k,x,y) 
-2 p\cdot k\,\frac{\partial}{\partial k^2}  \Delta(k,x,y)+\ldots
\end{align}
where all momenta are still Minkowskian and $\Delta$ can be any 
propagator. Using the identities for derivatives of Bessel functions 
we obtain  analytic expressions for the derivatives.
In the following we denote 
$\frac{\partial}{\partial k^2}  \Delta(k,x,y)$ by $\widetilde\Delta(k,x,y)$.
All expressions are only valid {\it after} the Wick rotation has been 
performed. The expressions for fermion propagators read:\footnote{To 
shorten the notation, we denote the bulk mass parameter by $c$, where 
$c=c_E$ in the SU(2)-singlet fermion propagators and $c=c_L$ for the 
doublet ones, as in (\ref{SingletPropagators}), (\ref{DoubletPropagators}).}
\begin{subequations}
\begin{align}
& \widetilde{F^+_E}(p, x, y) =  \nonumber\\
& - \frac{i k^4 x^{5/2}y^{5/2}\Theta(x - y)}{S_+(p,1/T,1/k,c)}
    \bigg( S'_+(p,x, 1/T , c) S_+(p,y,1/k , c)
         + S_+(p,x,1/T ,c) S'_+(p, y,1/k , c) \nonumber\\
& -\frac{ S_+(p,x,1/T,c)S_+(p,y,1/k,c) S'_+(p,1/T, 1/k , c)}
        {S_+(p,1/T, 1/k,c)}\bigg)\nonumber\\
& - \frac{i k^4 x^{5/2}y^{5/2}\Theta(y - x)}{S_+(p,1/T,1/k,c)}
        \bigg( S'_+(p, y, 1/T , c) S_+(p,x,1/k,c)  
              +S_+(p,y,1/T,c) S'_+(p,x,1/k,c)\nonumber\\
& -\frac{S_+(p,y,1/T , c) S_+(p,x,1/k,c) S'_+(p, 1/T, 1/k , c)}
        {S_+(p,1/T,1/k , c)}\bigg)\\[0.2cm]
%%%%%%%%%%%%%%%%%%%%%%%%%%%%%%%
 &\widetilde{F^-_E}(p, x, y)=  \nonumber\\
 &\frac{i k^4 x^{5/2}y^{5/2}\Theta(x - y)}{S_+(p,1/T, 1/k,c)}
     \bigg(\tilde{S}'_-(p,x,1/T , c) \tilde{S}_-(p,y, 1/k,c) 
             + \tilde{S}_-(p,x,1/T , c)  \tilde{S}'_-(p,y, 1/k,c) 
  \nonumber\\
& - \frac{\tilde{S}_- (p,x,1/T , c) \tilde{S}_-(p,y, 1/k,c) S'_+(p,1/T, 1/k,c)}
                {S_+(p,1/T, 1/k,c)}\bigg)\nonumber\\
&+\frac{i k^4 x^{5/2}y^{5/2}\Theta(y - x)}{S_+(p,1/T, 1/k,c)}
   \bigg( \tilde{S}'_-(p, y, 1/T , c) \tilde{S}_-(p,x,1/k , c)
       + \tilde{S}_-(p, y, 1/T , c) \tilde{S}'_-(p,x,1/k , c) 
\nonumber\\
& - \frac{\tilde{S}_-(p, y, 1/T , c)\tilde{S}_-(p,x,1/k , c)
             S'_+(p,1/T, 1/k,c)}{S_+(p,1/T, 1/k,c)} \bigg)
\end{align}

\begin{align}
%%%%%%%%%%%%%%%%%%%%%%%%%%%%%%%
&\widetilde{F^-_L}(p, x, y) = \nonumber\\
&-\frac{i k^4 x^{5/2}y^{5/2}\Theta(x - y)}{S_-(p,1/T, 1/k,c)}
      \bigg(S'_-(p, x, 1/T , c) S_-(p,y, 1/k , c) 
       + S_-(p, x, 1/T , c) S'_-(p,y, 1/k , c) 
\nonumber\\
&  - \frac{S_-(p, x, 1/T , c) S_-(p,y, 1/k , c)
             S'_-(p,1/T, 1/k,c)}{S_-(p,1/T, 1/k,c)}\bigg)
\nonumber\\
&-\frac{i k^4 x^{5/2}y^{5/2} \Theta(y - x)}{S_-(p,1/T, 1/k,c)}
     \bigg(   S'_-(p,y,1/T, c) S_-(p,x,1/k,c) 
            + S_-(p,y,1/T, c)  S'_-(p,x,1/k,c) 
\nonumber\\
& - \frac{S_-(p,y,1/T, c)S_-(p,x,1/k,c)
                   S'_-(p,1/T, 1/k,c)}{S_-(p,1/T, 1/k,c)}\bigg)\\[0.2cm]
%%%%%%%%%%%%%%%%%%%%%%%%%%%%%%%
&\widetilde{F^+_L}(p, x, y) = \nonumber\\
&\frac{i k^4 x^{5/2}y^{5/2}\Theta(x - y)}{S_-(p,1/T, 1/k,c)}
   \bigg( \tilde{S}'_+(p x, 1/T , c) \tilde{S}_+(p,y, 1/k,c)
          + \tilde{S}_+(p x, 1/T , c)  \tilde{S}'_+(p,y, 1/k,c)
\nonumber\\
& - \frac{\tilde{S}_+(p x, 1/T , c)\tilde{S}_+(p,y, 1/k,c)
                  S'_-(p,1/T, 1/k,c)}{S_-(p,1/T, 1/k,c)}\bigg)
\nonumber\\
& +  \frac{i k^4 x^{5/2}y^{5/2}\Theta(y - x)}{S_-(p,1/T, 1/k,c)} 
       \bigg(  \tilde{S}'_+(p,y,1/T,c) \tilde{S}_+(p,x,1/k,c)
           +   \tilde{S}_+(p,y,1/T,c) \tilde{S}'_+(p,x,1/k,c)
\nonumber\\
&  - \frac{ \tilde{S}_+(p,y,1/T,c) \tilde{S}_+(p,x,1/k,c)
             S'_-(p,1/T, 1/k,c)}{S_-(p,1/T, 1/k,c)}\bigg)
\end{align}
\end{subequations}

\begin{subequations}
\begin{align}
&\widetilde{d^-F^+_E}(p,x,y) =
 \nonumber\\
& \frac{i p k^4 x^{5/2}y^{5/2} \Theta(x - y)}{S_+(p,1/T, 1/k,c)}
      \bigg(  \tilde{S}'_-(p,x, 1/T , c) S_+(p,y,1/k,c)
           +  \tilde{S}_-(p,x,1/T,c) S'_+(p,y,1/k,c)
\nonumber\\
&   -\frac{\tilde{S}_-(p,x, 1/T , c) S_+(p,y,1/k,c){S}'_+(p,1/T, 1/k,c)}
                      {S_+(p,1/T, 1/k,c)}\bigg)
\nonumber \\
& +\frac{i p k^4 x^{5/2}y^{5/2}\Theta(y - x)}{S_+(p,1/T, 1/k,c)}
       \bigg(S'_+(p,y, 1/T,c) \tilde{S}_-(p,x,1/k,c)
              + S_+(p,y, 1/T,c)  \tilde{S}'_-(p,x,1/k,c)
\nonumber\\
& - \frac{S_+(p,y, 1/T,c) \tilde{S}_-(p,x,1/k,c) S'_+(p,1/T, 1/k,c)}
         {S_+(p,1/T, 1/k,c)}\bigg)
\nonumber \\
& -  \frac{1}{2 p^2} d^-F^+_E(p,x,y)\\[0.2cm]
%%%%%%%%%%%%%%%%%%%%%%%
& \widetilde{d^+F^-_E}(p,x,y) =\nonumber \\
& \frac{i p k^4 x^{5/2}y^{5/2}\Theta(x - y)}{S_+(p,1/T, 1/k,c)}
     \bigg( S'_+(p,x,1/T,c) \tilde{S}_-(p,y,1/k,c)
          + S_+(p,x,1/T,c)  \tilde{S}'_-(p,y,1/k,c)
\nonumber\\
& -\frac{ S_+(p,x,1/T,c)\tilde{S}_-(p,y,1/k,c)
          S'_+(p,1/T, 1/k,c)}{S_+(p,1/T, 1/k,c)}\bigg)
\nonumber\\
&+ \frac{i p k^4 x^{5/2}y^{5/2} \Theta(y - x)}{S_+(p,1/T, 1/k,c)}
        \bigg(  \tilde{S}'_-(p,y,1/T,c) S_+(p,x,1/k,c)
               +\tilde{S}_-(p,y,1/T,c) S'_+(p,x,1/k,c) 
\nonumber\\
& -\frac{\tilde{S}_-(p,y,1/,c) S_+(p,x,1/k,c) S'_+(p,1/T, 1/k,c)}
                       {S_+(p,1/T, 1/k,c)}\bigg) 
\nonumber\\
&-  \frac{1}{2 p^2} d^+F^-_E(p, x, y)
\end{align}

\begin{align}
&\widetilde{d^+F^-_L}(p,x,y)=  
\nonumber\\
 &- \frac{i p k^4 x^{5/2}y^{5/2} \Theta(x - y)}{S_-(p,1/T, 1/k,c)}
      \bigg( \tilde{S}'_+(p, x, 1/T,c) S_-(p,y,1/k, c) 
           + \tilde{S}_+(p, x, 1/T,c) S'_-(p,y,1/k, c)
\nonumber\\
 & - \frac{\tilde{S}_+(p, x,1/T,c) S_-(p,y, 1/k,c) S'_-(p, 1/T,1/k,c)}
                     {S_-(p,1/T, 1/k,c)}\bigg) 
\nonumber\\
& -\frac{i p k^4 x^{5/2}y^{5/2} \Theta(y - x)}{S_-(p,1/T, 1/k,c)}
        \bigg(    S'_-(p,y, 1/T, c) \tilde{S}_+(p,x,1/k,c)
                + S_-(p,y, 1/T, c) \tilde{S}'_+(p,x,1/k,c)
\nonumber\\
& -\frac{S_-(p,y, 1/T, c) \tilde{S}_+(p,x,1/k,c)
                        S'_-(p,1/T, 1/k,c)}{S_-(p,1/T, 1/k,c)}\bigg)
\nonumber\\
&    -\frac{1}{2 p^2}d^+F^-_L(p,x, y)\\[0.2cm]
%%%%%%%%%%%%%%%%%%%%%%%%%%%%%%%%%%%
&\widetilde{d^-F^+_L}(p, x,y) =
\nonumber\\
& -\frac{i p k^4 x^{5/2}y^{5/2} \Theta(x - y)}{S_-(p,1/T, 1/k,c)}
   \bigg( S'_-(p,x, 1/T,c) \tilde{S}_+(p ,y ,1/k ,c)
          + S_-(p,x, 1/T,c)  \tilde{S}'_+(p ,y ,1/k ,c)
\nonumber\\
& - \frac{S_-(p,x, 1/T,c) \tilde{S}_+(p ,y ,1/k ,c)
          S'_-(p,1/T, 1/k,c)}{S_-(p,1/T, 1/k,c)}\bigg) 
\nonumber\\
&-\frac{i k^4 x^{5/2}y^{5/2} \Theta(y - x)}{S_-(p,1/T, 1/k,c)}
      \bigg( \tilde{S}'_+(p,y,1/T,c) S_-(p,x,1/k,c)
           + \tilde{S}_+(p,y,1/T,c) S'_-(p,x,1/k,c)
\nonumber\\
& -\frac{\tilde{S}_+(p,y,1/T,c) S_-(p,x,1/k,c) S'_-(p,1/T, 1/k,c)}
                    {S_-(p,1/T, 1/k,c)}\bigg) 
\nonumber\\
& -  \frac{1}{2 p^2} d^-F^+_L(p, x,y)\\[-0.2cm]
\nonumber
\end{align}
\end{subequations}
For the gauge boson propagators we need:
\begin{eqnarray}
\widetilde{\Delta}_\perp(p,x,y) &=&  i k x y\, \Theta(x - y) \nn \\[0.1cm]
&& \hspace*{-2cm} \times  
\bigg(\frac{ \tilde{S}^\prime_+(p, x, 1 /T , 1/2)\tilde{S}_+(l,y, 1/k , 1/2)}
{{S}_-(p,1/T, 1/k, 1/2)} 
+ \frac{\tilde{S}_+(p, x, 1 /T , 1/2) \tilde{S}^\prime_+(l,y, 1/k , 1/2)}
{{S}_-(p,1/T, 1/k, 1/2)} \nn \\
&&  \hspace*{-1.3cm}  -\frac{ \tilde{S}_+(p, x, 1 /T , 1/2) 
\tilde{S}_+(l,y, 1/k , 1/2) {S}^\prime_-(p,1/T, 1/k, 1/2)}
                      {{S}_-(p,1/T, 1/k, 1/2)^2}\bigg) \nn \\
&&  \hspace*{-2cm} + \,\big\lbrace x\leftrightarrow  y \big\rbrace  
\end{eqnarray}
%Finally, the expanded propagator for the fifth boson component
%is given by
\begin{eqnarray}
\widetilde{\Delta}_5(p,x,y) &=& \,\frac{i k x y}{\xi^2} 
\Theta(x-y) \nonumber \\[0.1cm]
&& \hspace*{-2cm} \times  
\bigg(\frac{S^\prime_-(\tilde p,x,1/T,1/2)S_-(\tilde p,y,1/k,1/2)}
{S_-(\tilde p,1/T,1/k,1/2)} +
\frac{S_-(\tilde p,x,1/T,1/2)S^\prime_-(\tilde p,y,1/k,1/2)}
{S_-(\tilde p,1/T,1/k,1/2)}  \nn \\
 &&  \hspace*{-1.3cm}
-\frac{S_-(\tilde p,x,1/T,1/2)S_-(\tilde p,y,1/k,1/2)S^\prime_-(\tilde p,1/T,1/k,1/2)}
{S_-(\tilde p,1/T,1/k,1/2)^2} \bigg)\nonumber \\
&& \hspace*{-2cm} + \left\lbrace x\leftrightarrow  y \right\rbrace
\end{eqnarray}

%%%%%%%%%%%%%%%%%%%%%%%%%%%%%%%%%%%%%%%%%%%%%%%%%%%%%%%%%%%%%%%%%%%

\section{Explicit expressions for the diagrams}
\label{ap:diags}

In this section we provide explicit expressions for the diagrams 
relevant to the matching of the dipole operator. To simplify the 
expressions, we consider the diagrams with an external photon, 
that is the linear combination $c_W B^\mu+s_W \frac{\tau^3}{2}
W^{3,\mu}$ of diagrams with an external hypercharge and SU(2) gauge 
boson. We furthermore assume that the external Higgs field 
(represented by a grey square) has a vanishing upper isospin 
component, that is we pick up the lower component only, since 
this is relevant when $\Phi$ is replaced by its vacuum expectation
value. These simplifications allow us to drop diagrams, which 
are proportional to $c_W g_5^\prime -s_w g_5=0$. Using 
$e_5 = c_W g_5^\prime = s_w g_5$ it is always possible to extract 
the 5D electromagnetic coupling. The electric charge of a fermion 
$\psi$ in units of positron charge is denoted by $Q_\psi$. In our 
convention the hypercharges are given by $Y_L=-1$, $Y_e=-2$, 
$Y_\Phi=1$, and $Q_\mu =[\frac{Y_L}{2}+\frac{\tau^3}{2}]_{22}=-1$. 
We use the short-hand $\hat{p}=p-l$ and
$\hat{p}^\prime=p^\prime-l$ for momenta in the loops.

The labels of the diagrams correspond to those given in
figure \ref{fig:diagsall}. The first expression for each
diagram does not make use of any simplification.
Vertex factors and propagators are written in the form as given in 
appendix~\ref{ap:5Drules} and no terms have been omitted.
The second expression already includes some simplifications. 
In particular, the terms that vanish due to the presence of 
chiral projectors after multiplying out the 
chiral components of the fermion propagators have been 
dropped. This second representation is exact only if the diagram 
does not contain an insertion of the Higgs field into an external line, 
since we applied the replacement (\ref{extfirstterm}).
The (small, see discussion in the main text) ``off-shell'' terms 
present in these diagrams is not shown explicitly and must be 
added. These contributions can be readily obtained from the
unsimplified expressions by replacing the external fermion
propagator by the zero-mode propagator only, e.g.,
in B2a one would replace
\begin{equation}
\Delta^L_i(p,x,1/T)\to 
f^{(0)}_{L_i}(x)\frac{i \slashed{p}}{p^2}f^{(0)}_{L_i}(1/T).
\end{equation}
One then applies standard Dirac algebra and simplifications as for 
the other term shown explicitly, including an expansion in the 
external momenta $p$, $p^\prime$ to the appropriate order.

\newpage
%%%%%%%%%%%%%%%%%%%%%%%%%%%%%%%%%%%%%%%%%%%%%%%%%
%% B_mu diagrams start here
%%%%%%%%%%%%%%%%%%%%%%%%%%%%%%%%%%%%%%%%%%%%%%%%%
\begin{figure}[htbp]
\begin{center}
\parbox{12cm}{\epsfig{file=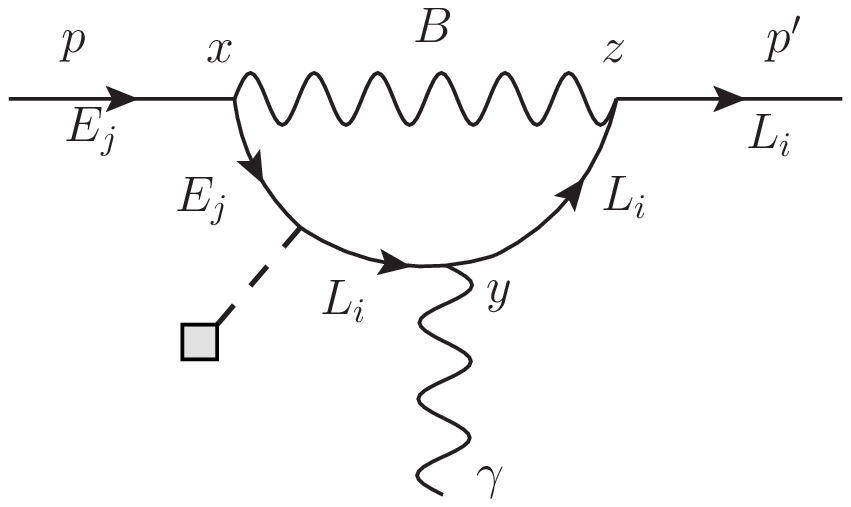,width=5.85cm,angle=0} 
\hspace*{5pt}
\epsfig{file=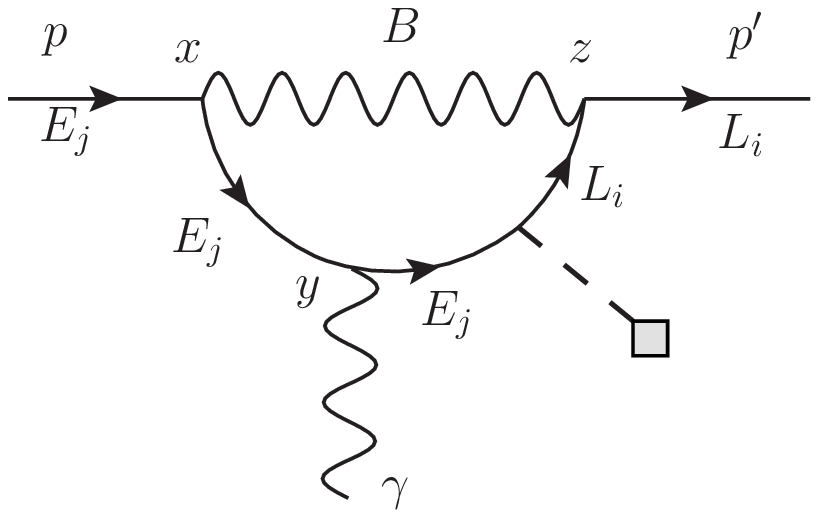,width=5.65cm,angle=0}
}
\end{center}
\end{figure}
\vspace*{-15pt}
%% Dia1 %%
\begin{eqnarray}
  \label{diag1}
  {\rm {\bf B1a}} &=& (ig'_5)^2(ie_5Q_\mu)\left(\frac{-iT^3}{k^3}\right)y^{\rm (5D)}_{ij} ~ \frac{Y_L}{2}\frac{Y_E}{2} \int^{1/T}_{1/k}
  \!\!\!\frac{dx}{(kx)^4} \int^{1/T}_{1/k}
  \!\!\!\frac{dy}{(ky)^4} \int^{1/T}_{1/k}
  \!\!\!\frac{dz}{(kz)^4}\int \!\!\frac{d^4l}{(2\pi)^4} \nonumber \\
  &&  f^{(0)}_{L_i}(z) f^{(0)}_\gamma(y) g^{(0)}_{E_j}(x)\epsilon^{*\mu} 
  \Delta^{\rho\nu}_{\mbox{\tiny ZMS}}(l,x,z)
  \nonumber \\
  &&  \times \bar{L_{i}}(p^\prime)P_{R} \gamma_\rho
  \Delta^{L}_{i}({p^{\,\prime}-l},z,y) \gamma_\mu
  \Delta^{L}_{i}({p-l},y,1/T) \Delta^{E}_{j}({p-l},1/T,x)
  \gamma_\nu P_{R} E_{j}(p)  \nonumber \\ 
  &=&  \frac{g'^2_5 e_5 Q_\mu Y_{L} Y_{E}
    y^{\rm(5D)}_{{ij}}T^3}{4k^3} \int^{1/T}_{1/k}
  \!\!\!\frac{dx}{(kx)^4} \int^{1/T}_{1/k}
  \!\!\!\frac{dy}{(ky)^4} \int^{1/T}_{1/k}
  \!\!\!\frac{dz}{(kz)^4}\int
  \!\!\frac{d^4l}{(2\pi)^4}  \nonumber \\
  &&  f^{(0)}_{L_i}(z) 
  f^{(0)}_\gamma(y) g^{(0)}_{E_j}(x) \epsilon^{*\mu}  \Delta^{\rho\nu}_{\mbox{\tiny ZMS}}(l,x,z)
  \nonumber \\
  && \times \bar{L_{i}}(p^{\,\prime}) \Big[
  F^+_{L_i}({\hat{p}^\prime},z,y) F^+_{L_i}({\hat{p}},y,1/T)
  F^-_{E_j}({\hat{p}},1/T,x) 
 \left\{ \gamma_\rho (\not\!p^{\,\prime}-\!\!\!\not l) \gamma_\mu \gamma_\nu\right\}(p-l)^2 + \nonumber \\  
  && d^+F^-_{L_i}({\hat{p}^{\,\prime}},z,y) d^-F^+_{L_i}({\hat{p}},y,1/T) F^-_{E_j}({\hat{p}},1/T,x) 
  \left\{ \gamma_\rho \gamma_\mu (\not\!p-\!\!\!\not l) \gamma_\nu
  \right\} \!\Big] P_{R} E_{{j}}(p)
\end{eqnarray}
%% Dia2 %%
\begin{eqnarray}
  \label{diag2}
  {\rm {\bf B1b}} &=&
  (ig'_5)^2(ie_5Q_\mu)\left(\frac{-iT^3}{k^3}\right)y^{\rm(5D)}_{ij}~
  \frac{Y_L}{2}\frac{Y_E}{2} \int^{1/T}_{1/k}
  \!\!\!\frac{dx}{(kx)^4} \int^{1/T}_{1/k}
  \!\!\!\frac{dy}{(ky)^4} \int^{1/T}_{1/k}
  \!\!\!\frac{dz}{(kz)^4}\int \!\!\frac{d^4l}{(2\pi)^4} \nonumber \\
  &&  f^{(0)}_{L_i}(z) f^{(0)}_\gamma(y) g^{(0)}_{E_j}(x)\epsilon^{*\mu} 
  \Delta^{\rho\nu}_{\mbox{\tiny ZMS}}(l,x,z)
  \nonumber \\
  && \times \bar{L_{i}}(p^{\prime})P_{R} \gamma_\rho
  \Delta^{L}_{i}({p^{\,\prime}-l},z,1/T) \Delta^{E}_{j}({p^{\,\prime}-l},1/T,y)
  \gamma_\mu \Delta^{E}_{j}({p-l},y,x) \gamma_\nu P_{R} E_{j}(p)
  \nonumber \\ 
  &=&  \frac{g'^2_5 e_5 Q_\mu Y_{L} Y_{E}
    y^{\rm(5D)}_{{ij}}T^3}{4k^3} \int^{1/T}_{1/k}
  \!\!\!\frac{dx}{(kx)^4} \int^{1/T}_{1/k}
  \!\!\!\frac{dy}{(ky)^4} \int^{1/T}_{1/k}
  \!\!\!\frac{dz}{(kz)^4}\int
  \!\!\frac{d^4l}{(2\pi)^4} \nonumber \\ 
  && f^{(0)}_{L_i}(z) f^{(0)}_\gamma(y) g^{(0)}_{E_j}(x) \epsilon^{*\mu}  \Delta^{\rho\nu}_{\mbox{\tiny ZMS}}(l,x,z)
  \nonumber \\
  && \times \bar{L_{i}}(p^{\prime}) \Big[ F^+_{L_i}({\hat{p}^\prime},z,1/T) F^-_{E_j}({\hat{p}^\prime},1/T,y) F^-_{E_j}({\hat{p}},y,x) 
  \left\{ \gamma_\rho \gamma_\mu (\not\!p-\!\!\!\not l) \gamma_\nu\right\}(p^{\,\prime}-l)^2 + \nonumber \\  
  && F^+_{L_i}({\hat{p}^\prime},z,1/T) d^-F^+_{E_j}({\hat{p}^\prime},1/T,y) d^+F^-_{E_j}({\hat{p}},y,x) 
  \left\{ \gamma_\rho (\not\!p^{\,\prime}-\!\!\!\not l) \gamma_\mu \gamma_\nu \right\} \!\Big] P_{R} E_{{j}}(p)
\end{eqnarray}
%% Dia3 %%
\newpage
\vspace*{-30pt}
\begin{figure}[htbp]
\begin{center}
\parbox{12cm}{\epsfig{file=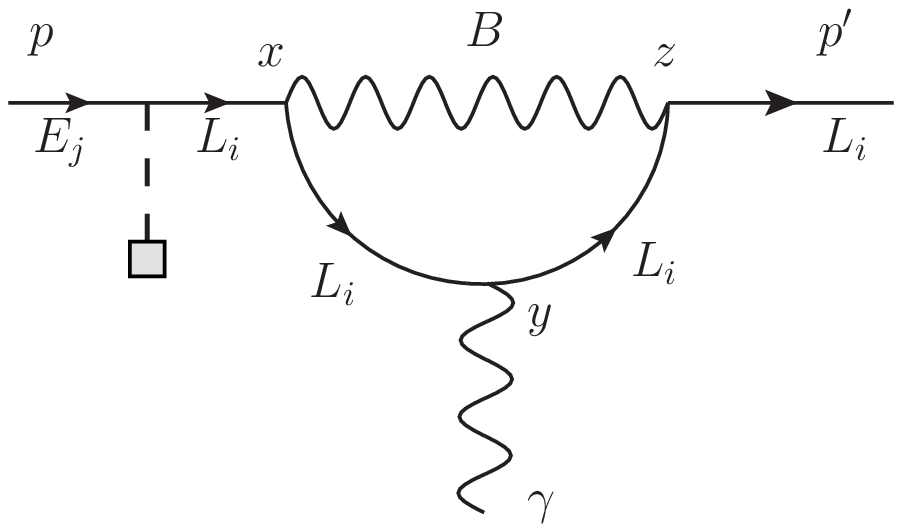,width=6.15cm,angle=0} 
\hspace*{5pt}
\epsfig{file=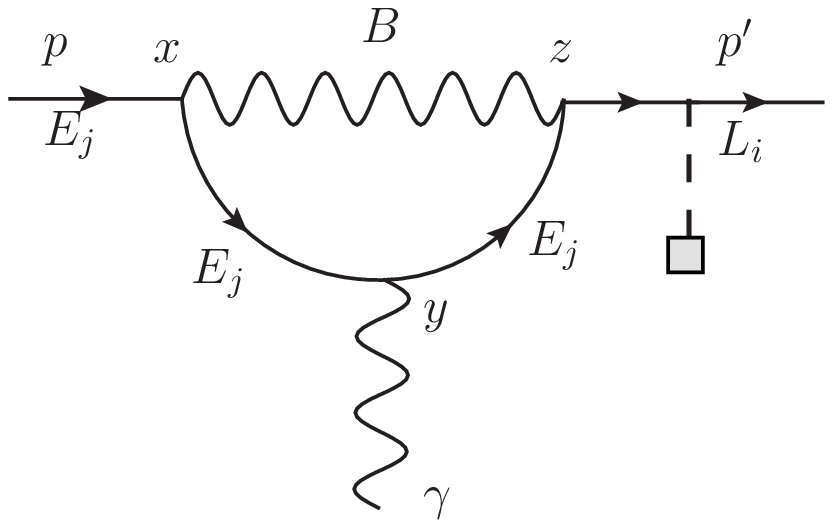,width=5.65cm,angle=0}
}
\end{center}
\end{figure}
\vspace*{-15pt}
\begin{eqnarray}
  \label{diag3}
  {\rm {\bf B2a}} &=& (ig'_5)^2(ie_5Q_\mu)\left(\frac{-iT^3}{k^3}\right)y^{\rm (5D)}_{ij}\frac{Y_L}{2}\frac{Y_L}{2} \int^{1/T}_{1/k}
  \!\!\!\frac{dx}{(kx)^4} \int^{1/T}_{1/k}
  \!\!\!\frac{dy}{(ky)^4} \int^{1/T}_{1/k}
  \!\!\!\frac{dz}{(kz)^4}\int \!\!\frac{d^4l}{(2\pi)^4} \nonumber \\
  &&  f^{(0)}_{L_i}(z) f^{(0)}_\gamma(y) g^{(0)}_{E_j}(1/T)\epsilon^{*\mu} 
  \Delta^{\rho\nu}_{\mbox{\tiny ZMS}}(l,x,z)
  \nonumber \\
  && \times \bar{L_{i}}(p^{\prime})P_{R} \gamma_\rho
    \Delta^{L}_{i}({p^{\,\prime}-l},z,y) \gamma_\mu \Delta^{L}_{i}({p-l},y,x)
    \gamma_\nu \Delta^{L}_{i}({p},x,1/T) P_{R} E_{j}(p)
   \nonumber \\[0.1cm] 
  &=&  \frac{g'^2_5 e_5 Q_\mu Y^2_{L}
    y^{\rm(5D)}_{{ij}}T^3}{4k^3} \int^{1/T}_{1/k}
  \!\!\!\frac{dx}{(kx)^4} \int^{1/T}_{1/k}
  \!\!\!\frac{dy}{(ky)^4} \int^{1/T}_{1/k}
  \!\!\!\frac{dz}{(kz)^4}\int
  \!\!\frac{d^4l}{(2\pi)^4} \nonumber \\
  && f^{(0)}_{L_i}(z) f^{(0)}_\gamma(y) g^{(0)}_{E_j}(1/T) \epsilon^{*\mu}  \Delta^{\rho\nu}_{\mbox{\tiny ZMS}}(l,x,z)
   \nonumber \\
  && \times \bar{L_{i}}(p^{\prime}) \Big[ F^+_{L_i}({\hat{p}^\prime},z,y) d^+F^-_{L_i}({\hat{p}},y,x) d^-F^+_{L_i}({p},x,1/T) 
  \left\{ \gamma_\rho (\not\!p^{\,\prime}-\!\!\!\not l) \gamma_\mu \gamma_\nu \right\} + \nonumber \\
  && d^+F^-_{L_i}({\hat{p}^\prime},z,y) F^-_{L_i}({\hat{p}},y,x) d^-F^+_{L_i}({p},x,1/T) 
  \left\{ \gamma_\rho \gamma_\mu (\not\!p-\!\!\!\not l) \gamma_\nu \right\} 
  \Big] P_{R} E_{{j}}(p)
\end{eqnarray}
%% Dia4 %%
\begin{eqnarray}
  \label{diag4}
  {\rm {\bf B2b}} &=& (ig'_5)^2(ie_5Q_\mu)\left(\frac{-iT^3}{k^3}\right)y^{\rm (5D)}_{ij}\frac{Y_E}{2}\frac{Y_E}{2} \int^{1/T}_{1/k}
  \!\!\!\frac{dx}{(kx)^4} \int^{1/T}_{1/k}
  \!\!\!\frac{dy}{(ky)^4} \int^{1/T}_{1/k}
  \!\!\!\frac{dz}{(kz)^4}\int \!\!\frac{d^4l}{(2\pi)^4} \nonumber \\
  &&  f^{(0)}_{L_i}(1/T) f^{(0)}_\gamma(y) g^{(0)}_{E_j}(x)\epsilon^{*\mu} 
  \Delta^{\rho\nu}_{\mbox{\tiny ZMS}}(l,x,z)
  \nonumber \\ 
  && \times \bar{L_{i}}(p^{\prime})P_{R} \Delta^{E}_{j}({p^{\,\prime}},1/T,z)
  \gamma_\rho \Delta^{E}_{j}({p^{\,\prime}-l},z,y) \gamma_\mu
  \Delta^{E}_{j}({p-l},y,x) \gamma_\nu P_{R} E_{j}(p)
  \nonumber \\[0.1cm]
  &=&  \frac{g'^2_5 e_5 Q_\mu Y^2_{E}
    y^{\rm(5D)}_{{ij}}T^3}{4k^3} \int^{1/T}_{1/k}
  \!\!\!\frac{dx}{(kx)^4} \int^{1/T}_{1/k}
  \!\!\!\frac{dy}{(ky)^4} \int^{1/T}_{1/k}
  \!\!\!\frac{dz}{(kz)^4}\int
  \!\!\frac{d^4l}{(2\pi)^4} \nonumber \\ 
  && f^{(0)}_{L_i}(1/T) f^{(0)}_\gamma(y) g^{(0)}_{E_j}(x) \epsilon^{*\mu} \Delta^{\rho\nu}_{\mbox{\tiny ZMS}}(l,x,z)
  \nonumber \\
  && \times \bar{L_{i}}(p^{\prime}) \Big[ d^-F^+_{E_j}({p^{\,\prime}},1/T,z) d^+F^-_{E_j}({\hat{p}^\prime},z,y) F^-_{E_j}({\hat{p}},y,x) 
  \left\{ \gamma_\rho \gamma_\mu (\not\!p-\!\!\!\not l) \gamma_\nu \right\} + 
  \nonumber \\  
  && d^-F^+_{E_j}({p^{\,\prime}},1/T,z) F^+_{E_j}({\hat{p}^\prime},z,y) d^+F^-_{E_j}({\hat{p}},y,x) 
  \left\{ \gamma_\rho (\not\!p^{\,\prime}-\!\!\!\not l) \gamma_\mu \gamma_\nu \right\} 
  \Big] P_{R} E_{{j}}(p)
\end{eqnarray}
%%%%%%%%%%%%%%%%%%%%%%%%%%%%%%%%%%%%%%%%%%%%%%%%%
%% B_5 diagrams start here
%%%%%%%%%%%%%%%%%%%%%%%%%%%%%%%%%%%%%%%%%%%%%%%%%
\newpage
\begin{figure}[htbp]
\begin{center}
\parbox{12cm}{\epsfig{file=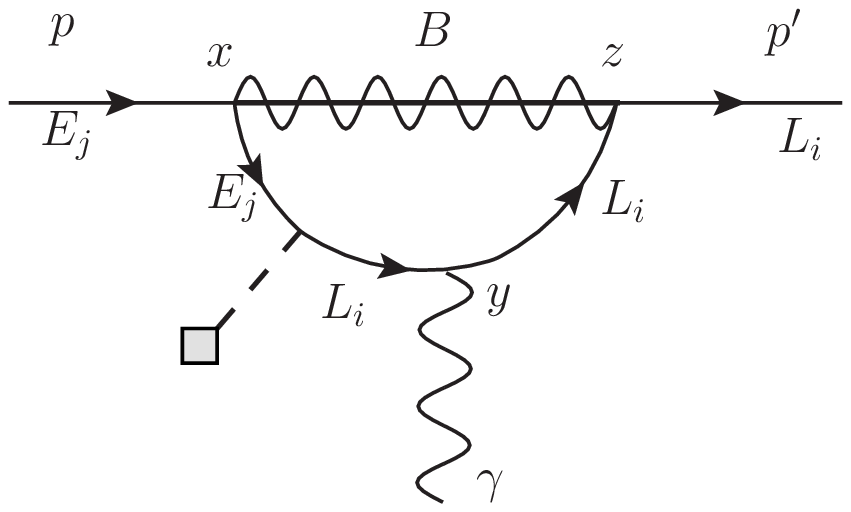,width=5.75cm,angle=0} 
\hspace*{5pt}
\epsfig{file=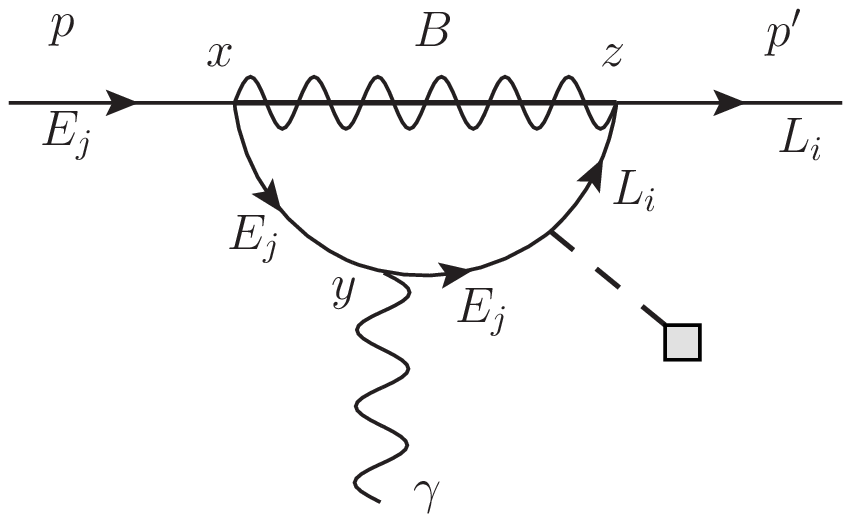,width=5.65cm,angle=0}
}
\end{center}
\end{figure}
\vspace*{-15pt}
%% Dia5 %%
\begin{eqnarray}
  \label{diag5}
  {\rm {\bf B3a}} &=& (g'_5)^2(ie_5Q_\mu)\left(\frac{-iT^3}{k^3}\right)y^{\rm (5D)}_{ij}~ \frac{Y_L}{2}\frac{Y_E}{2} \int^{1/T}_{1/k}
  \!\!\!\frac{dx}{(kx)^4} \int^{1/T}_{1/k}
  \!\!\!\frac{dy}{(ky)^4} \int^{1/T}_{1/k}
  \!\!\!\frac{dz}{(kz)^4}\int \!\!\frac{d^4l}{(2\pi)^4} \nonumber \\
  &&  f^{(0)}_{L_i}(z) f^{(0)}_\gamma(y) g^{(0)}_{E_j}(x)\epsilon^{*\mu} 
  \Delta_{5}(l,x,z) \nonumber \\
  &&  \times \bar{L_{i}}(p^{\prime})P_{R} \gamma_5
  \Delta^{L}_{i}({p^{\,\prime}-l},z,y) \gamma_\mu
  \Delta^{L}_{i}({p-l},y,1/T) \Delta^{E}_{j}({p-l},1/T,x)
  \gamma_5 P_{R} E_{j}(p)  \nonumber \\[0.1cm]
  &=& - \frac{g'^2_5 e_5 Q_\mu Y_{L} Y_{E}
    y^{\rm(5D)}_{{ij}}T^3}{4k^3} \int^{1/T}_{1/k}
  \!\!\!\frac{dx}{(kx)^4} \int^{1/T}_{1/k}
  \!\!\!\frac{dy}{(ky)^4} \int^{1/T}_{1/k}
  \!\!\!\frac{dz}{(kz)^4}\int
  \!\!\frac{d^4l}{(2\pi)^4}  \nonumber \\
  &&  f^{(0)}_{L_i}(z) f^{(0)}_\gamma(y) g^{(0)}_{E_j}(x) \epsilon^{*\mu} \Delta_{5}(l,x,z)
  \nonumber \\
  && \times \bar{L_{i}}(p^{\prime}) \Big[ d^-F^+_{L_i}({\hat{p}^\prime},z,y) F^+_{L_i}({\hat{p}},y,1/T) d^-F^+_{E_j}({\hat{p}},1/T,x) 
  \left\{ \gamma_\mu (\not\!p-\!\!\!\not l) \right\}+ \nonumber \\  
  && F^-_{L_i}({\hat{p}^\prime},z,y) d^-F^+_{L_i}({\hat{p}},y,1/T) d^-F^+_{E_j}({\hat{p}},1/T,x) 
  \left\{ (\not\!p^{\,\prime}-\!\!\!\not l) \gamma_\mu \right\} \Big] P_{R} E_{{j}}(p)
\end{eqnarray}
%% Dia6 %%
\begin{eqnarray}
  \label{diag6}
  {\rm {\bf B3b}} &=&
  (g'_5)^2(ie_5Q_\mu)\left(\frac{-iT^3}{k^3}\right)y^{\rm(5D)}_{ij}~
  \frac{Y_L}{2}\frac{Y_E}{2} \int^{1/T}_{1/k}
  \!\!\!\frac{dx}{(kx)^4} \int^{1/T}_{1/k}
  \!\!\!\frac{dy}{(ky)^4} \int^{1/T}_{1/k}
  \!\!\!\frac{dz}{(kz)^4}\int \!\!\frac{d^4l}{(2\pi)^4} \nonumber \\
  &&  f^{(0)}_{L_i}(z) f^{(0)}_\gamma(y) g^{(0)}_{E_j}(x) \epsilon^{*\mu}
  \Delta_{5}(l,x,z) \nonumber \\
  && \times \bar{L_{i}}(p^{\prime})P_{R} \gamma_5
  \Delta^{L}_{i}({p^{\,\prime}-l},z,1/T) \Delta^{E}_{j}({p^{\,\prime}-l},1/T,y)
  \gamma_\mu \Delta^{E}_{j}({p-l},y,x) \gamma_5 P_{R} E_{j}(p)
  \nonumber \\[0.1cm] 
  &=& - \frac{g'^2_5 e_5 Q_\mu Y_{L} Y_{E}
    y^{\rm(5D)}_{{ij}}T^3}{4k^3} \int^{1/T}_{1/k}
  \!\!\!\frac{dx}{(kx)^4} \int^{1/T}_{1/k}
  \!\!\!\frac{dy}{(ky)^4} \int^{1/T}_{1/k}
  \!\!\!\frac{dz}{(kz)^4}\int
  \!\!\frac{d^4l}{(2\pi)^4}  \nonumber \\ 
  && f^{(0)}_{L_i}(z) f^{(0)}_\gamma(y) g^{(0)}_{E_j}(x) \epsilon^{*\mu} \Delta_{5}(l,x,z)
  \nonumber \\
  && \times \bar{L_{i}}(p^{\prime}) \Big[ d^-F^+_{L_i}({\hat{p}^\prime},z,1/T) d^-F^+_{E_j}({\hat{p}^\prime},1/T,y) F^+_{E_j}({\hat{p}},y,x) 
  \left\{ \gamma_\mu (\not\!p-\!\!\!\not l)\right\} + \nonumber \\  
  && d^-F^+_{L_i}({\hat{p}^\prime},z,1/T) F^-_{E_j}({\hat{p}^\prime},1/T,y) d^-F^+_{E_j}({\hat{p}},y,x) 
  \left\{ (\not\!p^{\,\prime}-\!\!\!\not l) \gamma_\mu \right\} \Big] P_{R} E_{{j}}(p)
\end{eqnarray}
\newpage
\vspace*{-30pt}
\begin{figure}[htbp]
\begin{center}
\parbox{12cm}{\epsfig{file=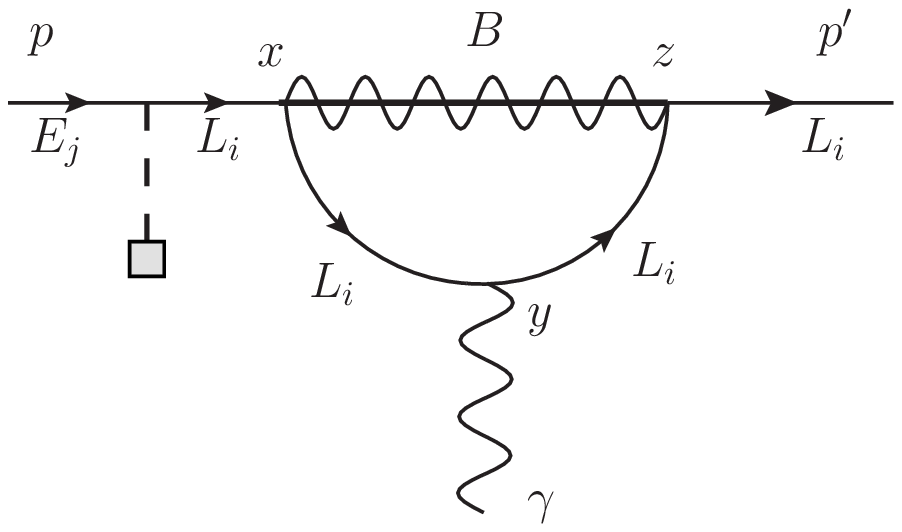,width=6.15cm,angle=0} 
\hspace*{5pt}
\epsfig{file=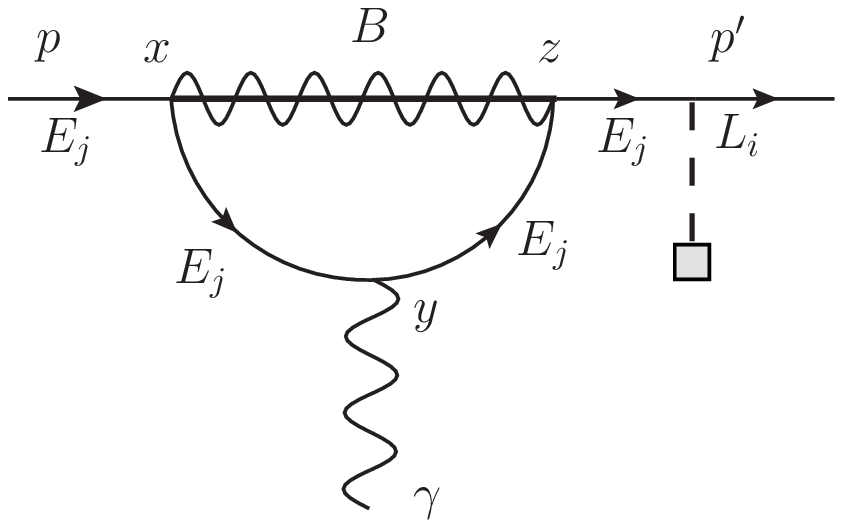,width=5.65cm,angle=0}
}
\end{center}
\end{figure}
\vspace*{-15pt}
%% Dia7 %%
\begin{eqnarray}
  \label{diag7}
  {\rm {\bf B4a}} &=& (g'_5)^2(ie_5Q_\mu)\left(\frac{-iT^3}{k^3}\right)y^{\rm (5D)}_{ij}\frac{Y_L}{2}\frac{Y_L}{2} \int^{1/T}_{1/k}
  \!\!\!\frac{dx}{(kx)^4} \int^{1/T}_{1/k}
  \!\!\!\frac{dy}{(ky)^4} \int^{1/T}_{1/k}
  \!\!\!\frac{dz}{(kz)^4}\int \!\!\frac{d^4l}{(2\pi)^4} \nonumber \\
  &&  f^{(0)}_{L_i}(z) f^{(0)}_\gamma(y) g^{(0)}_{E_j}(1/T) \epsilon^{*\mu} 
  \Delta_{5}(l,x,z) \nonumber \\
  && \times \bar{L_{i}}(p^{\prime})P_{R} \gamma_5
  \Delta^{L}_{i}({p^{\,\prime}-l},z,y) \gamma_\mu \Delta^{L}_{i}({p-l},y,x)
  \gamma_5 \Delta^{L}_{i}({p},x,1/T) P_{R} E_{j}(p)
  \nonumber \\[0.1cm] 
  &=& - \frac{g'^2_5 e_5 Q_\mu Y^2_{L}
    y^{\rm(5D)}_{{ij}}T^3}{4k^3} \int^{1/T}_{1/k}
  \!\!\!\frac{dx}{(kx)^4} \int^{1/T}_{1/k}
  \!\!\!\frac{dy}{(ky)^4} \int^{1/T}_{1/k}
  \!\!\!\frac{dz}{(kz)^4}\int
  \!\! \frac{d^4l}{(2\pi)^4}  \nonumber \\ && 
  f^{(0)}_{L_i}(z) f^{(0)}_\gamma(y) g^{(0)}_{E_j}(1/T) \epsilon^{*\mu} \Delta_{5}(l,x,z)
  \nonumber \\
  && \times \bar{L_{i}}(p^{\prime}) \Big[ d^-F^+_{L_i}({\hat{p}^\prime},z,y) F^+_{L_i}({\hat{p}},y,x) d^-F^+_{L_i}({p},x,1/T) 
  \left\{ \gamma_\mu (\not\!p-\!\!\!\not l) \right\} + \nonumber \\
  && F^-_{L_i}({\hat{p}^\prime},z,y) d^-F^+_{L_i}({\hat{p}},y,x) d^-F^+_{L_i}({p},x,1/T) 
  \left\{ (\not\!p^{\,\prime}-\!\!\!\not l) \gamma_\mu \right\} 
  \Big] P_{R} E_{{j}}(p)
\end{eqnarray}
%% Dia8 %%
\begin{eqnarray}
  \label{diag8}
  {\rm {\bf B4b}} &=& (g'_5)^2(ie_5Q_\mu)\left(\frac{-iT^3}{k^3}\right)y^{\rm (5D)}_{ij}\frac{Y_E}{2}\frac{Y_E}{2} \int^{1/T}_{1/k}
  \!\!\!\frac{dx}{(kx)^4} \int^{1/T}_{1/k}
  \!\!\!\frac{dy}{(ky)^4} \int^{1/T}_{1/k}
  \!\!\!\frac{dz}{(kz)^4}\int \!\!\frac{d^4l}{(2\pi)^4} \nonumber \\
  &&  f^{(0)}_{L_i}(1/T) f^{(0)}_\gamma(y) g^{(0)}_{E_j}(x) \epsilon^{*\mu} 
  \Delta_{5}(l,x,z) \nonumber \\ 
  && \times \bar{L_{i}}(p^{\prime})P_{R} \Delta^{E}_{j}({p^{\,\prime}},1/T,z)
  \gamma_5 \Delta^{E}_{j}({p^{\,\prime}-l},z,y) \gamma_\mu
  \Delta^{E}_{j}({p-l},y,x) \gamma_5 P_{R} E_{j}(p)
  \nonumber \\[0.1cm] 
  &=& - \frac{g'^2_5 e_5 Q_\mu Y^2_{E}
    y^{\rm(5D)}_{{ij}}T^3}{4k^3} \int^{1/T}_{1/k}
  \!\!\!\frac{dx}{(kx)^4} \int^{1/T}_{1/k}
  \!\!\!\frac{dy}{(ky)^4} \int^{1/T}_{1/k}
  \!\!\!\frac{dz}{(kz)^4}\int
  \!\!\frac{d^4l}{(2\pi)^4} \nonumber \\ && f^{(0)}_{L_i}(1/T) 
  f^{(0)}_\gamma(y) g^{(0)}_{E_j}(x) \epsilon^{*\mu} \Delta^{\rho\nu}_{\mbox{\tiny ZMS}}(l,x,z)
  \nonumber \\
  && \times \bar{L_{i}}(p^{\prime}) \Big[ d^-F^+_{E_j}({p^{\,\prime}},1/T,z) d^-F^+_{E_j}({\hat{p}^\prime},z,y) F^+_{E_j}({\hat{p}},y,x) 
  \left\{ \gamma_\mu (\not\!p-\!\!\!\not l) \right\} + 
  \nonumber \\  
  && d^-F^+_{E_j}({p^{\,\prime}},1/T,z) F^-_{E_j}({\hat{p}^\prime},z,y) d^-F^+_{E_j}({\hat{p}},y,x) 
  \left\{ (\not\!p^{\,\prime}-\!\!\!\not l) \gamma_\mu \right\} 
  \Big] P_{R} E_{{j}}(p)
\end{eqnarray}
%%%%%%%%%%%%%%%%%%%%%%%%%%%%%%%%%%%%%%%%%%%%%%%%%%%%%%%%%%%%%%%%%%%%%%
%% abelian higgs diagrams start here %%%
%%%%%%%%%%%%%%%%%%%%%%%%%%%%%%%%%%%%%%%%%%%%%%%%%%%%%%%%%%%%%%%%%%%%%%
\newpage
\begin{figure}[htbp]
\begin{center}
\parbox{12cm}{\epsfig{file=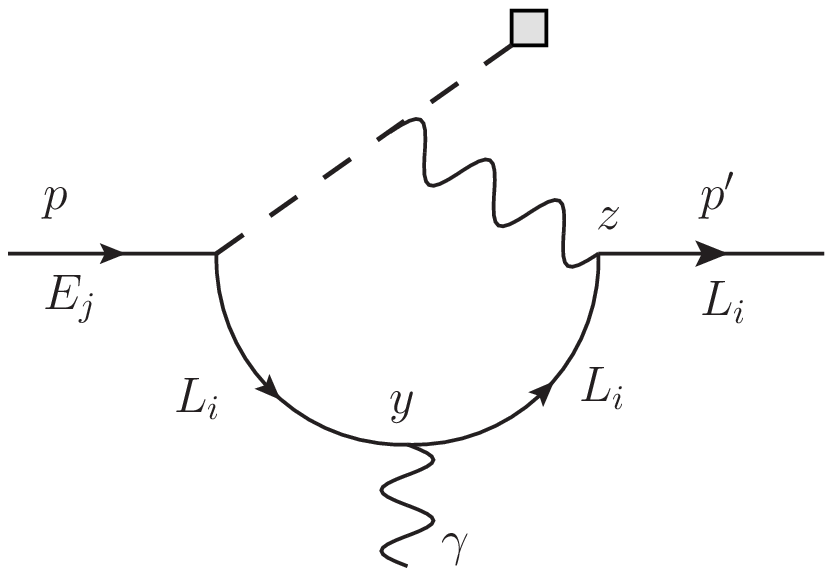,width=5.65cm,angle=0} 
\hspace*{5pt}
\epsfig{file=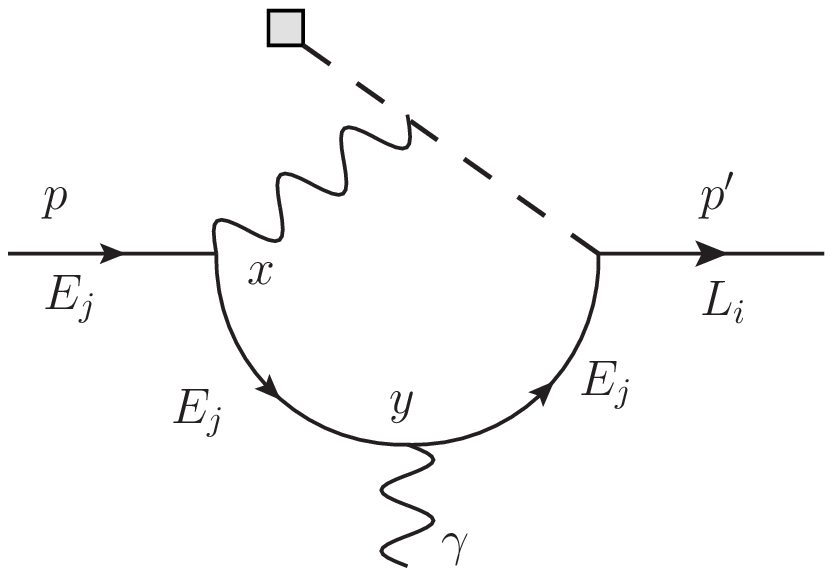,width=5.55cm,angle=0}
}
\end{center}
\end{figure}
\vspace*{-15pt}
%% Dia19 %%
\begin{eqnarray}
  \label{diag19}
  {\rm {\bf B5a}} &=&
(ig'_5)^2(ie_5Q_\mu)\left[\frac{Y_L}{2}\frac{Y_{\Phi}}{2}\right] \left(\frac{-iT^3}{k^3}\right)
  y^{\rm(5D)}_{ij}~\int^{1/T}_{1/k}
  \!\!\!\frac{dy}{(ky)^4}\int^{1/T}_{1/k}
  \!\!\!\frac{dz}{(kz)^4}\int \!\!\frac{d^4l}{(2\pi)^4} \nonumber \\
  &&  f^{(0)}_{L_i}(z) f^{(0)}_\gamma(y) g^{(0)}_{E_j}(1/T) \epsilon^{*\mu} 
  \Delta_{H}(l) \Delta^{\alpha\beta}_{\mbox{\tiny ZMS}}(l,1/T,z) 
  \nonumber \\
  && \times \bar{L_{i}}(p^{\prime})P_{R} \gamma_\alpha \Delta^{L}_{i}({p^{\,\prime}-l},z,y)\gamma_\mu \Delta^{L}_{i}({p-l},y,1/T) (-l_\beta) P_{R} E_{j}(p)
  \nonumber \\
  &=& \frac{g'^2_5 e_5 Q_\mu Y_{\Phi}Y_L y^{\rm(5D)}_{{ij}}T^3}{4k^3} \int^{1/T}_{1/k}
  \!\!\!\frac{dy}{(ky)^4}\int^{1/T}_{1/k}
  \!\!\!\frac{dz}{(kz)^4}\int
  \!\!\frac{d^4l}{(2\pi)^4} \nonumber \\ && f^{(0)}_{L_i}(z) f^{(0)}_\gamma(y) g^{(0)}_{E_j}(1/T) \epsilon^{*\mu} 
  \Delta_{H}(l) \Delta^{\alpha\beta}_{\mbox{\tiny ZMS}}(l,1/T,z) 
  \nonumber \\
  && \times \bar{L_{i}}(p^{\prime}) \Big[d^+F^-_{L_i}(\hat{p^{\,\prime}},z,y) d^-F^+_{L_i}(\hat{p},y,1/T) \left(\gamma_\alpha\gamma_\mu\right)l_\beta + \nonumber \\
  && F^+_{L_i}(\hat{p^{\,\prime}},z,y)
  F^+_{L_i}(\hat{p},y,1/T)\left\{\gamma_\alpha(\not\!p^{\,\prime}-\!\!\!\not l)\gamma_\mu(\not\!p-\!\!\!\not l)\right\}l_\beta\Big] P_{R} E_{{j}}(p) 
\end{eqnarray}
%% Dia20 %%
\begin{eqnarray}
  \label{diag20}
  {\rm {\bf B5b}} &=& 
 (ig'_5)^2(ie_5Q_\mu)\left[\frac{Y_E}{2}\frac{Y_{\Phi}}{2}\right] \left(\frac{-iT^3}{k^3}\right)
  y^{\rm(5D)}_{ij}~\int^{1/T}_{1/k}
  \!\!\!\frac{dx}{(kx)^4}\int^{1/T}_{1/k}
  \!\!\!\frac{dy}{(ky)^4}\int \!\!\frac{d^4l}{(2\pi)^4} \nonumber \\
  &&  f^{(0)}_{L_i}(1/T) f^{(0)}_\gamma(y) g^{(0)}_{E_j}(x) \epsilon^{*\mu} 
  \Delta_{H}(l) \Delta^{\alpha\beta}_{\mbox{\tiny ZMS}}(l,x,1/T) 
  \nonumber \\
  && \times \bar{L_{i}}(p^{\prime})P_{R} \Delta^{E}_{j}({p^{\,\prime}-l},1/T,y)\gamma_\mu \Delta^{E}_{j}({p-l},y,x)\gamma_\alpha l_\beta P_{R} E_{j}(p)
  \nonumber \\ 
  &=& -\frac{g'^2_5 e_5 Q_\mu Y_{\Phi}Y_E y^{\rm(5D)}_{{ij}}T^3}{4k^3} \int^{1/T}_{1/k}
  \!\!\!\frac{dx}{(kx)^4}\int^{1/T}_{1/k}
  \!\!\!\frac{dy}{(ky)^4}\int
  \!\!\frac{d^4l}{(2\pi)^4} \nonumber \\ && f^{(0)}_{L_i}(1/T) f^{(0)}_\gamma(y) g^{(0)}_{E_j}(x) \epsilon^{*\mu} 
  \Delta_{H}(l) \Delta^{\alpha\beta}_{\mbox{\tiny ZMS}}(l,x,1/T) 
  \nonumber \\
  && \times \bar{L_{i}}(p^{\prime}) \Big[d^-F^+_{E_j}(\hat{p^{\,\prime}},1/T,y) d^+F^-_{E_j}(\hat{p},y,x) \left(\gamma_\mu\gamma_\alpha\right)l_\beta + \nonumber \\
  && F^-_{E_j}(\hat{p^{\,\prime}},1/T,y)
  F^-_{E_j}(\hat{p},y,x)\left\{(\not\!p^{\,\prime}-\!\!\!\not
    l)\gamma_\mu(\not\!p-\!\!\!\not
    l)\gamma_\alpha\right\}l_\beta\Big] P_{R} E_{{j}}(p) 
\end{eqnarray}
%%%%%%%%%%%%%%%%%%%%%%%%%%%%%%%%%%%%%%%%%%%%%%%%%
%% W_mu diagrams start here
%%%%%%%%%%%%%%%%%%%%%%%%%%%%%%%%%%%%%%%%%%%%%%%%%
%% Dia9 %%
\newpage
\vspace*{20pt}
\begin{figure}[htbp]
\begin{center}
\parbox{12cm}{\epsfig{file=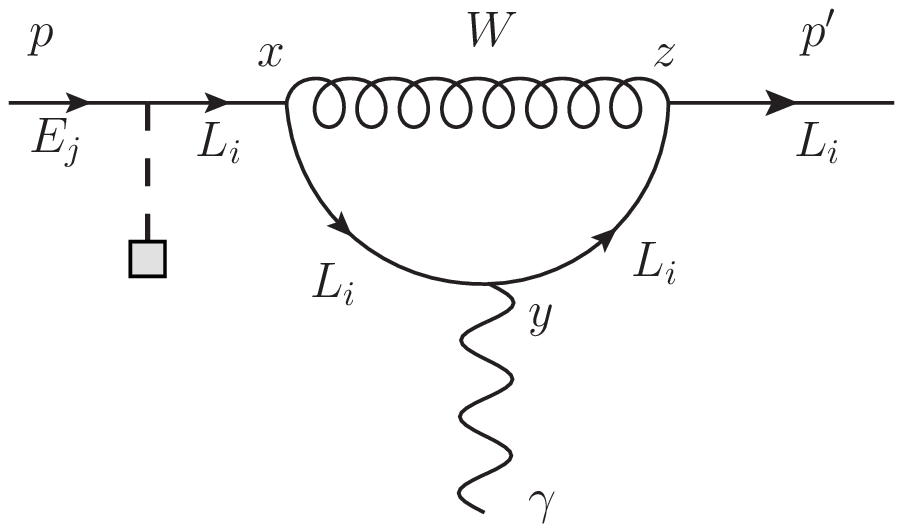,width=5.55cm,angle=0} 
\hspace*{5pt}
\epsfig{file=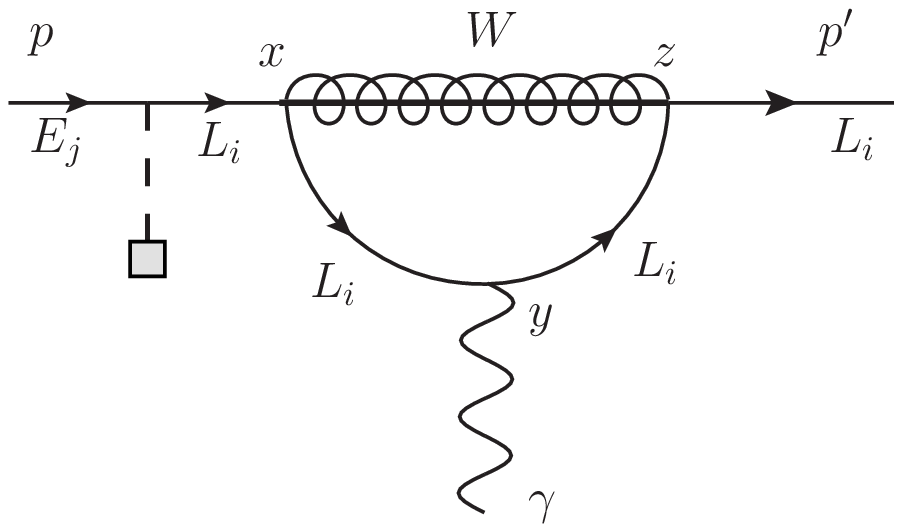,width=5.75cm,angle=0}
}
\end{center}
\vspace*{10pt} The amplitudes for the two diagrams above, ${\rm {\bf
    W1}}$ and ${\rm {\bf W2}}$, can be obtained from the amplitudes
for diagrams ${\rm {\bf B2a}}$ and ${\rm {\bf B4a}}$, respectively, by
replacing $g^{\prime \,2}_5\,Y^2_L$ with $g^2_5$ in (\ref{diag3}) and
(\ref{diag7}).
\end{figure}
\vspace*{20pt}
\begin{figure}[htbp]
\begin{center}
\parbox{6cm}{\epsfig{file=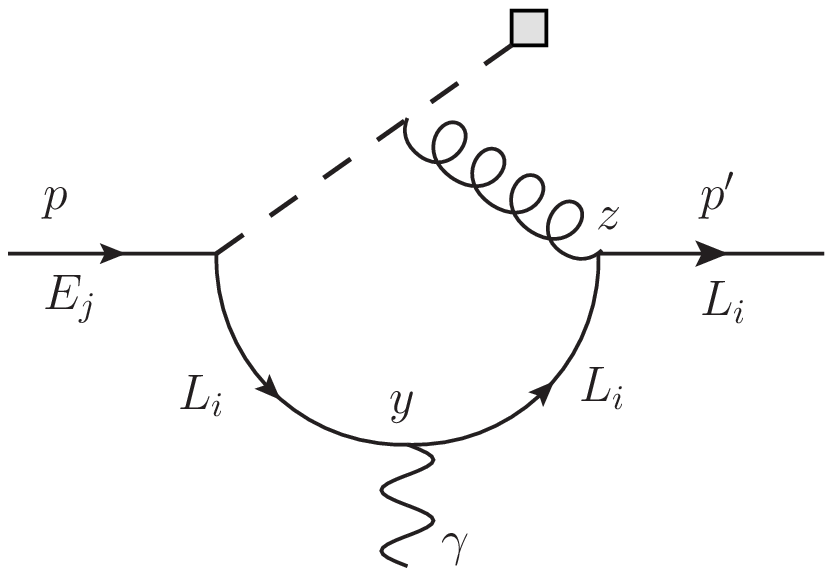,width=5.55cm,angle=0} 
}
\end{center}
\vspace*{10pt} The amplitude for the diagram above, ${\rm {\bf W3}}$,
can be obtained from the amplitude for diagram ${\rm {\bf B5a}}$ by replacing
$g^{\prime \,2}_5Y_{\Phi}Y_L$ with $g^2_5$ in (\ref{diag19}).
\end{figure}
\newpage
%%%%%%%%%%%%%%%%%%%%%%%%%%%%%%%%%%%%%%%%%%%%%%%%%
%% Other W-diagrams start here 
%%%%%%%%%%%%%%%%%%%%%%%%%%%%%%%%%%%%%%%%%%%%%%%%%
%% Dia11 %%
\newpage
\begin{figure}[htbp]
\begin{center}
\parbox{12cm}{\epsfig{file=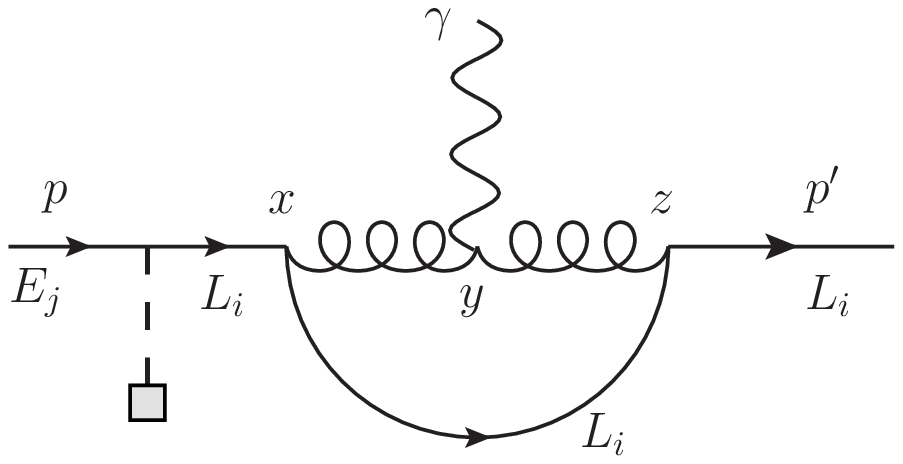,width=5.65cm,angle=0} 
\hspace*{5pt}
\epsfig{file=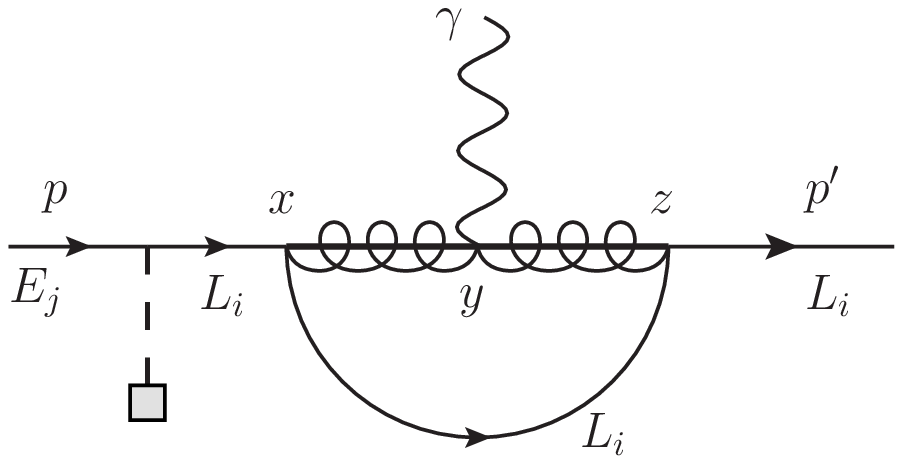,width=5.75cm,angle=0}
}
\end{center}
\end{figure}
\vspace*{-15pt}
\begin{eqnarray}
  \label{diag11}
  {\rm {\bf W4}} &=& (ig_5)^2(-e_5)\left(\frac{-iT^3}{k^3}\right)y^{\rm (5D)}_{ij}\left[\frac{\tau^a}{2}\frac{\tau^b}{2}\right]_{22} \epsilon^{ab3} \int^{1/T}_{1/k}
  \!\!\!\frac{dx}{(kx)^4} \int^{1/T}_{1/k}
  \!\!\!\frac{dy}{(ky)} \int^{1/T}_{1/k}
  \!\!\!\frac{dz}{(kz)^4}\int \!\!\frac{d^4l}{(2\pi)^4} \nonumber \\
  &&  f^{(0)}_{L_i}(z) f^{(0)}_\gamma(y) g^{(0)}_{E_j}(1/T) \epsilon^{*\mu} 
  \Delta^{\beta\lambda}_{\mbox{\tiny ZMS}}(p-l,x,y)\Delta^{\nu\alpha}_{\mbox{\tiny ZMS}}(p^{\,\prime}-l,y,z)
  \nonumber \\
  && \times
  \Big[\{(p^{\,\prime}-p)-(p-l)\}_\nu\eta_{\mu\lambda}+\{(p-l)-(l-p^{\,\prime})\}_\mu\eta_{\nu\lambda}  \nonumber \\
  && +\,\{(l-p^{\,\prime})-(p^{\,\prime}-p)\}_\lambda\eta_{\mu\nu}\Big] \times \bar{L_{i}}(p^{\prime})P_{R} \gamma_\alpha \Delta^{L}_{i}({l},z,x)
  \gamma_\beta \Delta^{L}_{i}({p},x,1/T) P_{R} E_{j}(p)
  \nonumber \\ 
  &=& - \frac{g^2_5 e_5
    y^{\rm(5D)}_{{ij}}T^3}{2k^3} \int^{1/T}_{1/k}
  \!\!\!\frac{dx}{(kx)^4} \int^{1/T}_{1/k}
  \!\!\!\frac{dy}{(ky)} \int^{1/T}_{1/k}
  \!\!\!\frac{dz}{(kz)^4}\int
  \!\!\frac{d^4l}{(2\pi)^4} \nonumber \\ && f^{(0)}_{L_i}(z) f^{(0)}_\gamma(y) g^{(0)}_{E_j}(1/T) \epsilon^{*\mu}  
  \Delta^{\beta\lambda}_{\mbox{\tiny ZMS}}(\hat{p},x,y)\Delta^{\nu\alpha}_{\mbox{\tiny ZMS}}(\hat{p}^\prime,y,z) \nonumber \\
  &&  \times \Big[(p^{\,\prime}-2p+l)_\nu\eta_{\mu\lambda}+(p-2l+p^{\,\prime})_\mu\eta_{\nu\lambda}+(l-2p^{\,\prime}+p)_\lambda\eta_{\mu\nu}\Big]
  \nonumber \\
  && \times \bar{L_{i}}(p^{\prime}) \Big[d^+F^-_{L_i}({l},z,x) d^-F^+_{L_i}(p,x,1/T) 
  \left\{\gamma_\alpha \gamma_\beta\right\}\Big] P_{R} E_{{j}}(p) 
\end{eqnarray}
%% Dia112 %%
\begin{eqnarray}
  \label{diag12}
  {\rm {\bf W5}} &=& (g_5)^2(e_5)\left(\frac{-iT^3}{k^3}\right)y^{\rm (5D)}_{ij}\left[\frac{\tau^a}{2}\frac{\tau^b}{2}\right]_{22} \epsilon^{ab3} \int^{1/T}_{1/k}
  \!\!\!\frac{dx}{(kx)^4} \int^{1/T}_{1/k}
  \!\!\!\frac{dy}{(ky)} \int^{1/T}_{1/k}
  \!\!\!\frac{dz}{(kz)^4}\int \!\!\frac{d^4l}{(2\pi)^4} \nonumber \\
  &&  f^{(0)}_{L_i}(z) f^{(0)}_\gamma(y) g^{(0)}_{E_j}(1/T) \epsilon^{*\mu} 
  \Delta_{5}(p-l,x,y)\Delta_{5}(p^{\,\prime}-l,y,z)
  \nonumber \\
  && \times \{(p-l)-(l-p^{\prime})\}_\mu \nonumber \\
  && \times \bar{L_{i}}(p^{\prime})P_{R} \gamma_5 \Delta^{L}_{i}({l},z,x)
  \gamma_5 \Delta^{L}_{i}({p},x,1/T) P_{R} E_{j}(p)
  \nonumber \\ 
  &=& - \frac{g^2_5 e_5 
    y^{\rm(5D)}_{{ij}}T^3}{2k^3} \int^{1/T}_{1/k}
  \!\!\!\frac{dx}{(kx)^4} \int^{1/T}_{1/k}
  \!\!\!\frac{dy}{(ky)} \int^{1/T}_{1/k}
  \!\!\!\frac{dz}{(kz)^4}\int
  \!\!\frac{d^4l}{(2\pi)^4} \nonumber \\ && f^{(0)}_{L_i}(z) f^{(0)}_\gamma(y) g^{(0)}_{E_j}(1/T) \epsilon^{*\mu} 
  \Delta_{5}(p-l,x,y)\Delta_{5}(p^{\,\prime}-l,y,z) \nonumber \\
  &&  \times (p-2l+p^{\prime})_\mu \, \bar{L_{i}}(p^{\prime}) \Big[d^-F^+_{L_i}({l},z,x) d^-F^+_{L_i}(p,x,1/T)\Big] P_{R} E_{{j}}(p) 
\end{eqnarray}
%%%%%%%%%%%%%%%
\newpage
\vspace*{-30pt}
\begin{figure}[htbp]
\begin{center}
\parbox{12cm}{\epsfig{file=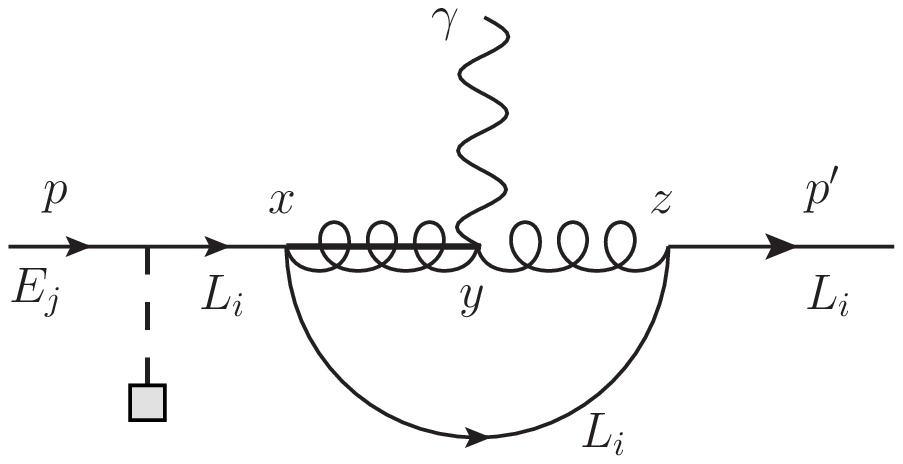,width=5.65cm,angle=0} 
\hspace*{5pt}
\epsfig{file=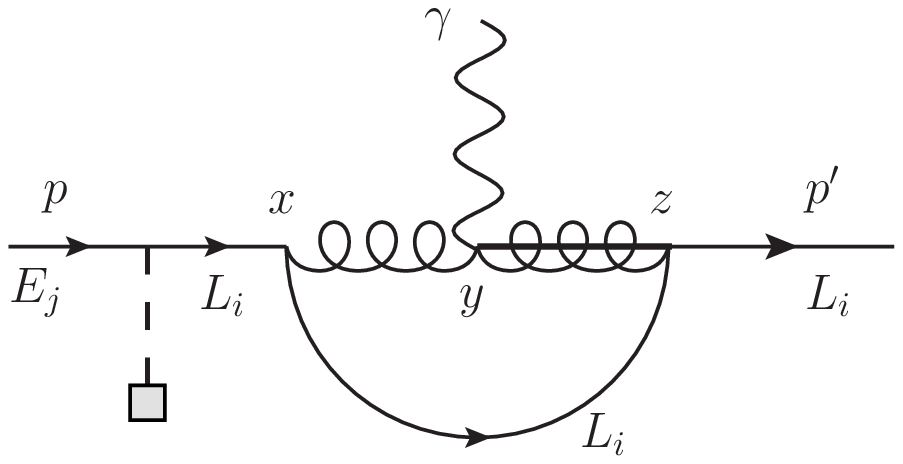,width=5.75cm,angle=0}
}
\end{center}
\end{figure}
\vspace*{-15pt}
%% Dia13 %%
\begin{eqnarray}
  \label{diag13}
  {\rm {\bf W6a}} &=& (ig_5) g_5 (-ie_5)\left(\!\frac{-iT^3}{k^3}\right)y^{\rm (5D)}_{ij}\left[\frac{\tau^a}{2}\frac{\tau^b}{2}\right]_{22} \!\!\epsilon^{b3a} \!\int^{1/T}_{1/k}
  \!\!\!\frac{dx}{(kx)^4} \int^{1/T}_{1/k}
  \!\!\!\frac{dy}{(ky)} \int^{1/T}_{1/k}
  \!\!\!\frac{dz}{(kz)^4}\int \!\!\frac{d^4l}{(2\pi)^4} \nonumber \\
  &&  f^{(0)}_{L_i}(z) g^{(0)}_{E_j}(1/T) \epsilon^{*\mu} 
  \Delta_{5}(p-l,x,y)
  \nonumber \\
  && \times \eta_{\beta\mu}\left[\Delta^{\alpha\beta}_{\mbox{\tiny ZMS}}(p^{\,\prime}-l,y,z) \frac{\partial}{\partial y} f^{(0)}_\gamma(y) 
    - f^{(0)}_\gamma(y) \frac{\partial}{\partial y} \Delta^{\alpha\beta}_{\mbox{\tiny ZMS}}(p^{\,\prime}-l,y,z) \right] 
  \nonumber \\
  && \times \bar{L_{i}}(p^{\prime})P_{R} \gamma_\alpha \Delta^{L}_{i}({l},z,x)
  \gamma_5 \Delta^{L}_{i}({p},x,1/T) P_{R} E_{j}(p)
  \nonumber \\[0.1cm] 
  &=& -\frac{g^2_5 e_5 
    y^{\rm(5D)}_{{ij}}T^3}{2k^3} \int^{1/T}_{1/k}
  \!\!\!\frac{dx}{(kx)^4} \int^{1/T}_{1/k}
  \!\!\!\frac{dy}{(ky)} \int^{1/T}_{1/k}
  \!\!\!\frac{dz}{(kz)^4}\int
  \!\!\frac{d^4l}{(2\pi)^4} \nonumber \\ && f^{(0)}_{L_i}(z)f^{(0)}_\gamma(y)g^{(0)}_{E_j}(1/T) \epsilon^{*}_{\mu} 
  \Delta_{5}(p-l,x,y)\times 
  \left[ \frac{\partial}{\partial y} \Delta^{\alpha\mu}_{\mbox{\tiny ZMS}}(p^{\,\prime}-l,y,z) \right]
  \nonumber \\
  && \times \bar{L_{i}}(p^{\prime}) \Big[F^+_{L_i}(l,z,x) d^-F^+_{L_i}(p,x,1/T) \left(\gamma_\alpha\!\!\!\not l\right)\Big] P_{R} E_{{j}}(p) 
\end{eqnarray}
%% Dia14 %%
\begin{eqnarray}
  \label{diag14}
  {\rm {\bf W6b}} &=& g_5 (ig_5)(-ie_5)\left(\!\frac{-iT^3}{k^3}\right)y^{\rm (5D)}_{ij}\left[\frac{\tau^a}{2}\frac{\tau^b}{2}\right]_{22} \!\!\epsilon^{a3b} \!\int^{1/T}_{1/k}
  \!\!\!\frac{dx}{(kx)^4} \int^{1/T}_{1/k}
  \!\!\!\frac{dy}{(ky)} \int^{1/T}_{1/k}
  \!\!\!\frac{dz}{(kz)^4}\int \!\!\frac{d^4l}{(2\pi)^4} \nonumber \\
  &&  f^{(0)}_{L_i}(z) g^{(0)}_{E_j}(1/T) \epsilon^{*\mu} 
  \Delta_{5}(p^{\,\prime}-l,y,z)
  \nonumber \\
  && \times \eta_{\beta\mu}\left[\Delta^{\alpha\beta}_{\mbox{\tiny ZMS}}(p-l,x,y) \frac{\partial}{\partial y} f^{(0)}_\gamma(y) 
    - f^{(0)}_\gamma(y) \frac{\partial}{\partial y} \Delta^{\alpha\beta}_{\mbox{\tiny ZMS}}(p-l,x,y) \right] 
  \nonumber \\
  && \times \bar{L_{i}}(p^{\prime})P_{R} \gamma_5 \Delta^{L}_{i}({l},z,x)
  \gamma_{\alpha} \Delta^{L}_{i}({p},x,1/T) P_{R} E_{j}(p)
  \nonumber \\[0.1cm] 
  &=& \frac{g^2_5 e_5 
    y^{\rm(5D)}_{{ij}}T^3}{2k^3} \int^{1/T}_{1/k}
  \!\!\!\frac{dx}{(kx)^4} \int^{1/T}_{1/k}
  \!\!\!\frac{dy}{(ky)} \int^{1/T}_{1/k}
  \!\!\!\frac{dz}{(kz)^4}\int
  \!\!\frac{d^4l}{(2\pi)^4} \nonumber \\ && f^{(0)}_{L_i}(z)f^{(0)}_\gamma(y) g^{(0)}_{E_j}(1/T) \epsilon^{*}_{\mu} 
  \Delta_{5}(p^{\,\prime}-l,y,z)\times 
  \left[\frac{\partial}{\partial y} \Delta^{\alpha\mu}_{\mbox{\tiny ZMS}}(p-l,x,y) \right] 
  \nonumber \\
  && \times \bar{L_{i}}(p^{\prime}) \Big[F^-_{L_i}(l,z,x) d^-F^+_{L_i}(p,x,1/T) \left(\!\!\not l\gamma_\alpha\right)\Big] P_{R} E_{{j}}(p) 
\end{eqnarray}
%%%%%%%%%%%%%%%%%%%%%%%%%%%%%%%%%%%%%%%%%%%%%%%%%
%% H-diagrams start here 
%%%%%%%%%%%%%%%%%%%%%%%%%%%%%%%%%%%%%%%%%%%%%%%%%
\newpage
\begin{figure}[htbp]
\begin{center}
\parbox{12cm}{\epsfig{file=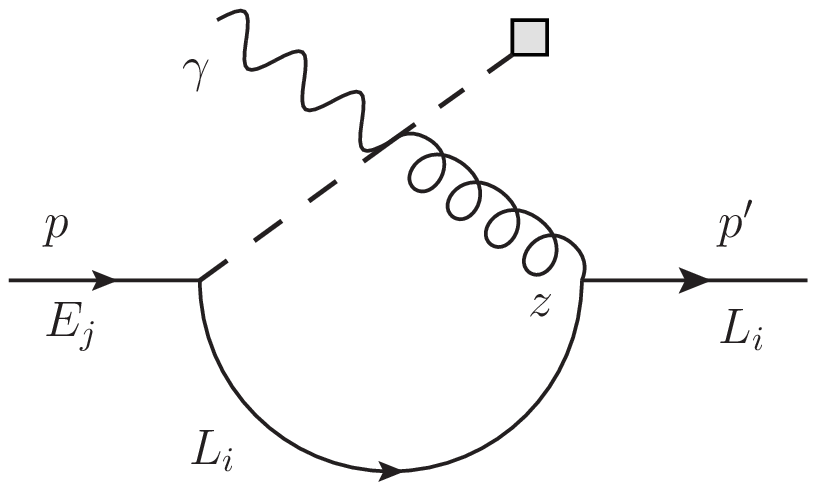,width=5.55cm,angle=0} 
\hspace*{5pt}
\epsfig{file=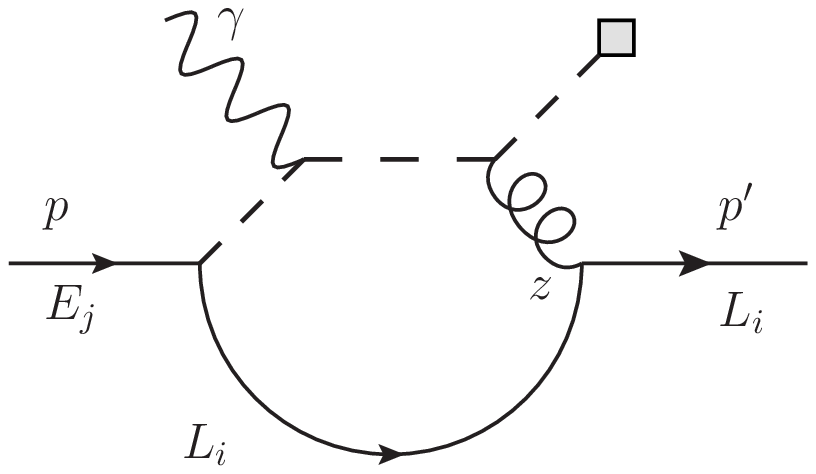,width=5.65cm,angle=0}
}
\end{center}
\end{figure}
\vspace*{-15pt}
%% Dia15 %%
\begin{eqnarray}
  \label{diag15}
  {\rm {\bf W7}} &=& (ig_5)^2(e_5) \left(\tau^a \left[\frac{Y_{\Phi}}{2}\frac{\tau^a}{2} + 
    \frac{1}{4}\delta^{a3}\right] \right)_{22}
  \left(\frac{-iT^3}{k^3}\right)y^{\rm (5D)}_{ij}\int^{1/T}_{1/k}
  \!\!\!\frac{dz}{(kz)^4}\int \!\!\frac{d^4l}{(2\pi)^4} \nonumber \\
  &&  f^{(0)}_{L_i}(z) f^{(0)}_\gamma(1/T) g^{(0)}_{E_j}(1/T) \epsilon^{*}_{\mu} 
  \Delta_{H}(p-l) \Delta^{\mu\nu}_{\mbox{\tiny ZMS}}(p^{\,\prime}-l,1/T,z) 
  \nonumber \\
  && \times \bar{L_{i}}(p^{\prime})P_{R} \gamma_\nu \Delta^{L}_{i}({l},z,1/T) P_{R} E_{j}(p)
  \nonumber \\ 
  &=& -i\frac{g^2_5 e_5 y^{\rm(5D)}_{{ij}}T^3}{2k^3} \int^{1/T}_{1/k}
  \!\!\!\frac{dz}{(kz)^4}\int
  \!\!\frac{d^4l}{(2\pi)^4} \nonumber \\ && f^{(0)}_{L_i}(z) f^{(0)}_\gamma(1/T) g^{(0)}_{E_j}(1/T) \epsilon^{*}_{\mu} 
  \Delta_{H}(p-l) \Delta^{\mu\nu}_{\mbox{\tiny ZMS}}(p^{\,\prime}-l,1/T,z) 
  \nonumber \\
  && \times \bar{L_{i}}(p^{\prime}) \Big[F^+_{L_i}(l,z,1/T) \left(\gamma_\nu\!\!\!\not l\right)\Big] P_{R} E_{{j}}(p) 
\end{eqnarray}
%% Dia16 %%
\begin{eqnarray}
  \label{diag16}
  {\rm {\bf W8}} &=& (ig_5)^2 (ie_5)
\left(\frac{\tau^a}{2} \left[\frac{Y_{\Phi}}{2}+\frac{\tau^3}{2}\right]\frac{\tau^a}{2}\right)_{22} 
\left(\frac{-iT^3}{k^3}\right)y^{\rm (5D)}_{ij}\int^{1/T}_{1/k}
  \!\!\!\frac{dz}{(kz)^4}\int \!\!\frac{d^4l}{(2\pi)^4} \nonumber \\
  &&  f^{(0)}_{L_i}(z) f^{(0)}_\gamma(1/T) g^{(0)}_{E_j}(1/T) \epsilon^{*\mu} 
  \Delta_{H}(p-l) \Delta_{H}(p^{\,\prime}-l) \Delta^{\alpha\beta}_{\mbox{\tiny ZMS}}(p^{\,\prime}-l,1/T,z) 
  \nonumber \\
  && \times \bar{L_{i}}(p^{\prime})P_{R} \gamma_\alpha \Delta^{L}_{i}({l},z,1/T) \,(l-p^{\prime})_\beta 
  \left\{(l-p)+(l-p^{\prime})\right\}_\mu P_{R} E_{j}(p)
  \nonumber \\ 
  &=& \frac{g^2_5 e_5 y^{\rm(5D)}_{{ij}}T^3}{2k^3} \int^{1/T}_{1/k}
  \!\!\!\frac{dz}{(kz)^4}\int
  \!\!\frac{d^4l}{(2\pi)^4} \nonumber \\ && f^{(0)}_{L_i}(z) f^{(0)}_\gamma(1/T) g^{(0)}_{E_j}(1/T) \epsilon^{*\mu} 
  \Delta_{H}(p-l) \Delta_{H}(p^{\,\prime}-l) \Delta^{\alpha\beta}_{\mbox{\tiny ZMS}}(p^{\,\prime}-l,1/T,z) 
  \nonumber \\
  && \times \bar{L_{i}}(p^{\prime}) \Big[F^+_{L_i}(l,z,1/T) \left(\gamma_\alpha\!\!\!\not l\right)(p^{\,\prime}-l)_\beta 
  \left(p+p^{\,\prime}-2l\right)_\mu\Big] P_{R} E_{{j}}(p) 
\end{eqnarray}
\newpage
\vspace*{-30pt}
\begin{figure}[htbp]
\begin{center}
\parbox{12cm}{\epsfig{file=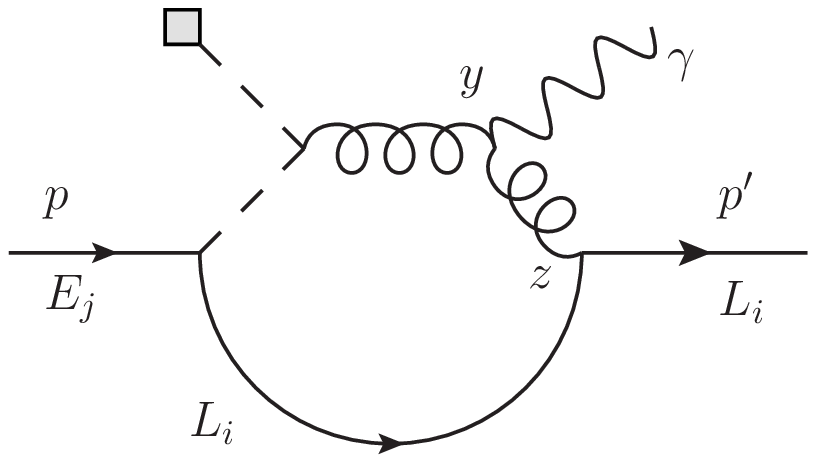,width=5.55cm,angle=0} 
\hspace*{5pt}
\epsfig{file=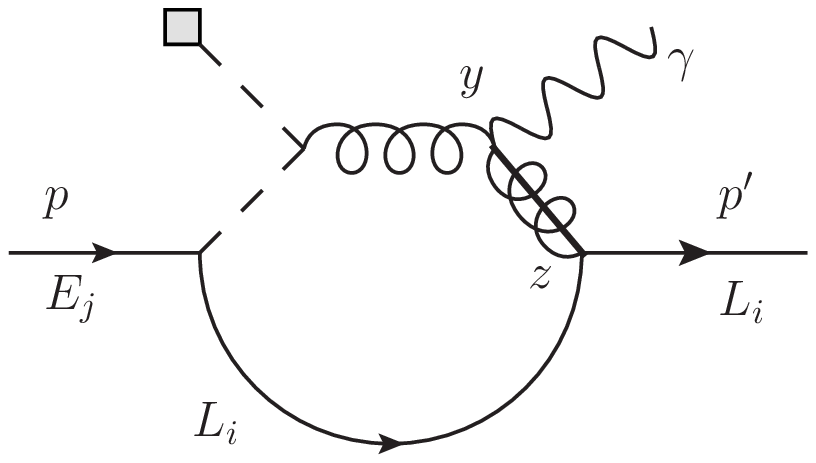,width=5.65cm,angle=0}
}
\end{center}
\end{figure}
\vspace*{-15pt}
%% Dia17 %%
\begin{eqnarray}
  \label{diag17}
  {\rm {\bf W9}} &=& (ig_5)^2(-e_5)\left[\frac{\tau^a}{2}\frac{\tau^b}{2}\right]_{22} \epsilon^{ab3} 
  \left(\frac{-iT^3}{k^3}\right)y^{\rm(5D)}_{ij}~\int^{1/T}_{1/k}
  \!\!\!\frac{dy}{(ky)}\int^{1/T}_{1/k}
  \!\!\!\frac{dz}{(kz)^4}\int \!\!\frac{d^4l}{(2\pi)^4} \nonumber \\
  &&  f^{(0)}_{L_i}(z) f^{(0)}_\gamma(y) g^{(0)}_{E_j}(1/T) \epsilon^{*\mu} 
  \Delta_{H}(p-l) \Delta^{\beta\lambda}_{\mbox{\tiny ZMS}}(p-l,1/T,y) \Delta^{\nu\alpha}_{\mbox{\tiny ZMS}}(p^{\,\prime}-l,y,z)
  \nonumber \\
  && \times
  \Big[\{(p^{\,\prime}-p)-(p-l)\}_\nu\eta_{\mu\lambda}+\{(p-l)-(l-p^{\,\prime})\}_\mu\eta_{\nu\lambda}
 \nonumber\\
 && +\,\{(l-p^{\,\prime})-(p^{\,\prime}-p)\}_\lambda\eta_{\mu\nu}\Big]
\times \bar{L_{i}}(p^{\prime})P_{R} \gamma_\alpha \Delta^{L}_{i}({l},z,1/T)(l-p)_\beta P_{R} E_{j}(p)
  \nonumber \\[0.1cm] 
  &=& \frac{g^2_5 e_5  y^{\rm(5D)}_{{ij}}T^3}{2k^3} \int^{1/T}_{1/k}
  \!\!\!\frac{dy}{(ky)}\int^{1/T}_{1/k}
  \!\!\!\frac{dz}{(kz)^4}\int
  \!\!\frac{d^4l}{(2\pi)^4} \nonumber \\ && f^{(0)}_{L_i}(z) f^{(0)}_\gamma(y) g^{(0)}_{E_j}(1/T) \epsilon^{*\mu} 
  \Delta_{H}(p-l) \Delta^{\beta\lambda}_{\mbox{\tiny ZMS}}(p-l,1/T,y) \Delta^{\nu\alpha}_{\mbox{\tiny ZMS}}(p^{\,\prime}-l,y,z) 
  \nonumber \\
  && \times \Big[(p^{\,\prime}-2p+l)_\nu\eta_{\mu\lambda}+(p-2l+p^{\,\prime})_\mu\eta_{\nu\lambda}+(l-2p^{\,\prime}+p)_\lambda\eta_{\mu\nu}\Big]
  \nonumber \\
  && \times \bar{L_{i}}(p^{\prime}) \Big[F^+_{L_i}(l,z,1/T) \left(\gamma_\alpha\!\!\!\not l\right)(l-p)_\beta\Big] P_{R} E_{{j}}(p) 
\end{eqnarray}
%% Dia18 %%
\begin{eqnarray}
  \label{diag18}
  {\rm {\bf W10}} &=& (g_5)(ig_5)(-ie_5)\left(\frac{-iT^3}{k^3}\right)y^{\rm (5D)}_{ij}\left[\frac{\tau^a}{2}\frac{\tau^b}{2}\right]_{22} \epsilon^{a3b} \int^{1/T}_{1/k}
  \!\!\!\frac{dy}{(ky)} \int^{1/T}_{1/k}
  \!\!\!\frac{dz}{(kz)^4}\int \!\!\frac{d^4l}{(2\pi)^4} \nonumber \\
  &&  f^{(0)}_{L_i}(z) g^{(0)}_{E_j}(1/T) \epsilon^{*\mu} 
  \Delta_{H}(p-l) \Delta_{5}(p^{\,\prime}-l,y,z)
  \nonumber \\
  && \times \eta_{\lambda\mu}\left[\Delta^{\beta\lambda}_{\mbox{\tiny ZMS}}(p-l,1/T,y) \frac{\partial}{\partial y} f^{(0)}_\gamma(y) 
    - f^{(0)}_\gamma(y) \frac{\partial}{\partial y} \Delta^{\beta\lambda}_{\mbox{\tiny ZMS}}(p-l,1/T,y) \right] 
  \nonumber \\
  && \times \bar{L_{i}}(p^{\prime})P_{R} \gamma_5 \Delta^{L}_{i}({l},z,1/T)(l-p)_\beta P_{R} E_{j}(p)
  \nonumber \\ 
  &=& -\frac{g^2_5 e_5 
    y^{\rm(5D)}_{{ij}}T^3}{2k^3} \int^{1/T}_{1/k}
  \!\!\!\frac{dy}{(ky)} \int^{1/T}_{1/k}
  \!\!\!\frac{dz}{(kz)^4}\int
  \!\!\frac{d^4l}{(2\pi)^4} \nonumber \\ && f^{(0)}_{L_i}(z) f^{(0)}_\gamma(y) g^{(0)}_{E_j}(1/T) \epsilon^{*\mu} 
  \Delta_{H}(p-l) \Delta_{5}(p^{\,\prime}-l,y,z)
  \nonumber \\
  && \times \left[\frac{\partial}{\partial y} \Delta^{\beta\mu}_{\mbox{\tiny ZMS}}(p-l,1/T,y) \right] \!\times \bar{L_{i}}(p^{\prime}) \Big[d^-F^+_{L_i}(l,z,1/T)(l-p)_\beta\Big] P_{R} E_{{j}}(p) 
\end{eqnarray}

%%%%%%%%%%%%%%%%%%%%%%%%%%%%%%%%%%%%%%%%%%%%%%%%%%%%%%%%%%%%%%%%%%%

\end{document}